\documentclass[11pt,letter]{article}
\newtheorem{theorem}{Theorem}[section]
\newtheorem{definition}{Definition}[section]

\newtheorem{lemma}{Lemma}[section]
\newtheorem{proposition}{Proposition}[section]

\newtheorem{corollary}{Corollary} [section]

\usepackage{epsfig}

\usepackage{mathrsfs}
\usepackage{amsfonts}
\usepackage{bbding}
\usepackage{pifont}

\usepackage{amsmath}
\usepackage{amssymb}

\usepackage{color}
\definecolor{mygray}{gray}{.6}

\usepackage{multirow}

\begin{document}



\title{\vspace{-1cm} A New Family of  Practical Non-Malleable  Diffie-Hellman Protocols}



\author{Andrew  C.   Yao \footnote{Institute for Interdisciplinary Information Sciences,   Tsinghua University, Beijing,
China. \quad \texttt{andrewcyao@tsinghua.edu.cn}}
 \and  Yunlei Zhao\footnote{Contact author.
 Software School,   Fudan University, Shanghai 200433,
 China.   \quad \texttt{ylzhao@fudan.edu.cn}}}

\date{}



\date{}          
\maketitle

\vspace{-1cm}
\begin{abstract}
Cryptography algorithm standards play a key role  both to  the practice 
 of  information security and  to cryptography theory research.  Among them, the MQV and HMQV protocols ((H)MQV, in short) are a family of (implicitly authenticated)  Diffie-Hellman key-exchange (DHKE) protocols that are widely standardized and deployed.
In this work, from some new perspectives and approaches and under some  new design rationales and insights, we develop  a new family of practical 
implicitly authenticated DHKE protocols, 
which 
enjoy notable
 performance among security, privacy, efficiency and easy deployment. 
 We make detailed comparisons between our new DHKE protocols and (H)MQV, 
  showing that the newly developed  protocols outperform HMQV in most aspects. Along the way, guided by our new design rationales, we also identify a new vulnerability  of (H)MQV, which brings some new perspectives (e.g., session-key 
   computational fairness) to the literature.

\end{abstract}

\vspace{-0.3cm}
\section{Introduction}       
\vspace{-0.1cm}

Diffie-Hellman key-exchange (DHKE)
 protocols \cite{DH76}  are at
the root of public-key cryptography, and  are one of 
 the main pillars of
both theory and practice of
cryptography \cite{CK01}. Among them, the (H)MQV  
 protocols \cite{MQV95,LMQSV03,K05,M05} 
 are among  the most efficient 
 DHKE protocols that provide (implicit) mutual authentications based upon public-key cryptography, and are widely standardized  \cite{A01,A63,I00,I02,N03,N04,S05}. In particular, it  has  been
 announced by the US National Security Agency  as the key
 exchange mechanism underlying ``the next generation cryptography
 to protect US government information", 
 including  the  protection of
 ``classified or mission critical national security
 information" \cite{N04,K05}.


 Despite its   seemingly
conceptual simplicity, 
designing ``\emph{sound}" and ``\emph{right}" DHKE protocols turns
out to be extremely error prone and can be notoriously subtle, particularly witnessed by the evolution  history of (H)MQV \cite{MQV95,K01,LMQSV03,K05,M05}.  Also, the analysis of even a simple cryptographic protocol in  intricate adversarial settings like the Internet can be a luxury and dauntingly complex task \cite{C06,K05}.  
The reason for this is the high
system complexity and enormous number of subtleties surrounding the design, definition and analysis  of DHKE protocols.
Given the intensive investigation  of (H)MQV both from cryptography theory  research and from industrial engineering, it may be commonly suggested that the state-of-the-art of (H)MQV, commonly viewed  as the best available  in the integrity of security and protocol efficiency,  should hardly be broken.   


In this work, we start with  investigating  highly practical mechanisms in the random oracle (RO) model, referred to as non-malleable joint proof-of-knowledge (NMJPOK) for presentation simplicity,  for proving DH-knowledges, say both the  secret-key and the DH-exponent, \emph{jointly} and  \emph{non-malleably}  in concurrent settings like the Internet.
 In light of this line of investigations, we develop  a new family of 
practical 
implicitly authenticated        DHKE protocols, referred to as OAKE   \footnote{There are two acronym  interpretations of OAKE. One  interpretation is:  (Online) Optimal (implicitly) Authenticated (Diffie-Hellman) Key-Exchange. Another interpretation is:  (Toward) Optimally-balanced  (implicitly) Authenticated (Diffie-Hellman) Key-Exchange   (in the integrity of protocol efficiency, security, privacy and easy deployment).} and single-hash OAKE protocols,
which enjoy notable  performance among security, privacy, efficiency and easy deployment. 
 For presentation simplicity, we refer to the newly developed DHKE protocols as (s)OAKE.
  We then  compare and justify (s)OAKE protocols with (H)MQV in detail, which shows that our new protocols outperform HMQV in most aspects.
  Detailed comparisons are listed  in Section \ref{SecYZvsHMQV}  after motivating the design rationales and building tools and after  presenting  the detailed OAKE specifications. 
  Guided by our new design rationales, in this work we particularly   identify a new vulnerability of  (H)MQV  beyond the Canetti-Krawczyk (CK) framework, which brings some new perspectives (e.g., session-key  computational fairness) to the literature.
  We do not know how to fix (H)MQV against this newly identified vulnerability  without sacrificing the provable security in the CK framework and many more other advantages enjoyed by (s)OAKE (with details referred to Section \ref{SecBeyondCK}),  
  which also further justifies and highlights the careful design of (s)OAKE.

%

%
%

 We suggest the developed (s)OAKE protocols are themselves a clear witness to the usefulness of the new design rationales and building  tool with NMJPOK, as (s)OAKE  aims for an alternative of (H)MQV that is widely standardized and deployed and as with the new design rationales we can identify some new vulnerabilities  bringing new perspectives to the literature of DHKE.
 But at the same time,  the new design rationales and building tools, developed for (s)OAKE, can also  be of independent interest, and may trigger more 
   applications. In particular, based on this work,  in a subsequent separate work we present the definition and candidates of non-malleable extractable one-way functions (NME-OWF), which can be viewed as pairing-based NMJPOK without random oracles, and demonstrate the applications of NME-OWF to both theory (e.g., 3-round concurrent non-malleable zero-knowledge, etc)  and applications (e.g., ID-based cryptography, etc) of cryptography.

\vspace{-0.2cm}
\section{Preliminaries}
\vspace{-0.1cm}
\textbf{Notations:}
If \emph{A} is
 a probabilistic algorithm, then $A(x_1, x_2, \cdots; r)$ is the result of running \emph{A} on inputs
 $x_1, x_2, \cdots$ and coins $r$. We let $y\leftarrow A(x_1, x_2, \cdots; r)$ denote the experiment of picking $r$ at
 random and letting $y$ be $A(x_1, x_2, \cdots; r)$. If $S$ is a finite set then $x\leftarrow S$,  sometimes also written  as $x\in_{\textup{R}} S$, is the operation of
 picking an element uniformly from $S$. If $\alpha$ is neither an algorithm nor a set then $x\leftarrow \alpha$ is a
 simple assignment statement.


Let $G^{\prime}$  be a finite Abelian group
of order $N$,   $G$ be
 a subgroup  of prime  order
$q$ in $G^{\prime}$. Denote by $g$ a generator of $G$, by $1_G$ the identity
 element, by $G\setminus 1_G=G-\{1_G\}$ the set of elements of $G$ except $1_G$ 
 and by $t=\frac{N}{q}$ the cofactor. 
In this work, we use multiplicative notation for the group
 operation in $G^{\prime}$.
 We assume the computational Diffie-Hellman (CDH) assumption holds over $G$, which says that given $X=g^x, Y=g^y \leftarrow G$ (i.e., each of $x$ and $y$ is taken uniformly at random from $Z_q$) no efficient (say, probabilistic polynomial-time) algorithm can compute $CDH(X,
Y)=g^{xy}$. 
       Let $(A=g^a,a)$ (resp., $(X=g^x,x)$) be the public-key and secret-key (resp., the
DH-component and DH-exponent)  of  
 player $\hat{A}$,  
  and $(B=g^b,b)$ (resp., $(Y=g^y,y)$) be the public-key and secret-key (resp., the  DH-component and DH-exponent) of 
   player $\hat{B}$, 
  where $a,  x, b, y$ are taken randomly and
independently  from
$Z^*_q$. 
 (H)MQV is
 recalled in Figure \ref{OAKE-MQVec} (page \pageref{OAKE-MQVec}), and the (H)MQV  variants are recalled in
 Appendix \ref{hmqvvariants}, where on a security parameter $k$
 $H_K$ (resp., $h$)  is  a hash function of $k$-bit (resp., $l$-bit) output and   $l$ is set to be $|q|/2$.

\textbf{Gap Diffie-Hellman (GDH) assumption \cite{OP01}.}
Let $G$ be a cyclic group generated by an element $g$, and a
decision predicate algorithm  $\mathcal{O}$  be a  (\emph{full})
\textsf{Decisional Diffie-Hellman (DDH) Oracle} for the group $G$
and generator $g$ such that on input  $(U, V, Z)$, for \emph{arbitrary} $(U, V) \in
G^2$, oracle $\mathcal{O}$ outputs 1 if and only if $Z=CDH(U, V)$.
We say the GDH assumption holds in $G$ if for any polynomial-time
CDH solver for $G$, the probability that on a pair of random elements  $(X, Y)\leftarrow G$ the solver computes the correct value
$CDH(X, Y)$ is negligible, even when the algorithm is provided
with the (full)  DDH-oracle $\mathcal{O}$ for $G$. The probability
is taken over the random coins of the solver, and the choice of
$X, Y$ (each one of them is taken uniformly  at random in $G$).

%
%
%
%


\textbf{Knowledge-of-Exponent Assumption (KEA).} Informally speaking, the KEA assumption says that, suppose on input $(g, C=g^c)$, where  $c$ is taken uniformly at random from $Z^*_q$,  a probabilistic polynomial-time (PPT) algorithm $\mathcal{A}$ outputs $(Y, Z=Y^c)\in G^2$, then the discrete logarithm $y$ of $Y=g^y$ can be efficiently extracted from the input $(g,C)$ and the random coins used by $\mathcal{A}$. The formal definition is referred to Definition \ref{DefKEA} (page \pageref{DefKEA}).  In other words, given $(g,C=g^c)$   the  ``only way" to produce $(Y,Z=Y^c)$ is to choose $y$ and compute $(Y=g^y, Z=C^y)$. The KEA assumption is derived from the CDH assumption, and  is a
\emph{non-black-box} assumption by nature \cite{BP04c}.
 The KEA assumption was introduced in
\cite{D91}, and has been used in many subsequent works (e.g.,
\cite{HT98,BP04a,BP04c,DG05,K05,D06,DGK06}, etc). In particular, the
KEA assumption  plays a critical role for provable deniability of
authentication and key-exchange (e.g., \cite{DG05,K05,DGK06}).

\textbf{Simultaneous exponentiation.}  Given two generators $g_1,g_2\in G$ and two values $x,y \in Z_q$, the computation of $g_1^xg_2^y$ amounts to about 1.3 exponentiations by the simultaneous exponentiation techniques \cite{MOV95,G98,DJM00}.

\vspace{-0.2cm}
\section{Design of (s)OAKE:  Motivation, Discussion and Specification} 
\label{SSJPOKsec}
\vspace{-0.1cm}



We consider an adversarial setting, where polynomially many instances (i.e., sessions) of a Diffie-Hellman protocol $\langle \hat{A}, \hat{B}\rangle$  are run concurrently over an asynchronous network  like the Internet.
To distinguish concurrent
sessions,  each session   run
at the side of an uncorrupted  player is  labeled by a tag, which
is  the concatenation, 
in the order of session initiator and
then session
responder,
of players' identities/public-keys and   
   DH-components available from the session transcript.  A session-tag is complete if it consists of a complete set of all these components.

In this work, we study the mechanisms, in the random oracle (RO) model,  for  \emph{non-malleably} and \emph{jointly}   proving the knowledge  of  both $b$ and $y$  w.r.t. a challenge DH-component $X$ between the prover $\hat{B}$ (of public-key $B=g^b$ and DH-component $Y=g^y$) and the verifier  $\hat{A}$ (who presents the challenge DH-component $X=g^x$), where $b,y,x\in Z^*_q$. For presentation simplicity, such protocol mechanism is referred to as $JPOK{(b,y)}$.
Moreover, we look for solutions  of $JPOK_{(b,y)}$ such that $JPOK_{(b,y)}$ can be efficiently computed with  one single exponentiation by the knowledge prover.
 Note that the tag for a complete session of  $JPOK_{(b,y)}$ 
  is $(\hat{A},\hat{B},B,X,Y)$. 
 The possibility of  $JPOK_{(b,y)}$  without ROs (based upon pairings) is left  to be studied in a subsequent separate paper. 
 Throughout this work, we use a hash function $h$, which is modeled as a random oracle, and we denote  by the output length,  i.e., $l$, of $h$ as the security parameter.

 One naive solution of $JPOK_{(b,y)}$ is just to set $JPOK_{(b,y)}=X^b\cdot X^y=X^{b+y}$. But, such a naive solution is totally insecure, for example, an adversary $\mathcal{A}$  can easily impersonate the prover
  $\hat{B}$ and pre-determine 
   $JPOK_{(b,y)}$ to be $1_G$, by  setting $Y=B^{-1}$. The underlying reason is:   $\mathcal{A}$
can malleate $B$ and $Y$ into $X^{y+b}$ \emph{by maliciously
correlating the values of $y$ and $b$}, but actually without
knowing  either of them.   A further remedy of  this situation is to mask the exponents $b$ and $y$ by
some random values. In this case, the proof is denoted as $JPOK_{(b,
y)}=X^{db+ey}$, where $d$ and $e$ are random values  (e.g.,  $d=h(X, \hat{B})$  and $e=h(Y, \hat{A})$ as in HMQV in the RO model).
  The
intuition with this remedy solution is: since $d$ and $e$ are
random values, 
 $db$ and $ey$ are also random (even if the values  $Y$ and $B$, and thus the values
of $y$ and $b$, may be  maliciously correlated). 
  This  intuition however
 turns out also to be wrong \emph{in general}. With the values $d=h(B,\hat{A})$  and $e=h(X,\hat{B})$
 as an illustrative example, after receiving $X$ an adversary $\mathcal{A}$ can  generate and send
$Y=B^{-d/e}$, and in this case $JPOK_{(b, y)}=X^{db+ey}=1_G$. This  shows that masking  $b$ and $y$
 by random values is also not sufficient 
 for ensuring the non-malleability
 of  $JPOK_{(b,y)}$.  
 The key point here is that the values $db$ and $ey$
  are \emph{not} necessarily  \emph{independent}, and thus a malicious prover can still  make the values $db$ and $ey$ correlated. This line of investigations bring us to the following two candidates for  non-malleable joint proof-of-knowledge (NMJPOK) of  both $b$ and $y$ w.r.t. $X$, under the preference of on-line efficiency and minimal use of RO. More details are referred to Appendix  \ref{AppSSJPOK}. 

  \vspace{-0.15cm}
\begin{itemize} \item NMJPOK:
$NMJPOK_{(b,y)}=X^{db+ey}$, where $d=h(B, X)$
and $e=h(X, Y)$;
\vspace{-0.15cm}
\item   Single-hash NMJPOK (sNMJPOK):
$sNMJPOK_{(b,y)}=X^{db+ey}$, where $d=1$ and  $e=h(B,  X, Y)$.

\end{itemize}
\vspace{-0.2cm}

  Below, we provide some informal justifications of $NMJPOK$ and $sNMJPOK$, by avoiding introducing and employing some cumbersome  terminologies for easier  interpretation. Formal treatments are referred to Appendix  \ref{AppSSJPOK}.
 Informally speaking, the underlying rationale of $NMJPOK_{(b,y)}$ is: given a random challenge $X$,
 no matter how a malicious $\hat{B}$ chooses the values $Y=g^y$ and $B=g^b$ (where 
  $y$ and $b$ can be arbitrarily correlated), it actually has no control over the values $db$ and $ey$ in the RO model (by the birthday paradox). That is, it is infeasible for a malicious $\hat{B}$ to set $db$ (resp., $ey$) to some predetermined value, which may be determined by $ey$ (resp., $db$) via some predetermined polynomial-time computable relation $\mathcal{R}$,
 with non-negligible probability  in the RO model in order to make the values $db$ and $ey$ correlated.  Alternatively speaking, given a random challenge $X$, it is infeasible for a malicious $\hat{B}$  to output $B=g^b$ and $Y=g^y$  such that the values $db$ and $ey$ satisfy some predetermined 
 relation $\mathcal{R}$ with non-negligible probability in the RO model.

 The situation with $sNMJPOK_{(b,y)}$ is a bit different. Though as in $NMJPOK_{(b,y)}$, the malicious  $\hat{B}$ is infeasible to set $ey$ to a  predetermined value,   $\hat{B}$ can always set the value $db=b$ at its wish as $d=1$ for $sNMJPOK_{(b,y)}$. But,  $\hat{B}$ is still infeasible to set the value $b$ correlated to $ey=h(B,X,Y)y$, particularly because the value $B$ is put into the input of $e$. Specifically, for any value $B=g^b$ 
   set by $\hat{B}$, with the goal of making $b$ and $ey$ correlated, the probability that the values $ey=h(B,X,Y)y$ and $b$  satisfy some predetermined (polynomial-time computable) relation $\mathcal{R}$  is negligible in the RO model (by the birthday paradox).  In particular, the probability that $\Pr[b=f(ey)]$  or $\Pr[f(b)=ey]$, where $f$ is some predetermined polynomial-time computable function (that is in turn determined by the predetermined  relation $\mathcal{R}$), is negligible  in the RO model, no matter how the malicious $\hat{B}$ does.

\begin{figure}[t]
\vspace{-6.8cm}
\begin{center}

\centerline{\includegraphics[width=1.1\textwidth]{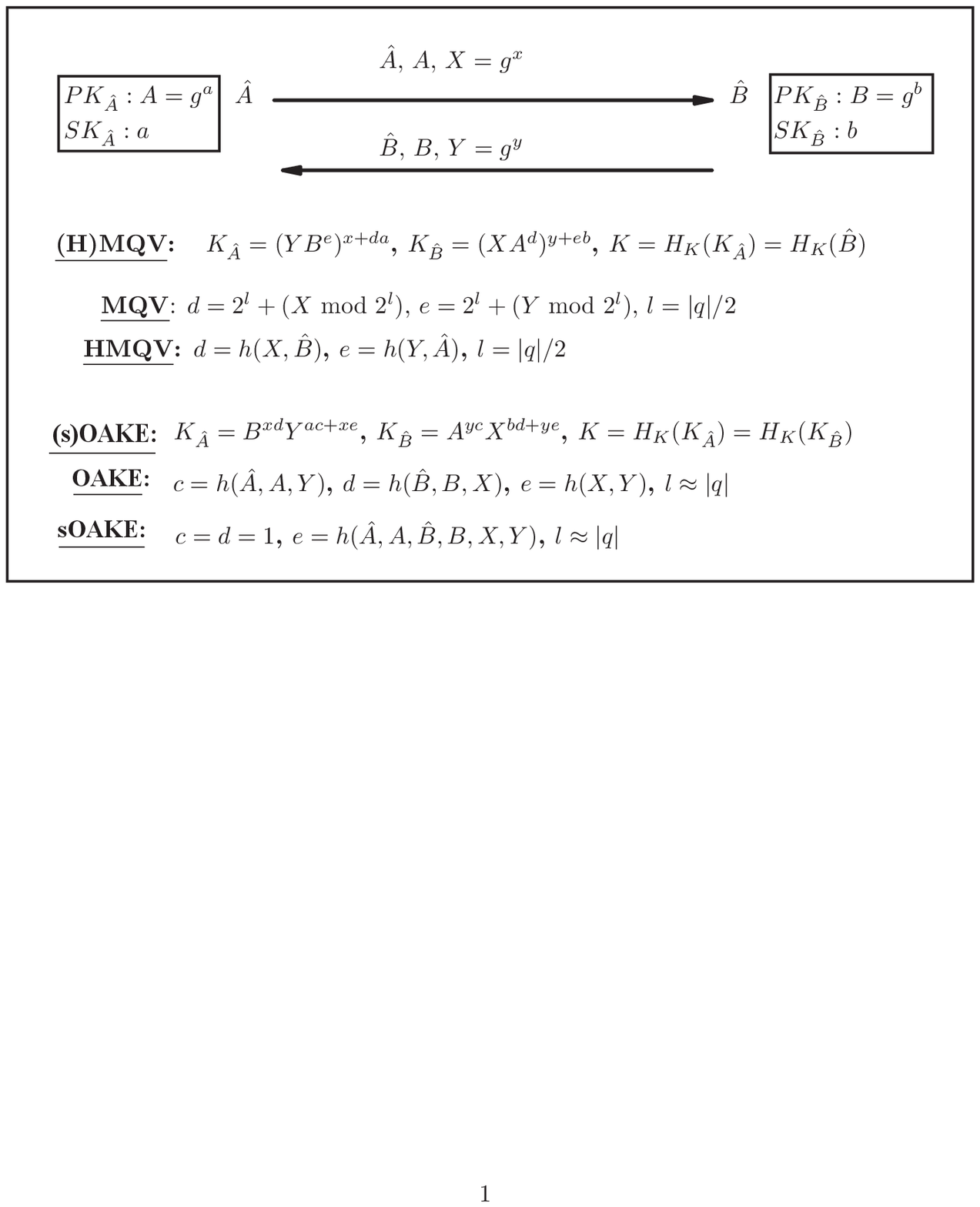}}
\vspace{-12.6cm}
 \caption{\label{OAKE-MQVec}
 Specifications of (H)MQV and (s)OAKE }

\end{center}
\vspace{-1.3cm}
\end{figure}

 Note that $NMJPOK_{(b, y)}=X^{db+ey}=(B^{d}Y^{e})^x$, where $d=h(B,X)$ and
 $e=h(X, Y)$,  
  actually
 can be used to demonstrate the knowledge of $x$. 
  The key observation now  is: in order for $\hat{A}$  to
 additionally prove  the knowledge of its secret-key $a$, we can
 multiply  $X^{db+ey}$ 
  by another POK $Y^{ca}$ for $c=h(A,Y)$.
  This  yields $K_{\hat{A}}=B^{dx}Y^{ca+ex}
  =A^{cy}X^{db+ey}=K_{\hat{B}}$, where  $K_{\hat{A}}$ (resp., $K_{\hat{B}}$)
   is computed  by $\hat{A}$ (resp., $\hat{B}$) respectively.
   As we aim for secure DHKE protocols in concurrent settings like the Internet, we let the values $K_{\hat{A}}$ and  $K_{\hat{B}}$ commit to the complete session tag by putting users' identities into the inputs of $d$ and/or $e$, which particularly ensures the ``key-control" property of  \cite{LMQSV03} for DHKE.
   All the observations are   boiled  down to the OAKE protocol, which  is
depicted in Figure \ref{OAKE-MQVec}.  The version derived from sNMJPOK, referred to as single-hash OAKE (sOAKE), is also depicted in Figure \ref{OAKE-MQVec}. Note that the output length of $h$, i.e., $l$, is set to be $|q|/2$ in (H)MQV,  but approximately $|q|$ in OAKE and sOAKE protocols.  
In particular,
with the (s)OAKE protocol family, $h$ and  $H_K$ (that is used for deriving the session-key $K$)  can be
identical.
%
Also note that,  for (s)OAKE,  $\hat{A}$ (resp., $\hat{B}$) can offline pre-compute $X$ and $B^{dx}$ (resp., $Y$ and $A^{cy}$). Some  (s)OAKE variants are given in Appendix \ref{NOvariants}. 
We also highlight another property, called tag-based self-seal (TBSS),  of (s)OAKE in the RO model: given any complete session tag $(\hat{A},A,\hat{B},B, X,Y)$ and any $\alpha\in G\setminus 1_G$, $\Pr[K_{\hat{A}}=K_{\hat{B}}=\alpha]\leq
\frac{1}{2^l-1}$, where the probability is taken over the choice of the random function of $h$ (see more discussions on TBSS in Appendix \ref{AppSSJPOK}).


\textbf{Notes on subgroup tests in (s)OAKE.} 
  The basic technique to check the
DH-component, e.g. $X$, is in $G$ is to verify $X^q=1_G$ (and
$X\in G^{\prime}\setminus 1_G$) that
needs performing one modular exponentiation. But, 
if the cofactor $t$ is small, e.g., $G^{\prime}=Z^*_N$ such that
$N=2q+1$ or $G$ is the subgroup of an elliptic curve over a finite
field (in this case the cofactor $t$ is usually a  small constant), 
 the subgroup test of $X$  can be essentially reduced to: (1)
check $X\in
G^{\prime}$; (2). $X^t\neq 1_G$. In general, 
checking $X\in G^{\prime}$ and  $X^t\neq 1_G$ guarantees that $X$ is not in
a (small) subgroup of $G^{\prime}$ with the order  that is a
factor of $t$, but it does not fully guarantee $X\in G$ (e.g.,
considering  that $X=-g^x$). This leads to the following (s)OAKE 
variant with embedded subgroup tests, 
in which the values $K_{\hat{A}}, K_{\hat{B}}$ are set to be:
 $K_{\hat{A}}=B^{dxt}Y^{cat+ext}$ and
 $K_{\hat{B}}=A^{cyt}X^{dbt+eyt}$. 
  The subgroup test is performed as follows: each player first verifies that its peer's
  DH-component is in $G^{\prime}$, and then acts in accordance with one of the following two cases.
 \vspace{-0.2cm}
 \begin{description}
 \item [Case-1.]If $B^{dxt}$ and $Y^{cat+ext}$ (resp., $A^{cyt}$ and $X^{dbt+eyt}$) are computed
 separately, \emph{particularly when $B^{dxt}$ (resp., $A^{cyt}$) is offline pre-computed by $\hat{A}$ (resp., $\hat{B}$)},  $\hat{A}$ (resp., $\hat{B}$) checks that $Y^{cat+ext}\neq 1_G$
 (resp., $X^{dbt+eyt}\neq 1_G$);
 \vspace{-0.2cm}
 \item [Case-2.] In case of no separate computation, $\hat{A}$ (resp., $\hat{B}$) verifies
 $K_{\hat{A}}\neq 1_G$  (resp.,  $K_{\hat{B}}\neq 1_G$). Note that  the
 checking of $K_{\hat{A}}\neq 1_G$ and $K_{\hat{B}}\neq 1_G$,
 as done in 
 MQV, does not \emph{fully}
 guarantee  $X^t\neq 1_G$ or $Y^t\neq 1_G$, 
  but it still  provides reasonable assurance in the elliptic curve setting as clarified above.
   \end{description}
   \vspace{-0.2cm}
   We remark that the embedded subgroup test in Case-1, well supported by (s)OAKE,  provides stronger security guarantee than that in Case-2 as done in (H)MQV. Note that (H)MQV cannot offline pre-compute the values $B^{e}$ and $A^{d}$ to ease the more robust  Case-1 embedded subgroup test.
We note that the damage caused by  ignoring the subgroup test of
peer's DH-component (but still  with  the supergroup $G^{\prime}$ membership
check) can be much   relieved (and even waived),   if the ephemeral
private values generated 
 within  the protocol
run are well-protected.
 More notes on the subgroup test, and on the ephemeral private values that can be exposed to adversary, are referred to Appendix \ref{morespecifications}.
\vspace{-0.1cm}
\section{Advantageous Features of (s)OAKE}\label{SecYZvsHMQV}
\vspace{-0.1cm}

\textbf{Efficiency advantages.} The online computational complexity of (s)OAKE can remarkably be only 1 exponentiation at each player side (with embedded subgroup test), which is optimal for DHKE. Specifically,  the value  $B^{dxt}$ (resp., $A^{cyt}$) can be  offline pre-computed by $\hat{A}$ (resp., $\hat{B}$).  In comparison,   (H)MQV cannot offline pre-compute the values $B^{e}$ and $A^{d}$ to improve online efficiency, and thus the online efficiency of (H)MQV is about 1.3 exponentiations.

 The total computational complexity of (s)OAKE   is essentially the same as that of (H)MQV, with sOAKE being still  slightly more efficient than HMQV. In particular, by the simultaneous exponentiation techniques \cite{MOV95,G98,DJM00},   each player in (H)MQV and (s)OAKE performs  about 1.3 exponentiations in computing $K_{\hat{A}}$ or $K_{\hat{B}}$.
But, the computation of $K_{\hat{A}}$ (resp., $K_{\hat{B}}$) of HMQV is still slightly more inefficient than that of  sOAKE with a single hash. For example, to compute $K_{\hat{A}}$, besides the same other operations needed for simultaneous exponentiations, HMQV (resp., sOAKE) needs to compute $\{d,e,x+da,e(x+da)\}$ (resp., only $\{e,a+xe\}$).


On the same subgroup order $q$, (s)OAKE ensures more robust resistance to collision attacks against the underlying hash function $h$ than HMQV, as the output length of $h$, i.e., $l$, is set to be $|q|/2$ for HMQV but $|q|$ for (s)OAKE. To strengthen its security, some standards specify  larger subgroups (e.g., $|q|=255$ in \cite{N04}) to use for HMQV. 
However,  in memory-restricted environments (like smart-cards
or other portable electronic tokens), subgroup  size is an influential
parameter in  favor  of a given algorithmic solution.

 \textbf{Reasonable deniability.} For key-exchange protocols, both security and privacy are desired,
which would also have been being  one of the major criteria  underlying
the
evolution 
 of a list of important   industrial standards of DHKE (e.g., Internet key-exchange). 
  Among
privacy concerns, deniability  is an essential privacy property,
and has always been a central concern in personal and business
communications, with off-the-record communication serving as an
essential social and political tool \cite{DG05,DGK06}. The reader is referred to \cite{DG05,DGK06} for a list of  scenarios where  deniability is desirable. 
(Needless to say, there are special applications where
non-repudiable communication  is essential, but this is not the
case for most of our nowaday communications over Internet
\cite{DG05,DGK06} where deniable authentication  is  much more
desirable than non-repudiable  authentication.) 

A 2-round implicitly authenticated  DHKE protocol 
is defined to be  of \textsf{reasonable deniability}, if the session-key can be  computed merely from the ephemeral DH-exponents without involving any player's static secret-key. Note that we cannot count on  DHKE with implicit authentication, like (H)MQV and (s)OAKE, to enjoy  full-fledged deniability (zero-knowledge).
 It is clear that (s)OAKE enjoys reasonable deniability, as the session-key of (s)OAKE 
   can be computed  merely from
  the DH-exponents $x$ and $y$,  
   which  is useful 
  to preserve privacy for  both protocol players.
 Note that (H)MQV is not reasonably deniable, as  the use of the session-key of (H)MQV can be  traced back to the
group of the  two players particularly  in view that  the value  $g^{ab}$ is  involved in the session-key computation. 

\textbf{Modular, parallel and post-ID computability.}
First note that $B^{dx}$,  $Y^{ca+ex}$ and the explicit sub-group test $Y^q$
by $\hat{A}$ (resp.,   $A^{cy}$,  $X^{db+ey}$ and $X^q$
by $\hat{B}$) can be computed in a parallel, modular  and post-ID way, which allows  for various trade-offs among security, privacy and
efficiency  for the deployment of (s)OAKE in practice. 
 Specifically, the offline pre-computability of $B^{dx}$ and $A^{cy}$ 
   eases more efficient explicit subgroup test by computing   $Y^{ca+ex}$
  and $Y^q$ (resp., $X^{db+ey}$ and $X^q$) \emph{in parallel} that amounts to  about 1.2 exponentiations. Also, as clarified, offline pre-computability of $A^{cy}$ (resp., $B^{dx}$) allows the above  more robust Case-1 embedded subgroup test of $X^{dbt+ext}$ (resp., $Y^{cat+ext}$). 
Observe that, for OAKE,
   $Y^{ca+ex}$  (resp., $X^{db+ey}$)  can be computed before
  learning peer's identity and public-key information. Such a
  post-ID computability, besides reasonable deniability,  is useful  for privacy preserving \cite{CK02}. (H)MQV does not support such offline pre-computability and post-ID computability. 

    \textbf{Ease deployment with lower-power devices.}  As we shall see in Section \ref{SecBeyondCK} and Appendix \ref{publiccomputation},  (s)OAKE (with offline pre-computation to an almost maximum extent) well supports the public computation model
\cite{KP06} (while (H)MQV does not),   which is desirable for deploying KE protocols with authentication devices of limited computational ability  in hostile computing environments.  (s)OAKE allows   smaller  parameter $|q|$ than HMQV (in resistance  to collision attacks against $h$),
which is  important for deployment 
 with memory-restricted devices  (like smart-cards
or other portable electronic tokens).

\textbf{Minimal setup.} (s)OAKE does not
 mandate  proof of possession/knowledge (POP/K) of secret-key during  public-key registration, while POP/K is now  commonly assumed for
 MQV. POP/K is explicitly abandoned in HMQV, however  as we shall see,  there exists a way to maliciously \emph{asymmetrically} compute the session-key of HMQV  \emph{without knowing either static secret-key or ephemeral DH-exponent}.


\vspace{-0.2cm}
\subsection{Security in the CK-Framework}\label{SecWithinCK}
\vspace{-0.1cm}


At a high level, the design rationale of  (s)OAKE   is new, 
 with  NMJPOK  as the core building tool.
     The design of MQV is based
    on implicit signatures \cite{MQV95}.
     The  design of HMQV is  based on  Hashed Dual
challenge-Response (HDR)
 signatures and Hashed Challenge-Response (HCR) signatures,
 which are in turn based  on Dual  Challenge-Response
(DCR)  and eXponential Challenge-Response (XCR)
signatures. 
 To further justify the robustness of the NMJPOK-based (s)OAKE protocols, 
 we will show (in Section \ref{SecCKproof}) that (s)OAKE  
 can also be casted in terms of
HDR signatures. 
Moreover, in comparison with the HDR signature implied by HMQV (referred to as HMQV-HDR), the HDR signatures implied by (s)OAKE, referred to as (s)OAKE-HDR/HCR,
 are both \emph{online
efficient} (i.e., only one online exponentiation)
and \emph{strongly secure} (by providing  stronger secrecy exposure capability to the signature forger and posing more stringent forgery success condition).

In the CK-framework for a DHKE protocol, a concurrent man-in-the-middle (CMIM) adversary $\mathcal{A}$ controls all the communication channels among concurrent session runs of the  KE protocol. In addition, $\mathcal{A}$ is allowed access to secret information via the following three types of queries: (1) state-reveal queries for ongoing incomplete sessions; (2) session-key queries for completed sessions; (3) corruption queries upon which all information in the memory of the corrupted parties will be leaked to $\mathcal{A}$. A session $(\hat{A},\hat{B},X,Y)$ is called \emph{exposed},  if it or its matching session $(\hat{B},\hat{A},Y,X)$ suffers from any of these three queries.
The session-key security (SK-security) within the CK-framework is captured as follows: for any complete session $(\hat{A},\hat{B},X,Y)$ adaptively selected by $\mathcal{A}$, referred to as the \emph{test session}, as long as it is unexposed,  with overwhelming probability it holds   that (1) the session-key outputs of the test session and its matching session are identical; (2) $\mathcal{A}$ cannot distinguish the session-key output of the test session from a random value.

At a first glance, as (s)OAKE is of reasonable deniability (i.e., the session-key can be computed merely from $x$ and $y$), (s)OAKE may not be secure in the CK-framework. However, this does not pose a problem for probable security within the CK-framework, where the test-session is required to be unexposed. 
Actually, as we shall see, the provable security of (s)OAKE  within the CK-framework assumes much stronger secrecy exposure than HMQV. If one wants to sacrifice privacy for seemingly stronger security against exposure of both $x$ and $y$ even for the test-session, one can use the protocol  variant of robust (s)OAKE proposed in Appendix \ref{NOvariants} that is also provably secure in the CK-framework. The only difference between robust (s)OAKE and (s)OAKE is that, the values  $K_{\hat{A}}$ and $K_{\hat{B}}$ in robust (s)OAKE are set to be: $K_{\hat{A}}=B^{a+xd}Y^{ac+xe}$  and $K_{\hat{B}}=A^{b+yc}X^{bd+ye}$.  But, as discussed in Appendix \ref{AppExposedDH}, the security advantage of robust (s)OAKE over (s)OAKE is 
 insignificant, and from our view (s)OAKE achieves much better balance between security and privacy than the robust (s)OAKE variant.

 For provable SK-security within the
CK-framework,
denote  by  $(\hat{A}, \hat{B}, X, Y)$ the test-session, we show both OAKE and sOAKE  (actually their weaker  \emph{public-key free} variants with players' public-keys  removed from the inputs of $c, d, e$), \emph{with pre-computed  and exposed
DH-components, DH-exponents and the values $A^{cy}$'s and $B^{dx}$'s} 
(which renders much stronger secrecy exposure capability  to attacker than HMQV within the CK-framework),
 are  SK-secure in the RO model, under the following assumptions (with proof details referred to Section \ref{SecCKproof} and Appendix \ref{CKanalysis}): 
 \begin{itemize}
 \item  The GDH assumption, in case $\hat{A}\neq \hat{B}$ (which is also the most often case in practice).  We note that, 
     \emph{whenever the DH-exponent is generated and  exposed during a session-run without offline pre-computation  prior to the session run, OR, there exists an honest player whose public DH-component for a session is offline pre-computed and exposed prior to the session run (no matter whether the secret DH-exponent is   exposed   or not)}, the security of HMQV  is based on both the GDH assumption and the KEA assumption. That is, for this most often case of $\hat{A}\neq \hat{B}$, (s)OAKE not only allows more powerful secrecy leakage 
     but also is based on weaker assumptions than HMQV.

 \vspace{-0.1cm}
 \item The CDH assumption, in case $\hat{A}=\hat{B}$ and $X=Y$.
 \vspace{-0.1cm}
  \item  The GDH assumption and the KEA assumption, in case $\hat{A}=\hat{B}$ and $X\neq Y$ (the security of HMQV is based on the same assumptions in this  case). 
  \end{itemize}
   \vspace{-0.1cm}
   As stressed in \cite{K05}, security against exposed DH-exponents is deemed to be the main and prime concern  for any robust DHKE, and security against exposed offline pre-computed values (particularly, the DH-components) is important to both lower-power devices and to high volume servers \cite{K05}. The reason is, as pointed out in \cite{K05}, many applications in practice
will boost protocol performance by pre-computing and storing values  for later use in the
protocol.  In this case,
however, these stored values  are more vulnerable to leakage, 
particularly when DHKE is deployed in hostile environments with plagued
spyware or virus and in view of that the offline pre-computed DH-components are much less protected in practice as they are actually public values to be exchanged in plain.


  In addition, (s)OAKE enjoys the following security advantages: (1) tighter security reduction of sOAKE than HMQV (discussed in Appendix \ref{AppHDRsecurity} and \ref{basicanalysis});
(2) more robust  embedded subgroup test supported by offline pre-computability of $A^{cy}$ and $B^{dx}$ (as clarified above);
Due to space limitation, more discussions on the security of (s)OAKE vs. (H)MQV are given in Appendix \ref{AppExposedDH}.
  For (s)OAKE, putting public-keys into the input of $c,d,e$ are necessary in order to ensure  non-malleable joint proof-of-knowledge of both $(a,x)$ (resp., $(b,y)$) by player $\hat{A}$ (resp., $\hat{B}$), as clarified with the development of (s)OAKE based on  the  underlying building tool of NMJPOK in Section \ref{SSJPOKsec} and Appendix \ref{AppSSJPOK}. But, as we shall see below (by concrete attacks), the SK-security in accordance with the CK-framework does not ensure  joint proof-of-knowledge of $(a,x)$ or $(b,y)$. 
This is also the reason that we can prove the SK-security of (s)OAKE w.r.t. the public-key free variant. 
Next, we show that (s)OAKE also enjoys essential advantages over (H)MQV beyond the CK-framework.


\vspace{-0.2cm}
\subsection{Security Beyond the CK-Framework}\label{SecBeyondCK}
\vspace{-0.1cm}

\textbf{A new perspective to DHKE:}
\textbf{exponent-dependent attacks (EDA) on (H)MQV, and  the introduction of  computational fairness.} In this work, we identify  EDA attacks against  (H)MQV, which causes \emph{computational unfairness} between malicious users and honest users in the sense that  an adversary can compute the shared DH-secret with an honest player \emph{in an asymmetric way}. We then discuss the implications and damages caused by EDA attacks, and then introduce a new security notion called ``\emph{computational fairness}" for authenticated DHKE protocols.


Given a
value $X\in G$ for which the malicious player $\hat{A}$ (e.g., a client) does not
necessarily know the discrete logarithm of $X$, $\hat{A}$ computes $d$ and sets $A=X^{-d^{-1}}\cdot g^t$ where $t\in Z_q$ and  $d=h(X,
\hat{B})$ for HMQV or $d=2^l+ (X\mod 2^l)$ for MQV. Note that
$XA^d=X(X^{-d^{-1}}\cdot g^{t})^d=XX^{-1}g^{td}=g^{td}$, and the  shared DH-secret  now is $K_{\hat{A}}=(XA^d)^{y+eb}=g^{tdy}g^{tdeb}=Y^{td}B^{tde}$. We call such an attack exponent dependent attack.  If $\mathcal{A}$ sets $t=0$ then the shared DH-secret $K_{\hat{A}}$ is always $1_G$. If $\mathcal{A}$ sets $t=d^{-1}$, then $K_{\hat{A}}=YB^e$. For all these two specific cases, the value $K_{\hat{A}}$ can be \emph{publicly computed} (without involving any secret values).  In any case, the computational complexity in computing the shared DH-secret by the malicious $\hat{A}$ is much lesser than that by its peer $\hat{B}$, which clearly  indicates some unfairness. 
   In general,  the malicious $\hat{A}$ can honestly generate its public-key $A=g^a$  and  compute the session-keys, thus
explicitly requiring POP/K 
 of secret-key during public-key registration  and
explicit key-confirmation and mutual authentication (as required
by the 3-round (H)MQV)  do not prevent the above attacks. As  there are many choices of the value $t$ by the adversary in different sessions,  explicitly checking whether the shared DH-secret is  $YB^e$ also does not work. The above attacks can also be
trivially modified (actually simplified) to be against the
one-round HMQV variant. 
 \emph{ We stress that such  attacks do not violate the security analysis of HMQV in \cite{K05}, as they are beyond the CK framework.}


We note that MQV (with embedded subgroup membership test of peer's DH-component) explicitly checks  the shared DH-secret is not $1_G$, and thus the attack with $t=0$ does not work against MQV.  But, for (H)MQV with explicit subgroup tests of peer's public-key and DH-component, whether still  checking the shared DH-secret is $1_G$   is  however  unspecified. In particular, the basic version of HMQV \cite{K05} does not check whether the shared DH-secret is $1_G$ or not, and POP/K of secret-keys is explicitly abandoned in HMQV. 
 We also note the version of HMQV proposed in \cite{K06} does check  and ensure  the shared DH-secret is not $1_G$. But, (H)MQV does not  resist   the above attacks with $t\neq 0$.

Besides asymmetric computation, 
 such drawbacks also allow more effective DoS attacks. Though an adversary can send arbitrary messages to an honest party (say, player $\hat{B}$ in the above attacks) to issue DoS attacks, 
which however can be easily detected  by the authentication mechanism  of (the 3-round version of) (H)MQV. But, with our above attacks, the honest player $\hat{B}$ is hard to distinguish  and detect an attack from an honest execution of (H)MQV.

This motivates us to introduce a new notion for DHKE, called session-key computational fairness.
Roughly  speaking, we say that a DHKE protocol enjoys  \emph{session-key computational fairness}, if the session-key computation (for any successfully finished session between a possibly malicious player and an honest player) involves the same number of   \emph{non-malleably independent} \emph{dominant-operation values} for both the malicious player and the honest player. Here, dominant operation  is specific to  protocols, and for (s)OAKE and (H)MQV,  the dominant operation is defined  just to be  modular exponentiation.
Informally speaking, a set of dominant-operation values $\{V^I_1,\cdots,V^I_{m}\}$ for $m\geq 2$ are non-malleably independent, if any   polynomial-time  malicious player $I\in \{\hat{A},\hat{B}\}$ cannot make these values correlated under any predetermined polynomial-time computable relation (no matter how the malicious player does). More formally, for any complete session-tag $Tag$,  we say that a set of dominant-operation values $\{V^I_1,\cdots,V^I_{m}\}$ 
(w.r.t. $Tag$)
 are non-malleably independent, if they are indistinguishable from independent random values $\{U_1, \cdots, U_{m}\}$ or $\{U_1, \cdots,U_{j-1}, V^I_j, U_{j+1}, \cdots, U_{m}\}$  for at most one $j, 1\leq j\leq m$. 
  We then show  that (s)OAKE enjoys session-key computational fairness, while (H)MQV does not by the above concrete EDA attacks.
We also  propose some  HMQV variants,  \emph{just in the spirit of (s)OAKE and NMJPOK}, to prevent our EDA attacks. 
 The key point is to put $A$ (resp.,  $B$) into the input of $d$ (resp., $e$).  Unfortunately, we failed in providing provable security of these fixing approaches  in the CK-framework. In particular, we observed that it is hard  to extend the  security proof of  HMQV  \cite{K05} to any of  the proposed fixing solutions (indeed, HMQV was very carefully designed to enjoy provable security in the CK-framework). Besides lacking provable security in the CK-framework, many other advantageous features enjoyed by (s)OAKE are also lost with  these  fixing solutions.
To the best of our knowledge, we do not know how to achieve, besides the newly developed  (s)OAKE family,  implicitly authenticated DHKE protocols that enjoy all the following properties: (1) provable security in the CK-framework; (2) online optimal (i.e., only one exponentiation) efficiency  \emph{and/or} reasonable deniability; (3) session-key computational fairness. The surrounding issues are quite subtle and tricky, and indeed (s)OAKE was very carefully designed to achieve all these features (and much more as clarified above). Due to space limitation, the reader is referred to  Appendix \ref{AppFairness} for more details.

%

\textbf{On supporting the public computation model \cite{KP06}.}
The work  \cite{KP06} proposed the public computation model for KE protocols,
where an entity (performing a run of KE-protocol) is split
into two parts:  a trusted  authentication device (which enforces
the confidentiality of the authentication data), and an
untrusted  computing device (in which   some
computing operations  are \emph{publicly}
carried out). This
allows to use an authentication device with little computing
power, and to make computing devices independent from users
\cite{KP06}. Some  concrete applications  suggested in \cite{KP06}
 are: (1)  Mobile phones include
smart cards which store the user authentication data; the handsets
themselves are the computing devices. (2) PCs (corresponding to
the computing device) equipped with a crypto token (corresponding
to the authentication device)  have a lot more computing power
than the token itself, but may be plagued by spyware or virus. (H)MQV does not well support deployment with such public computation as shown in \cite{KP06}, 
while (s)OAKE well supports deployment in this model (see details in Appendix \ref{beyondCK}).
Specifically,  the natural split of authentication computation
and public computation for (s)OAKE  is as follows, with the
computation of $\hat{B}$ as an example: (1)  The authentication device generates  $(y, Y)$ and
possibly $A^{cy}$ (in case the authentication device has learnt the peer identity  $\hat{A}$),
 and then forwards $Y$ and possibly $A^{cy}$ to the
computation device; (2) After getting $X$ from the computation
device, the authentication device computes $s=db+ey$,
and then forwards
$s$ to the computation device; (3) After getting $s$ from the
authentication device, the computation device computes
$K_{\hat{B}}=A^{cy}X^s$ and  the session-key, and then
communicate with $\hat{A}$ with the session-key.
\emph{Note that
$y, Y, c, d, A^{cy}, db$ can be offline pre-computed by the
authentication device, and the authentication device can only
online compute $ey$ and $X^s$. }
%
%

More discussions of the security of (s)OAKE beyond CK-framework are referred to Appendix \ref{beyondCK}.
  The security  of (s)OAKE, in the CK-framework and beyond,  further justifies the soundness and robustness of the design rational and building tools 
   of  (s)OAKE.

 \vspace{-0.2cm}
\section{Casting (s)OAKE  in Terms of HDR Signatures} \label{SecCKproof}
 \vspace{-0.2cm}

Informally speaking, to distinguish the session-key output of the unexposed test-session from a random value, an efficient adversary $\mathscr{A}$  only has  two
 strategies in the RO model:

\vspace{-0.3cm}
 \begin{description}
\item [Key-replication attack.] $\mathscr{A}$ succeeds in forcing
the establishment of a session (other than the  test-session or
its matching session) that has the same session-key output as the
test-session. In this case, $\mathscr{A}$ can learn the
test-session key by simply querying the session to get the same
key. 

\vspace{-0.3cm}
\item [Forging attack.] At some point in its run, $\mathscr{A}$
queries  the RO $H_K$ with the value $K_{\hat{A}}$ or $K_{\hat{B}}$. This implies that
$\mathscr{A}$ succeeds in outputting 
  the value $K_{\hat{A}}$ or $K_{\hat{B}}$.

 \end{description}
\vspace{-0.2cm}

At high level, the possibility of key-replication attack against (s)OAKE  is  ruled out \emph{unconditionally} in the RO model by the NMJPOK and TBSS properties of (s)OAKE, which actually holds also for the public-key free variant of (s)OAKE (as matching sessions are defined without taking public-keys into account in the CK-framework).  Below, we  focus on ruling out the possibility of forging attack. Intuitively, by the NMJPOK property of (s)OAKE, an attacker can compute the DH-secret $K_{\hat{A}}$ or $K_{\hat{B}}$ of the test-session only if it does indeed ``know" both the corresponding  static secret-key and the ephemeral DH-exponent, which then violates the discrete logarithm assumption. But, turning this intuition into a formal proof needs introducing some non-standard non-black-box assumptions (though it  much simplifies the security analysis), which may not be very  favorable  and is left to a subsequent separate work (for analyzing (s)OAKE in more security models). In this work, we  mainly focus on the black-box analysis of (s)OAKE  in the CK-framework. 
In the rest, we show the forging attack can still be ruled out in a \emph{black-box} manner,    by casting  (s)OAKE in terms of \emph{online-efficient} and \emph{strongly secure} HDR signatures.
Full details (of this section) are given in  Appendix \ref{CKanalysis}.

 Informally speaking, a HDR signature scheme is an
\emph{interactive} signature scheme between two parties in the
public-key model, with the \emph{dual}
roles of signer and challenger.


 \vspace{-0.2cm}
\begin{definition}  [(s)OAKE-HDR signatures]Let $\hat{A}$,$\hat{B}$ be two parties with public-keys $A=g^a$, $B=g^b$, respectively. 
 Let  $m_{\hat{A}}$, $m_{\hat{B}}$ be two messages.
The \textsf{\textup{(s)OAKE-HDR}} 
signatures of $\hat{B}$ on messages
$(m_{\hat{A}},m_{\hat{B}},\hat{A},A, \hat{B},B,X,Y)$ 
 are defined as a vector  of values
(the signatures of $\hat{A}$ 
 are defined similarly):

 \vspace{-0.2cm}
\begin{description}

\item [OAKE-HDR.] $\{\hat{A}, A, m_{\hat{A}}, m_{\hat{B}}, X, Y,
HSIG^{OAKE}_{\hat{A},\hat{B}}(m_{\hat{A}}, m_{\hat{B}}, X,
Y)=H_K(A^{yc}X^{bd+ye})\}$, where $X=g^x$, $Y=g^y$ are chosen by
$\hat{A}$, $\hat{B}$ respectively as the random \emph{challenge}
and \emph{response}, $x, y \in_{\textup{R}} Z^*_q$,
$c=h(m_{\hat{A}}, \hat{A}, A, Y)$, $d=h(m_{\hat{B}}, \hat{B}, B,
X)$ and $e=h(X, Y)$.


\item [sOAKE-HDR.] $\{\hat{A}, A, m_{\hat{A}},  m_{\hat{B}}, X, Y,
HSIG^{sOAKE}_{\hat{A},\hat{B}}(m_{\hat{A}},  m_{\hat{B}}, X,
Y)=H_K(A^{yc}X^{bd+ye})\}$, where $c=d=1$,  $e=h(m_{\hat{A}},
m_{\hat{B}}, \hat{A}, A,
\hat{B}, B,  X, Y)$. 



\end{description}

\end{definition}
 \vspace{-0.2cm}

 \vspace{-0.3cm}
\begin{definition} [\emph{Strong} security  of HDR signatures (with off-line
pre-computation)\label{DCRdef-SEC}]
   We say a HDR
signature scheme (of $\hat{B}$) 
 is
\textsf{strongly}  secure,  if no polynomial-time machine
$\mathcal{F}$ can win the game in Figure \ref{HDR} with
non-negligible probability \emph{with respect to any uncorrupted
party  $\hat{A}$  of public-key $A=g^a$ such that 
 the secret-key $a$ was  not chosen by the attacker
$\mathcal{F}$}.
\end{definition}
 \vspace{-0.1cm}

 \begin{figure}[!t]
\begin{center}

\begin{tabular} {|c|}
 \hline 

\begin{minipage}[t] {6.2in} \small
\vspace{-0.3cm}

\begin{enumerate}

\item Forger $\mathcal{F}$ is given values $B$, $X_0$, where $B,
X_0 \in_{\textup{R}} G$.
\vspace{-0.2cm}

\item $\mathcal{F}$ is given access to a signing oracle $\hat{B}$
(of public-key $B=g^b$ and secret-key $b$).
\vspace{-0.2cm}

\item Each signature query from $\mathcal{F}$ to $\hat{B}$
consists of the following interactions:
\vspace{-0.3cm}

\begin{enumerate}

\item $\mathcal{F}$ presents $\hat{B}$ with messages $(\hat{Z}, Z,
m_{\hat{Z}}, m_{\hat{B}})$.  Here, $\hat{Z}$ can be any (even
corrupted) party chosen by $\mathcal{F}$, and $Z=g^z\in G\setminus 1_G$ is the
public-key of $\hat{Z}$. Note that $\mathcal{F}$ may not
necessarily know the corresponding secret-key $z$ of $\hat{Z}$.
\vspace{-0.2cm}

\item  $\hat{B}$ generates $y\in_{\textup{R}}Z^*_q$ and $Y=g^y$,
 and computes $Z^{cy}$, where $c=h(m_{\hat{Z}}, \hat{Z}, Z, Y)$ for OAKE-HDR or  $c=1$ for sOAKE-HDR.
 Then, $\hat{B}$  responds with $(y, Y=g^y,
Z^{cy})$ to $\mathcal{F}$ (\emph{which captures the powerful exposure
capability  to the forger}), and stores the vector  $(\hat{Z}, Z,
m_{\hat{Z}}, m_{\hat{B}}, y, Y, Z^{cy})$ as an ``incomplete
session". \emph{Here, $(y, Y, Z^{cy})$ can be offline pre-computed by
$\hat{B}$, and leaked to  $\mathcal{F}$ prior to the session involving 
 $(y, Y, Z^{cy})$.}
\vspace{-0.2cm}

\item $\mathcal{F}$  presents $\hat{B}$ with $(\hat{Z}, Z,
 m_{\hat{Z}}, m_{\hat{B}}, Y)$, and a \emph{challenge}
$X$.
\vspace{-0.2cm}

\item $\hat{B}$ checks that $X\in G\setminus 1_G$ (if not, it aborts) and
that $(\hat{Z}, Z, m_{\hat{Z}}, m_{\hat{B}}, Y)$ is in one of its
incomplete sessions (if not, it ignores). $\hat{B}$ then computes
$r=H_K(Z^{cy}X^{db+ey})$, where 
$d=h(m_{\hat{B}}, \hat{B}, B, X)$ and $e=h(X, Y)$ for OAKE-HDR
(resp., $d=1$ and $e=(m_{\hat{Z}}, m_{\hat{B}}, \hat{Z}, Z,
\hat{B}, B,  X, Y)$ for
sOAKE-HDR). 
  $\hat{B}$
responds $(\hat{Z}, Z, m_{\hat{Z}}, m_{\hat{B}}, X, Y,  r)$ to
$\mathcal{F}$, and marks the vector $(\hat{Z}, Z, m_{\hat{Z}},
m_{\hat{B}},
  y, Y, Z^{cy})$ as a ``\emph{complete session}", and
stores with it the signature
values $(\hat{Z}, Z, m_{\hat{Z}}, m_{\hat{B}}, X,  y,  Y,   r)$. 

\vspace{-0.2cm}
\end{enumerate}
\vspace{-0.2cm}
\item $\mathcal{F}$ is allowed a polynomial number of adaptive  queries to
$\hat{B}$ in \emph{arbitrarily}
interleaved order. 
\vspace{-0.2cm}

\item $\mathcal{F}$ halts with output ``fail" or with a
\emph{guess} in the form of a tuple  $(\hat{A}, A,  m_1, m_0, X_0,
Y_0, r_0)$. $\mathcal{F}$'s guess is called a successful
\emph{forgery} if the following two conditions hold:

\begin{enumerate}
\vspace{-0.3cm}
\item $(\hat{A}, A, m_1, m_0,  X_0, Y_0,  r_0)$  is a valid
HDR-signature of $\hat{B}$ on the messages $(m_1, m_0, \hat{A}, A,
\hat{B}, B,  X_0, Y_0)$, where $\hat{A}$
  is an uncorrupted player  of public-key $A=g^a$, $m_1$ corresponds to $m_{\hat{A}}$ (that is an arbitrary message sent  by the adversary $\mathcal{F}$ impersonating the signer $\hat{B}$ to the honest player $\hat{A}$), and $m_0$ corresponds  to
$m_{\hat{B}}$ (that is chosen by the honest player $\hat{A}$).   
 Note that the
value $X_0$ is the one received by $\mathcal{F}$ as input. 
\vspace{-0.1cm}

\item \emph{ $(\hat{A}, A,   m_1, m_0,  X_0, Y_0)$ did
not appear in any one  of the responses of $\hat{B}$ to $\mathcal{F}$'s
queries.}
\vspace{-0.3cm}
\end{enumerate}

We say $\mathcal{F}$ wins the game, if it outputs a successful
forgery (w.r.t. any  $A=g^a$ not chosen by $\mathcal{F}$).
\end{enumerate}

\end{minipage}
\\
\hline
\end{tabular}
\caption{\label{HDR}  Forgery game for (strongly secure)  (s)OAKE-HDR signatures (with offline
pre-computation)}
\end{center}
\vspace{-0.9cm}
\end{figure}

\vspace{-0.3cm}

More discussions on the above strong  HDR unforgeability security definition and the comparisons between (s)OAKE-HDR and HMQV-HDR are referred to Appendix \ref{AppHDRsecurity}.
 Due to space limitation, we only present the analysis sketch for OAKE-HDR here, the analysis for sOAKE-HDR is similar and actually much simpler.
 See Appendix \ref{CKanalysis} for full details.

 \vspace{-0.2cm}
\begin{theorem}\label{HDR-analysis-SEC}
 Under the GDH
assumption, (public-key free) OAKE-HDR  signatures of $\hat{B}$, with
offline pre-computed and exposable  $(y, Y, A^{cy})$,  are \emph{strongly}
secure in the random oracle model, 
 with respect to any
uncorrupted player other than the signer $\hat{B}$ itself
\emph{even if the forger is given the private keys of all
uncorrupted players in the system other than $b$ of
$\hat{B}$} 
\end{theorem}
 \vspace{-0.3cm}

\noindent \textbf{Proof} (sketch of Theorem \ref{HDR-analysis-SEC}).
The efficient solver $\mathcal{C}$ (who runs a supposed forger $\mathcal{F}$ as a subroutine) for the GDH problem is
presented in Figure \ref{OAKE-DCR} (page \pageref{OAKE-DCR}). It is easy to check, with overwhelming probability, the simulation of $\mathcal{O}$ is perfect in the RO model (with details referred to  Appendix \ref{CKanalysis}).

Here, we only highlight the analysis of  the probability that
$\mathcal{C}$ aborts at step F3. In the RO model, except for some negligible probability, $\mathcal{F}$ cannot succeed with undefined $c_0,d_0,e_0$.  Also, $\mathcal{F}$ can guess the value $r$ with negligible probability. The only left  way for $\mathcal{C}$ to abort at step F3 is: $r_0$ is  the value $r$ set  by
  $\mathcal{C}$ at one of S3.1 steps, where $r$ is supposed to be
  $H_K(\sigma)$ w.r.t. a stored vector  $(\hat{Z}, Z, m_{\hat{Z}}, m_{\hat{B}}, X,  y,
  Y, Z^{cy},
  r)$. 
  Recall that for the value $r$ set at   step S3.1,
  $\mathcal{C}$ does not know $\sigma$ (as it does not know $b$),
  and thus in this case both $\mathcal{C}$ and $\mathcal{F}$ may
  not   make  the RO-query $H_K(\sigma_0)=H_K(\sigma)$. In this
  case,  except for some negligible probability,   
  $\sigma_0=\sigma$, i.e., $A^{c_0y_0}X_0^{d_0b+e_0y_0}=Z^{cy}X^{db+ey}$,
  where $c=h(m_{\hat{Z}}, \hat{Z}, Z, Y)$,  $d=h(m_{\hat{B}}, \hat{B}, B,  X)$,
  $e=h(X, Y)$, $c_0=h(m_1, \hat{A}, A, Y_0)$,  $d_0=h(m_0, \hat{B}, B,
  X_0)$,      $e_0=h(X_0, Y_0)$, and $(m_0, m_1, \hat{A}, A, \hat{B}, B, X_0, Y_0)\neq
(m_{\hat{A}}, m_{\hat{B}}, \hat{Z}, Z, \hat{B}, B,  X, Y)$.  However, by the
NMJPOK and TBSS properties  of OAKE, for any value $\sigma\in
G\setminus 1_G$ and any  $(m_1, m_0, \hat{A}, A,  \hat{B},  B, X_0,
Y_0)$, the probability
$\Pr[A^{c_0y_0}X_0^{d_0b+e_0y_0}=\sigma]\leq \frac{1}{2^l-1}$,
where $X_0$  is the given random element in $G\setminus 1_G$, $\hat{A}$ and
$\hat{B}$ are uncorrupted players. This is true,  even if  the
public-key $A$ (resp., $B$) is removed from $c_0$ (resp.,
$d_0$), as the public-keys $A$ and $B$ are generated by the
uncorrupted players  $\hat{A}$ and $\hat{B}$ independently at
random, and $X_0$ is the given random DH-component (not generated
  by the attacker).

Finally, by applying  a slightly  extended version of the forking lemma in \cite{PS00}, which is referred to as  divided forking lemma and  is  presented  in Section \ref{dividedforking}, we have that, provided that $\mathcal{F}$ succeeds with non-negligible probability  in the first run  of $\mathcal{C}$, with non-negligible probability $\mathcal{F}$ will also succeed in  the repeat experiment C1 or C2. In this case, the output of
$\mathcal{C}$ is the just correct value of $CDH(X_0, B)$. \hfill $\square$

%
%
%
%

 \begin{figure}[!p]
\begin{center}

\begin{tabular} {|c|}
 \hline 
 \textbf{Building the CDH solver $\mathcal{C}$ from the OAKE-HDR forger $\mathcal{F}$}\\ 

\begin{minipage}[t] {6.7in} \footnotesize

\textbf{Setup:} The inputs to $\mathcal{C}$ are random elements
$U=g^u, V=g^v$ in $G$, and its goal is to compute $CDH(U,
V)=g^{uv}$ with oracle access to a DDH oracle $\mathcal{O}$.
To this end, $\mathcal{C}$  sets $B=V$ and $X_0=U$, and  sets the
public-keys and secret-keys for all other uncorrupted players in
the system.  $\mathcal{C}$  runs the forger $\mathcal{F}$ on input
$(B, X_0)$  against the  signer $\hat{B}$ of public-key $B$.
 $\mathcal{C}$ provides $\mathcal{F}$ with a random
tape, and provides the secret-keys of all uncorrupted players
other than the signer $\hat{B}$ itself (the attacker $\mathcal{F}$
may register arbitrary public-keys for corrupted players, based on
the public-keys and secret-keys of uncorrupted players).

\textbf{Signature query simulation:} Each time $\mathcal{F}$
queries $\hat{B}$ for a signature on values $(\hat{Z}, Z,
m_{\hat{B}}, m_{\hat{A}})$, $\mathcal{C}$ answers the query for
$\hat{B}$ as follows (note that $\mathcal{C}$ does not know $b$):

\begin{description}
\item [S1.]  $\mathcal{C}$ generates $y\in_{\textup{R}}Z^*_q$,
$Y=g^y$  
and  $Z^{cy}$, where $c=h(m_{\hat{Z}}, \hat{Z}, Z,  Y)$ (that may
be pre-defined, otherwise
$\mathcal{C}$ defines  $c$ with  the RO $h$). 
Actually, $(y, Y, Z^{cy})$ can be pre-computed by $\mathcal{C}$
and leaked to $\mathcal{F}$ prior to the session. Then,
$\mathcal{C}$ responds  $(y, Y=g^y, Z^{cy})$ to $\mathcal{F}$, and
stores the vector $(\hat{Z}, Z, m_{\hat{Z}}, m_{\hat{B}}, y, Y,
A^{cy})$ as an ``incomplete session".

\item [S2.]  $\mathcal{F}$  presents $\mathcal{C}$ with $(\hat{Z},
Z, m_{\hat{Z}}, m_{\hat{B}}, Y)$, and a \emph{challenge} $X$.

\item [S3.]  $\hat{B}$ checks that $X\in G\setminus 1_G$ (if not, it
aborts) and that $(\hat{Z}, Z, m_{\hat{Z}}, m_{\hat{B}}, Y)$ is in
one of its incomplete sessions (if not, it ignores the query).  
  Then,
$\mathcal{C}$ checks  for every value $\sigma\in G\setminus 1_G$
\emph{previously used by $\mathcal{F}$} as input to $H_K$ whether
$\sigma=Z^{cy}X^{bd+ye}$, where $d=h(m_{\hat{B}}, \hat{B}, B,  X)$ and $e=h(X, Y)$  
(in case $d,e$ undefined, $\mathcal{C}$ defines them with $h$): it
does so using the DDH-oracle $\mathcal{O}$, specifically, by
checking whether $CDH(X, B)=(\sigma/Z^{cy}X^{ye})^{d^{-1}}$. If
the answer is positive, then $\mathcal{C}$ sets $r$ to the already
determined value of $H_K(\sigma)$.

\begin{description} \item [S3.1.]In any other cases, 
  $r$ is set to be a random value in $\{0, 1\}^k$, where
$k$ is the output length of $H_K$. Note that, in this case,
$\mathcal{C}$ does not know $\sigma=Z^{cy}X^{db+ey}$, as it does
not know $b$, which also implies that $\mathcal{C}$ does not make
(actually realize) the RO-query $H_K(\sigma)$ \emph{even if the
value $\sigma$ has been well-defined 
 and known to $\mathcal{F}$}.
\end{description}
Finally, $\mathcal{C}$ marks the vector  $(\hat{Z}, Z,
m_{\hat{Z}}, m_{\hat{B}}, X,  y, Y, Z^{cy})$ as a ``\emph{complete
session}", stores $(\hat{Z}, Z, m_{\hat{Z}}, m_{\hat{B}}, X,  y,
Y, Z^{cy}, r)$ and  responds  $(\hat{Z}, Z, m_{\hat{Z}},
m_{\hat{B}}, X,
 Y,  r)$ to $\mathcal{F}$.

\end{description}

\textbf{RO queries:} $\mathcal{C}$  provides random answers to
queries to  the
random oracles $h$ and $H_K$ (made by $\mathcal{F}$),  
under the limitation that if the same RO-query is presented more
than once, $\mathcal{C}$ answers it with the same response as in
the first time. But, for each \emph{new} query $\sigma$ to $H_K$,
$\mathcal{C}$ checks whether  $\sigma=Z^{cy}X^{db+ey}$ for any one
of the stored  vectors $(\hat{Z}, Z, m_{\hat{Z}}, m_{\hat{B}}, X,
y, Y, Z^{cy},  r)$ 
 (as before, this
check is done using the DDH-oracle). If equality holds then the
corresponding $r$ is returned as the predefined $H_K(\sigma)$,
otherwise a random $r$ is returned.

\textbf{Upon $\mathcal{F}$'s termination.} When $\mathcal{F}$
halts, $\mathcal{C}$ checks whether the following conditions hold:
\begin{description}

\item [F1.] $\mathcal{F}$ outputs a valid HDR-signature $(\hat{A},
A,  m_1, m_0, X_0, Y_0, r_0)$, where $\hat{A}\neq \hat{B}$ is an
uncorrupted player. In particular, it implies
that $r_0$ should be  $H_K(\sigma_0)$, where 
$\sigma_0=A^{y_0c_0}X_0^{bd_0+y_0e_0}$, 
 $Y_0=g^{y_0}$ (chosen by $\mathcal{F}$), $c_0=h(m_1, \hat{A}, A, Y_0)$,
$d_0=h(m_0, \hat{B}, B,  X_0)$ and $e_0=h(X_0, Y_0)$.

\item [F2.]  
  $(\hat{A}, A, m_1, m_0, X_0, Y_0)$ did
not appear in any of the above responses of the simulated OAKE-HDR
signatures. 



\item [F3.] The values  $c_0=h(m_1, \hat{A}, A, Y_0)$,
 $d_0=h(m_0, \hat{B}, B,  X_0)$  and $e_0=h(X_0, Y_0)$ were 
queried 
 from the RO $h$, and the value $H_K(\sigma_0)$ was queried  
 from  $H_K$ \emph{being posterior
 to
 the queries $c_0, d_0, e_0$}. 
  Otherwise,  $\mathcal{C}$ aborts.

\end{description}

If these three conditions hold, $\mathcal{C}$ proceeds to the
``repeat experiments" below, else it aborts. 

\textbf{The repeat experiments.} $\mathcal{C}$ runs $\mathcal{F}$
again for a second time, under the same input $(B, X_0)$ and using
the same coins for 
$\mathcal{F}$. 
 There are two cases according to the
order of the  queries of $h(m_0, \hat{B}, B,  X_0)$  and $h(X_0,
Y_0)$


\begin{description}
\item [C1.] $h(m_0, \hat{B}, B,  X_0)$ posterior to $h(X_0, Y_0)$:
$\mathcal{C}$ rewinds $\mathcal{F}$ to the point of making the RO
query $h(m_0, \hat{B}, B,  X_0)$, responds back a new independent
value $d^{\prime}_0 \in_{\textup{R}} \{0, 1\}^l$. All subsequent
actions of $\mathcal{C}$ (including random answers to subsequent
RO
queries) are independent of the first run. 
 If in
this repeated run $\mathcal{F}$ outputs a successful forgery
 $(\hat{A}^{\prime}, A^{\prime}, m^{\prime}_1, m_0, X_0, Y_0,  r^{\prime}_0)$ satisfying  the conditions F1-F3
 (otherwise, $\mathcal{C}$
 aborts), which particularly
 implies that $r^{\prime}_0=H_K(\sigma^{\prime}_0)$,
 $\sigma^{\prime}_0=A^{\prime y_0c^{\prime}_0}X_0^{bd^{\prime}_0+y_0e_0}$,  $\mathcal{C}$ computes
 $CDH(U, V)=CDH(X_0, B)=
 [(\sigma_0/Y_0^{ac_0})/(\sigma^{\prime}_0/Y_0^{a^{\prime}c^{\prime}_0})]^{(d_0-d^{\prime}_0)^{-1}}$,
 where $a$ and $a^{\prime}$ are the private keys of the \emph{uncorrupted}   $\hat{A}$ and $\hat{A}^{\prime}$
 (different from $\hat{B}$, which are assumed to be known to
 $\mathcal{C}$).  Note that $(\hat{A}^{\prime}, A^{\prime},
 m^{\prime}_1)$ need not necessarily to
  equal  $(\hat{A}, A, m_1)$.

\item [C2.] $h(X_0, Y_0)$ posterior  to $h(m_0, \hat{B}, B,
X_0)$: $\mathcal{C}$ rewinds   $\mathcal{F}$ to the point of
making the RO query $h(X_0, Y_0)$, responds back a new independent
value $e^{\prime}_0 \in_{\textup{R}} \{0, 1\}^l$.
 If in
this repeated run $\mathcal{F}$ outputs a successful forgery
$(\hat{A}^{\prime}, A^{\prime}, m^{\prime}_1, m_0, X_0, Y_0,
r^{\prime}_0)$ satisfying the conditions F1-F3
 (otherwise, $\mathcal{C}$
 aborts), which particularly
 implies that $r^{\prime}_0=H_K(\sigma^{\prime}_0)$,
  $\sigma^{\prime}_0=A^{\prime y_0c^{\prime}_0}X_0^{bd_0+y_0e^{\prime}_0}$, $\mathcal{C}$ computes
 $X_0^{y_0}=((\sigma_0/Y_0^{ac_0})/(\sigma^{\prime}_0/Y_0^{a^{\prime}c^{\prime}_0}))^{(e_0-e^{\prime}_0)}$,
 and then  $CDH(U, V)=CDH(X_0, B)=
(\sigma_0/((X_0^{y_0})^{e_0}\cdot
Y_0^{ac_0}))^{d_0^{-1}}$.


\end{description}

\end{minipage}
\\

\hline
\end{tabular}
\caption{\label{OAKE-DCR} Reduction from GDH to OAKE-HDR forgeries}
\end{center}

\vspace{-0.7cm}
\end{figure}

Now, we consider the case that the forger $\mathcal{F}$
is against the signer $\hat{B}$ itself (i.e., $\hat{A}=\hat{B}$).
We further distinguish two cases: (1) $Y_0\neq X_0$  and (2) $Y_0=X_0$.

\vspace{-0.2cm}
\begin{corollary}\label{HDR-signerXY-SEC}
 Under the GDH assumption, and additionally the KEA
 assumption, (public-key free) OAKE-HDR  signatures of $\hat{B}$, with
offline pre-computed and exposable  $(y, Y, A^{cy})$, are \emph{strongly}
secure in the random oracle model, 
 with respect to the signer $\hat{B}$ itself with $Y_0\neq X_0$.

\end{corollary}
\vspace{-0.2cm}

\noindent \textbf{Proof (sketch).}
 The main difference between the proof of Corollary  \ref{HDR-signerXY-SEC}
 and that of Theorem \ref{HDR-analysis-SEC} is that, here, the forger
 outputs with non-negligible probability a successful forgery of
 the form: $(m_1, m_0, \hat{B}, B, 
\hat{B},  B, X_0, Y_0, r_0)$, i.e., $\hat{A}=\hat{B}$, where $r_0=H_K(\sigma_0)$,
$\sigma_0=B^{c_0y_0}X_0^{d_0b+e_0y_0}$,  $c_0=h(m_1, \hat{B}, B,
Y_0)$, $d_0=h(m_0, \hat{B}, B,
  X_0)$,      $e_0=h(X_0, Y_0)$. The key point is that, by
  performing the rewinding experiments, we cannot directly output
  the $CDH(B, X_0)$, as we do not know the private key $b$ of
  $\hat{B}$.
Recall that, in this case, the uncorrupted player  and the signer are the same.

  We modify the algorithm $\mathcal{C}$ depicted in Figure
  \ref{OAKE-DCR}  as follows: the actions
  of $\mathcal{C}$ remain unchanged until the rewinding
  experiments;  but $\mathcal{C}$ performs the rewinding experiments
  according to the order of the RO-queries $c_0, d_0, e_0$.

\vspace{-0.3cm}
  \begin{description}

\item [$d_0$ posterior to $c_0, e_0$.] In this case, by rewinding
$\mathcal{F}$ to the point of making the query $d_0=h(m_0,
\hat{B}, B,   X_0)$, and redefines $h(m_0, \hat{B}, B,   X_0)$ to
be a new independent $d^{\prime}_0$,  $\mathcal{C}$ will get
$\sigma^{\prime}_0=B^{c_0y_0}X_0^{d^{\prime}_0b+e_0y_0}$. Then,
from $\sigma_0$ and $\sigma^{\prime}_0$, $\mathcal{C}$ gets that
$CDH(B,
X_0)=(\sigma/\sigma^{\prime}_0)^{(d_0-d^{\prime}_0)^{-1}}$.
\emph{Note that, in this case, $\mathcal{C}$ does not rely on the
KEA assumption for breaking the CDH assumption} (but still with
the DDH-oracle).

\vspace{-0.2cm}
\item [$c_0$ posterior to $d_0, e_0$.] In this case, by rewinding
$\mathcal{F}$ to the point of making the query $c_0=h(m_1,
\hat{B}, B,   Y_0)$, and redefines $h(m_1, \hat{B}, B,   Y_0)$ to
be a new independent $c^{\prime}_0$,  $\mathcal{C}$ will get
$\sigma^{\prime}_0=B^{c^{\prime}_0y_0}X_0^{d_0b+e_0y_0}$. Then,
from $\sigma_0$ and $\sigma^{\prime}_0$, $\mathcal{C}$ gets
$CDH(B,
Y_0)=B^{y_0}=(\sigma/\sigma^{\prime}_0)^{(c_0-c^{\prime}_0)^{-1}}$.
That is, given $B$, $\mathcal{C}$ can output $(Y_0, B^{y_0})$. By
the KEA assumption, it implies that $\mathcal{F}$ knows $y_0$
(which can be derived from the internal state of $\mathcal{F}$).
More formally, there exists an algorithm that, given $B$ and $X_0$ and the random coins of $\mathcal{C}$ and
$\mathcal{F}$,  can successfully output $y_0$. Now, with the
knowledge of $y_0$, $CDH(B, X_0)$ can be derived from $\sigma_0$
(or $\sigma^{\prime}_0$).

\vspace{-0.2cm}
\item [$e_0$ posterior to $c_0, d_0$.] In this case, by rewinding
$\mathcal{F}$ to the point of making the query $e_0=h(X_0, Y_0)$,
and redefines $h(X_0,   Y_0)$ to be a new independent
$e^{\prime}_0$,  $\mathcal{C}$ will get
$\sigma^{\prime}_0=B^{c_0y_0}X_0^{d_0b+e^{\prime}_0y_0}$. Then,
from $\sigma_0$ and $\sigma^{\prime}_0$, $\mathcal{C}$ gets
$CDH(X_0,
Y_0)=X_0^{y_0}=(\sigma/\sigma^{\prime}_0)^{(e_0-e^{\prime}_0)^{-1}}$.
Then, by the KEA assumption, the knowledge of $y_0$ can be
derived, with which $CDH(X_0, B)$ can then be computed
  \hfill $\square$


  \end{description}

\vspace{-0.3cm}
\begin{corollary}\label{HDR-signerXX}
 Under the computational Diffie-Hellman (CDH)
assumption, (public-key free) OAKE-HDR signatures of $\hat{B}$, with
offline pre-computed and exposable  $(y, Y, A^{cy})$, are \emph{strongly}
secure in the random oracle model, 
 with respect to the signer $\hat{B}$ itself with $Y_0=X_0$.

\end{corollary} 
\vspace{-0.2cm}

After establishing the \emph{strong} unforgeability security of (s)OAKE-HDR, similar to the analysis of HMQV, the analysis of (s)OAKE within the CK-framework is quite straightforward and  less interesting.
In particular,  the special  structure of  sOAKE-HDR
  also  much simplifies 
the security analysis of  sOAKE by only using  the standard forking lemma \cite{PS00},  and
tightens the security reductions.
%
Full details   are referred to 
   Appendix  \ref{basicanalysis}.


\vspace{0.5cm}

 \appendix


\vspace{-0.5cm}

\section{Variants of (H)MQV} \label{hmqvvariants}
  Three-round HMQV (resp., MQV)  adds
key confirmation as follows: 
 let $K_m=H_K(K_{\hat{A}}, 0)=H_K(K_{\hat{B}}, 0)$, 
 $\hat{B}$ uses $K_m$ as the MAC key to authenticate 0 (resp., $(2,
\hat{B}, \hat{A}, Y, X)$)  in the second-round of HMQV (resp.,
MQV);  and $\hat{A}$ uses $K_m$ to authenticate 1 (resp., $(3,
\hat{A}, \hat{B}, X, Y)$) in an additional third-round of HMQV
(resp., MQV). The session-key is set to be $K=H_K(K_{\hat{A}},
1)=H_K(K_{\hat{B}}, 1)$.

 In one-round HMQV, only $\hat{A}$ sends $X$, and the session-key is derived as follows:
 $K_{\hat{A}}=B^{x+da}$, $K_{\hat{B}}=(XA^d)^b$, $d=h(X, \hat{A},
\hat{B})$, $K=H_K(K_{\hat{A}})=H_K(K_{\hat{B}})$.




\vspace{-0.2cm}
\section{NMJPOK: Motivation, Formulation, and Implementations} 
\label{AppSSJPOK}
\vspace{-0.1cm}



We consider an adversarial setting, where polynomially many instances (i.e., sessions) of a Diffie-Hellman protocol $\langle \hat{A}, \hat{B}\rangle$  are run concurrently over an asynchronous network  like the Internet.
To distinguish concurrent
sessions,  each session   run
at the side of an uncorrupted  player is  labeled by a tag, which
is  the concatenation, 
in the order of session initiator and
then session
responder,
of players' identities/public-keys and   
   DH-components available from the session transcript.  A session-tag is complete if it consists of a complete set of all these components.

In this work, we study the mechanisms for  \emph{non-malleably} and \emph{jointly}   proving the knowledge  of  both $b$ and $y$  w.r.t. a challenge DH-component $X$ between the prover $\hat{B}$ (of public-key $B=g^b$ and DH-component $Y=g^y$) and the verifier  $\hat{A}$ (who presents the challenge DH-component $X=g^x$), where $b,y,x\in Z^*_q$. In particular, we investigate
joint proof-of-knowledge (JPOK) of the type  $JPOK_{(b,y)}=f^h_0(X^b,  aux_0)\cdot f^h_1(X^y,   aux_1)$ in the random oracle model,
 where $f^h_0$ and $f^h_1$ are some   functions from $\{0, 1\}^*$ to $G\setminus 1_G$ with  oracle access to an RO $h:\{0,1\}^*\rightarrow \{0,1\}^l$, $aux_0$ and $aux_1$ are some  public values. Moreover, we look for solutions  of $JPOK_{(b,y)}$ such that $JPOK_{(b,y)}$ can be efficiently computed with  one single exponentiation by the knowledge prover.
 Note that the tag for a complete session of  $JPOK_{(b,y)}$ 
  is $(\hat{A},\hat{B},B,X,Y)$. 
 The possibility of  NMJPOK without ROs (based upon pairings) is left  to be studied in a subsequent separate paper. 
 In the rest of this paper, we denote by the output length,  i.e., $l$, of $h$ as the security parameter.

 One naive solution of $JPOK_{(b,y)}$ is just to set $JPOK_{(b,y)}=X^b\cdot X^y=X^{b+y}$. But, such a naive solution is totally insecure, for example, an adversary $\mathcal{A}$  can easily impersonate the prover $\hat{B}$ and pre-determine the value of $JPOK_{(b,y)}$ to be $1_G$, by simply setting $Y=B^{-1}$. The underlying reason is:   $\mathcal{A}$
can malleate $B$ and $Y$ into $X^{y+b}$ \emph{by maliciously
correlating the values of $y$ and $b$}, but actually without
knowing  either of them.   A further remedy of  this situation is to mask the exponents $b$ and $y$ by
some random values. In this case, the proof is denoted as $JPOK_{(b,
y)}=X^{db+ey}$, where $d$ and $e$ are random values  (e.g.,  $d=h(X, \hat{B})$  and $e=h(Y, \hat{A})$ as in HMQV in the RO model).
  The
intuition with this remedy solution is: since $d$ and $e$ are
random values, the values of $db$ and $ey$ are also random (even if the values  $Y$ and $B$, and thus the values
of $y$ and $b$, may be  maliciously correlated). 
  This  intuition however
 turns out also to be wrong. With the values $d=h(B,\hat{A})$  and $e=h(X,\hat{B})$
 as an illustrative example, after receiving $X$ an adversary $\mathcal{A}$ can  generate and send
$Y=B^{-d/e}$, and in this case $JPOK_{(b, y)}=X^{db+ey}=1_G$. This  shows that masking  $b$ and $y$
 by random values is also not sufficient 
 for ensuring the non-malleability
 of  $JPOK_{(b,y)}$.  
 The key point here is that the values $db$ and $ey$
  are \emph{not} necessarily  \emph{independent}. A series of careful investigations bring us to the following principles for proving DH knowledges non-malleably  and jointly:

\textsc{Inside Computational Independence.}  Denote $S_0=\{X, B\}$, $Z_0=CDH(X,B)=g^{xb}$, $F_0=f^h_0(Z_0, aux_0)$,  $S_1=\{X, Y\}$, $Z_1=CDH(X, Y)$ and  $F_1=f^h_1(Z_1,  aux_1)$. The key principle is: the inside  multiplied components $F_0$ and $F_1$ of $JPOK_{(b, y)}$ 
should  be \emph{computationally independent}, no matter how a  malicious knowledge prover  $\hat{B}$ (of public-key $B=g^b\in G$) 
   does. That is, the
adversarial attempts  at $Z_{\delta}$ for any $\delta\in \{0, 1\}$  should be essentially
    \emph{sealed} 
 (i.e., 
   localized)   to $F_{\delta}$, 
   and are isolated (i.e., ``independent")
   from  the adversarial attempts at
  $Z_{1-\delta}$. 
This essentially ensures that no matter how the possibly malicious knowledge-prover $\hat{B}$
does,  to compute $JPOK_{(b,y)}$ $\hat{B}$ has to compute  two
``\emph{independent}" DH-secrets $F_0$ and $F_1$ 
 w.r.t. the
fresh challenge $X$, which    implies that $\hat{B}$ does indeed ``know" both
$b$ and $y$.

\vspace{-0.1cm}

 \begin{definition}[computational independence]\label{DefComInd}
 We formulate two types of ``computational independence" w.r.t. $JPOK_{(b, y)}$:

 \textsf{(1) Self-sealed computational independence.}  Given arbitrary values $(\alpha, \beta)\in (G\setminus 1_G)^2$, no matter how a malicious  $\hat{B}$  does, both $\Pr[F_0=\alpha]$ and $\Pr[F_1=\beta]$ are  negligible.

 \textsf{(2) Committed  computational independence.} There exists $\delta\in \{0, 1\}$ such that for any $\alpha \in G\setminus 1_G$ 
  $\Pr[F_\delta=\alpha]$ is negligible, no matter how a malicious $\hat{B}$  does. This captures the independence of $F_{\delta}$ on $F_{1-\delta}$, i.e., the infeasibility of adversarial attempts by a malicious prover  on setting $F_{\delta}$ to be correlated to $F_{1-\delta}$;
 On the other hand, the value $F_{1-\delta}$  is committed to $F_\delta$,  
     in the sense that
     \vspace{-0.2cm}
      \begin{itemize}
     \item $S_{1-\delta} \bigcup aux_{1-\delta} \subseteq aux_{\delta}$.
        \vspace{-0.1cm}
        \item Given  $(Z_\delta, aux_{\delta})$ that determines $F_\delta=f^h_\delta(Z_\delta,aux_\delta)$,  no efficient algorithm 
      can provide, with non-negligible probability, $(S^{\prime}_{1-\delta}, aux^{\prime}_{1-\delta})\subseteq  aux^\prime_{\delta}$ (w.r.t. the same challenge $X=S_{1-\delta}\cap S^{\prime}_{1-\delta}$ from $\hat{A}$ and $aux_\delta-aux_{1-\delta}=aux^\prime_\delta-aux^\prime_{1-\delta}$) such that 
      $S^{\prime}_{1-\delta}\bigcup aux^{\prime}_{1-\delta} \neq  S_{1-\delta}\bigcup aux_{1-\delta}$ but 
      $f^h_{\delta}(Z_\delta, aux_{\delta})=f^h_{\delta}(Z_\delta, aux^{\prime}_{\delta})$.
        That is, any adversarial attempt by a malicious prover on setting  $F_{1-\delta}$ to be  correlated to a given value $F_\delta$,  by changing   $\{S_{1-\delta}, aux_{1-\delta}\}$ into $\{S^{\prime}_{1-\delta},aux^{\prime}_{1-\delta}\}$ w.r.t. the same random challenge $X=S_{1-\delta}\cap S^{\prime}_{1-\delta}$ and $aux_\delta-aux_{1-\delta}=aux^\prime_\delta-aux^\prime_{1-\delta}$ (for example, by simply changing $B$ for the case of $\delta=1$ or $Y$ for the case of $\delta=0$), will cause the value $F_\delta$ itself changed that in turn  determines and commits to  the value $F_{1-\delta}$ (while $\Pr[F_\delta=\alpha]$ is negligible for any $\alpha\in G\setminus 1_G$). This implies the infeasibility of adversarial attempt on setting $F_{1-\delta}$ to be correlated to $F_\delta$, i.e.,  the ``computational independence" of $F_{1-\delta}$ on $F_{\delta}$.

\end{itemize}
\vspace{-0.2cm}

The probabilities are taken over  the random coins used by the malicious  $\hat{B}$ and the honest   $\hat{A}$, and the choice of the random function $h$ in the RO model.  

\end{definition}
\vspace{-0.1cm}

Informally speaking, the underlying rationale of $NMJPOK_{(b,y)}$ is: given a random challenge $X$,
 no matter how a malicious $\hat{B}$ chooses the values $Y=g^y$ and $B=g^b$ (where the values $y$ and $b$ can be arbitrarily correlated), it actually has no control over the values $db$ and $ey$ in the RO model. That is, by the birthday paradox it is infeasible for a malicious $\hat{B}$ to set $db$ (resp., $ey$) to some predetermined value
 with non-negligible probability  in the RO model (in order to make the values $db$ and $ey$ correlated).  Alternatively speaking, given a random challenge $X$, (by the birthday paradox) it is infeasible for a malicious $\hat{B}$  to output $B=g^b$ and $Y=g^y$  such that the values $db$ and $ey$ satisfy some predetermined (polynomial-time computable) relation with non-negligible probability in the RO model.

 The situation with $sNMJPOK_{(b,y)}$ is a bit different. Though as in $NMJPOK_{(b,y)}$, the malicious prover  $\hat{B}$ is infeasible to set $ey$ to a  predetermined value,   $\hat{B}$ can always set the value $db=b$ at its wish as $d=1$ for $sNMJPOK_{(b,y)}$. But,  $\hat{B}$ is still infeasible to set the value $b$ correlated to $ey=h(B,X,Y)y$, particularly because the value $B$ is put into the input of $e$. Specifically, for any value $B$ (that determines the value $b$)  set by $\hat{B}$, with the goal of making $b$ and $ey$ correlated, the probability that the values $ey=h(B,X,Y)y$ and $b$  satisfy some predetermined (polynomial-time computable) relation is negligible in the RO model (again by the birthday paradox).  In particular, the probability that $\Pr[b=f(ey)]$  or $\Pr[f(b)=ey]$, where $f$ is some predetermined polynomial-time computable function (that is in turn determined by some predetermined polynomial-time computable relation), is negligible  in the RO model, no matter how the malicious $\hat{B}$ does.

\textsf{Outside Non-Malleability.} As JPOK may be composed  with other protocols in practice, another principle is that  the JPOK  provided by one party
in a session  should be bounded to that session, in the sense
that the JPOK should not be malleated \emph{into} or \emph{from}
other sessions. This is captured by the following definition, which particularly implies the property   of ``key control" \cite{LMQSV03} for DHKE. 


\vspace{-0.1cm}
\begin{definition} [tag-binding self-seal (TBSS)]
For a DH protocol in the RO model, denote by   $Z_{Tag}$  the random variable  of
the shared  DH-secret in $G$ (say, JPOK or session-key) determined by a complete  session-tag $Tag$ (taken over the choice of the random function $h$ in the RO model).
 We say it is \emph{tag-binding self-sealed},
  if 
for
 any  $\alpha \in G\setminus 1_G$ and any  complete $Tag$, 
 $\Pr[Z_{Tag}=\alpha]\leq O(\frac{1}{2^l})$ where $l$ is the security parameter. 
 The probability is   taken over  the choice of the random function $h$ in the RO model. 





\end{definition}
\vspace{-0.1cm}

 The definition of TBSS
particularly implies  that: given an arbitrary yet  complete session-tag   $Tag$, by the birthday paradox  no efficient (polynomial-time)  algorithm can, \emph{with non-negligible probability}, 
 output a different $Tag^{\prime}\neq Tag$
such that $Z_{Tag^{\prime}}$ and $Z_{Tag}$ collide in the sense
$Z_{Tag^{\prime}}=Z_{Tag}$ in the RO model assuming $h$ is a random function. In more detail, by the birthday paradox,  the probability that an efficient algorithm  finds  two colliding tags $(Tag, Tag')$ such that $Z_{Tag}=Z_{Tag'}$ is bounded by $O(\frac{T^2}{2^l})$, where  $T=poly(l)$ is the running time of the algorithm.
In a sense, 
the
 DH-secret  determined by   a complete session-tag  is ``bounded" to this specific session, and  is essentially ``independent" of the outside world composed  concurrently with the current session. In particular, the shared DH-secret is random and unpredictable.   

\textsf{ TBSS vs. contributiveness}. 
The work \cite{ACCP09}  introduced the notion of ``contributiveness" property for password-authenticated group key exchange protocols, which roughly says that the distributions of session-keys are guaranteed to be random, as long as there are enough honest players in a session.
  We noted that our TBSS definition, originally presented in \cite{OAKE07,OAKE08}  independently  of \cite{ACCP09},   has similar security guarantee. 
    As we shall see,  (H)MQV lacks the TBSS property by the EDA attacks presented in Section \ref{SecBeyondCK}, which  implies also that the TBSS property is not captured by the CK-framework.  


We say that $JPOK_{(b,y)}$ is a non-malleable joint proof-of-knowledge (NMJPOK),  of the knowledges $(b,y)$  w.r.t. the random DH-component challenge $X$, if $JPOK_{(b,y)}$ satisfies both the above two principles.

\textbf{Preferable candidates for NMJPOK.}
 Guided by the  above  principles, we propose
two preferable solutions for NMJPOK  
in the RO model:

\vspace{-0.4cm}
\begin{itemize} \item Self-sealed JPOK (SSJPOK):
$SSJPOK_{(b,y)}=X^{db+ey}$, where $d=h(\hat{A}, \hat{B}, B, X)$
and $e=h(X, Y)$; Specifically, $aux_0=\{\hat{A}, \hat{B}, B, X\}$ and $aux_1=\{X,Y\}$, $F_0=f^h_0(X^b, aux_0)=X^{bh(aux_0)}$ and $F_1=f^h_1(X^y, aux_1)=X^{yh(aux_1)}$. Here, $h:\{0,1\}^*\rightarrow \{0,1\}^l/{0}\subseteq Z^*_q$ is a hash function  and $l\approx |q|$ (in the unlikely case that $h(x)=0$ for some $x$, the output of $h(x)$ can be defined by default to be  a value in $Z^*_q-\{0,1\}^l$).
\vspace{-0.3cm}
\item   Single-hash SSJPOK (sSSJPOK):
$sSSJPOK_{(b,y)}=X^{db+ey}$, where $d=1$ and  $e=h(\hat{A},
\hat{B},  B,  X, Y)$; Specifically, $aux_0$ is empty and $aux_1=\{\hat{A},
\hat{B},  B,  X, Y\}$, $F_0=f^h_0(X^b, aux_0)=X^b$ and $F_1=f^h_1(X^y, aux_1)=X^{yh(aux_1)}$.

\end{itemize}
\vspace{-0.2cm}

%
%
%
%

Needless to say, there are other NMJPOK candidates (e.g., $d=h(B,X)$ and $e=h(\hat{A},\hat{B}, X, Y)$, or $d=h(\hat{A},
\hat{B},  B,  X, Y)$ and $e=h(Y,X, \hat{B},\hat{A})$, etc). But the above
explicitly proposed  solutions enjoy the following
advantageous properties, which make them more desirable:
\vspace{-0.2cm}
\begin{itemize}
\item   Post-ID, modular and offline computability of SSJPOK. 
 Specifically, as the input of $e$ does not
include $\hat{A}$'s identity and public-key,   $\hat{A}$ can first
send $X$ without revealing its identity information. In this case,
$\hat{B}$ can first compute $X^{ey}$, and then $X^{db}$ only after
learning $\hat{A}$'s identity and public-key. Also, 
without inputting $Y$ into $d$  allows $\hat{A}$ to pre-compute
$B^{dx} (=X^{db})$ prior to the protocol run. 
\vspace{-0.2cm}
\item  sSSJPOK is preferable  because of its offline computability, more efficient computational
complexity and the less use of hash function $h$. 
\end{itemize}
\vspace{-0.2cm}





It is quite straightforward  to check that,  in the RO model,  SSJPOK (resp., sSSJPOK) satisfies self-sealed (resp., committed) computational independence, and both of them are tag-binding self-sealed.
 In more details, for SSJPOK,  for any given values $(B, Y)$ (which
determine $(b, y)$) output by a malicious prover $\hat{B}$  and any value $\hat{\beta} \in Z^*_q$
$\Pr[db=\hat{\beta}]$ (resp., $\Pr[ey=\hat{\beta}]$) is constant: either 0
or $\frac{1}{2^l-1}$ in the RO model (no matter how a malicious prover  $\hat{B}$ does).  
  The committed computational independence of sSSJPOK is from  the observation: $\{X, B\}$ (that determines $F_0=X^b$) are committed to $F_1=X^{yh(aux_1)}$  in the RO model as $\{X, B\}\subseteq aux_1$. The TBSS property of (s)SSJPOK can be derived by a
straightforward calculation. Proof details that (s)SSJPOK are NMJPOK in the RO model are given below.

\begin{proposition}\label{SSJPOKproof}
SSJPOK is NMJPOK in the RO Model.

\end{proposition}
\noindent \textbf{Proof.}
We first prove the self-sealed computational independence of SSJPOK in the RO model.
Note that for SSJPOK, $F_0=X^{db}=X^{h(\hat{A},\hat{B},B,Y)b}$ and $F_1=X^{ey}=X^{h(X,Y)y}$, where $b,y,x\in Z^*_q$.
For any given  challenge $X\in G\setminus 1_G$,  each pair of values $(B=g^b, Y=g^y)\in (G\setminus 1_G)^2$ (that
determine $(b, y) \in (Z^*_q)^2$)  
and any pair of given  values $\alpha=g^{\hat{\alpha}},\beta=g^{\hat{\beta}} \in (G\setminus 1_G)^2$, where $\hat{\alpha},\hat{\beta}\in Z^*_q$,   we consider the set of values that $F_0$ can be assigned in the RO model $S_{F_0}=\{X^{db}|0\leq d=h(\hat{A},\hat{B},B,Y)\leq 2^l-1\}$ and also the set of values that $F_1$ can be assigned in the RO model $S_{F_1}=\{X^{ey}|0\leq e=h(X,Y)\leq 2^l-1\}$.  If $\alpha \not\in S_{F_0}$ or $d=0$ (resp.,  $\beta \not\in S_{F_1}$ or $e=0$), then we have $\Pr[F_0=\alpha]=0$ (resp.,  $\Pr[F_1=\beta]=0$). If $\alpha \in S_{F_0}$ (resp.,  $\beta \in S_{F_1}$), then we have $\Pr[F_0=\alpha]=\frac{1}{2^l-1}$ (resp., $\Pr[F_1=\beta]=\frac{1}{2^l-1}$)  in the RO model.
As the malicious prover $\hat{B}$ is polynomial-time, we have that, no matter the polynomial-time malicious $\hat{B}$ does on a challenge $X$,  the probability that it outputs $B,Y$ such that $F_0=\alpha$ and $F_1=\beta$ is negligible. Specifically, suppose $N=2^l-1$ and $T=poly(l)$ is the running time of $\hat{B}$,  by the birthday paradox  the probability that on input $(X, \alpha, \beta)$ the malicious $\hat{B}$ outputs $(B,Y)$ such that  $F_0=\alpha$ or  $F_1=\beta$ is at most $\frac{T(T-1)}{2N}$ that is negligible (in $l$).

Next we prove the TBSS property of SSJPOK in the RO model, which is based on and can be easily derived from the NMJPOK property of OAKE.  For a complete session of $SSJPOK$, its tag is: $Tag=(\hat{A},\hat{B},B=g^b, X=g^x, Y=g^y)$, where $b,x,y\in Z^*_q$, we consider the value $Z_{Tag}=X^{db+ey}=X^{h(\hat{A},\hat{B},B,Y)b}\cdot X^{h(X,Y)y}$ in the RO model where $h$ is assumed to be a random oracle. As for each value $\alpha \in G\setminus 1_G$, $\Pr[X^{h(\hat{A},\hat{B},B,Y)b}=\alpha]\leq \frac{1}{2^l-1}$ and $\Pr[ X^{h(X,Y)y} =\alpha]\leq \frac{1}{2^l-1}$ in the RO model, we get (by straightforward calculation) that $\Pr[Z_{Tag}=\alpha] \leq O(\frac{1}{2^l})$. \hfill $\square$

\begin{proposition}\label{sSSJPOKproof}
sSSJPOK is NMJPOK in the RO model.

\end{proposition}
\noindent \textbf{Proof.} We first show  the committed computational independence property of  sSSJPOK. Similar to the analysis of Proposition \ref{SSJPOKproof}, for the case $\delta=1$ we have that for any given $\alpha\in G\setminus 1_G$ and any DH-component challenge $X$, and any $(B,Y)\in (G\setminus 1_G)^2$, 
$\Pr[F_\delta=X^{yh(aux_1)}=X^{yh(\hat{A},\hat{B},B,X,Y)}=\alpha]\leq \frac{1}{2^l-1}$ in the RO model, where $\delta=1$. As the malicious $\hat{B}$ is polynomial-time, we have the probability that the malicious $\hat{B}$ outputs $(B,Y)$, given a random challenge $X$ and a given value $\alpha \in G\setminus 1_G$, such that $F_1=\alpha$ is negligible in the RO model.\footnote{Specifically, by the birthday paradox, the probability is at most $O(\frac{T^2}{2^l})$, where $T=poly(l)$ is the running time of $\hat{B}$.}   Then, the committed computational independence of sSSJPOK is from  the following observation that $X^b$ is committed to $X^{yh(\hat{A},\hat{B},B,X,Y)}$. Specifically,
 \begin{itemize}
\item  $S_{1-\delta}=S_0=\{X, B\}\subseteq aux_{\delta}=aux_1=\{\hat{A},\hat{B},B,X,Y\}$. Note that the value $F_0=Z_0=X^b$ (resp., $F_1=f^h_1(Z_1,aux_1)=f^h_1(X^y,aux_1)=X^{yh(aux_1)}=X^{yh(\hat{A},\hat{B},B,X,Y)}$) is determined by $S_0=\{X,B\}$ (resp., $aux_1=\{\hat{A},\hat{B},B,X,Y\}$),  and $aux_0$ is empty for sSSJPOK.

\item  Given $Z_\delta=Z_1=X^y$ and $aux_{\delta}=aux_1=\{\hat{A},\hat{B},B,X,Y\}$, for any $B^\prime \neq B$ such that  $S^\prime_0=\{X, B^\prime \}\subseteq aux^\prime_1=\{\hat{A},\hat{B},B^\prime,X,Y\}$,   we get   $\Pr[f^h_1(Z_1, aux_1)=f^h_1(Z_1,aux^\prime_1)]=\Pr[X^{yh(\hat{A},\hat{B},B,X,Y)}=X^{yh(\hat{A},\hat{B},B^\prime,X, Y)]}\leq \frac{1}{2^l-1}$.  Thus for any polynomial-time algorithm, the probability that it, on input $Z_1,aux_1$, outputs $S^\prime_0=\{X, B^\prime \} $ for $B^\prime \neq B$ such that $X^{yh(\hat{A},\hat{B},B,X,Y)}=X^{yh(\hat{A},\hat{B},B^\prime,X, Y)}$ is negligible (again by the birthday paradox).

   %
%
\end{itemize}

Next, we show the TBSS property of sSSJPOK in the RO model, which is based on and can be easily derived from the NMJPOK property of OAKE. For the tag $Tag=(\hat{A},\hat{B},B,X,Y)$ of a complete session of sSSJPOK, we consider the value $Z_{Tag}=X^{b+yh(\hat{A},\hat{B},B,X,Y)}=X^b\cdot X^{yh(\hat{A},\hat{B},B,X,Y)}$. No matter what value $X^b$ is, for any value $\alpha\in G\setminus 1_G$ we have  $\Pr[X^{yh(\hat{A},\hat{B},B,X,Y)}=\alpha]\leq \frac{1}{2^l-1}$ in the RO model. Thus, for any value $\alpha\in G\setminus 1_G$ we have also that $\Pr[Z_{Tag}=\alpha]\leq \frac{1}{2^l-1}=O(\frac{1}{2^l})$.
\hfill $\square$

\section{Some Variants of (s)OAKE}\label{NOvariants}

\textbf{One-round   OAKE (oOAKE):}  The player $\hat{A}$ sends
$X=g^x$ to $\hat{B}$. Normally, $\hat{A}$ is a client machine and
$\hat{B}$ is a server machine.  Let $K_{\hat{A}}=B^{a+ex}$ and
$K_{\hat{B}}=A^bX^{eb}$,  
 where $e=h(\hat{A}, A,
\hat{B}, B, X)$ and the session-key is
$K=H_K(K_{\hat{A}})=H_K(K_{\hat{B}})$. For oOAKE, it is also
recommend to set the output length of $h$ to be  shorter, e.g.,
$|q|/2$, to ease the computation of
$K_{\hat{B}}=A^bX^{eb}=(AX^e)^b$ in some  application scenarios
(e.g., when the pre-computation of $A^b$ is inconvenient).

 Note that  the computational complexity of $\hat{A}$ is 2
 exponentiations in total and all the
computation of $\hat{A}$ can be offline.  To improve the on-line
efficiency of $\hat{B}$, the player $\hat{B}$ can pre-compute
$A^b$ in an off-line way (and store it in a database entry
corresponding to the client
$\hat{A}$),  
  and only on-line computes $X^{eb}$ and $X^q$ which amounts to about 1.2
exponentiations (it is recommended for  $\hat{B}$ to explicitly
check the   subgroup membership  of $X$). In case of embedded
subgroup test, $\hat{B}$ should explicitly check $X\in G^{\prime}$
and $X^{ebt}\neq 1_G$ (only checking $K_{\hat{B}}\neq 1_G$ is not
sufficient to prevent the small subgroup attack).   
 We remind   that oOAKE intrinsically suffers from the key compromise  impersonation (KCI) vulnerability in case $\hat{B}$'s static secret-key $b$ is compromised,  and lacks perfect forward secrecy (the same vulnerabilities hold also for one-round variant of HMQV). 



\textbf{Robust (s)OAKE:} The only difference between robust (s)OAKE and (s)OAKE is that, the values  $K_{\hat{A}}$ and $K_{\hat{B}}$ in robust (s)OAKE are set to be: $K_{\hat{A}}=B^{a+xd}Y^{ac+xe}$  and $K_{\hat{B}}=A^{b+yc}X^{bd+ye}$. Specifically, the values $K_{\hat{A}}$ and $K_{\hat{B}}$ in OAKE and sOAKE are now multiplied with the value $g^{ab}$ in robust OAKE and robust sOAKE.

We show in  Appendix \ref{robustHDR} that the provable security of (s)OAKE in the CK-framework can be easily  extended to robust (s)OAKE under the same complexity assumptions.

 \textbf{Adding (explicit) mutual authentication.}
  For adding mutual authentication to (s)OAKE, 
 besides the
session-key $K$ we also need a MAC-key $K_m$ to be used within the
protocol run (but erased after the protocol run). Both the session-key and MAC-key are derived from the shared DH-secret $K_{\hat{A}}=K_{\hat{B}}$, and  are independent in the random oracle model.
%
%
For (s)OAKE with  mutual authentication,  $\hat{B}$ sends an additional value
$t_B=MAC_{K_m}(1)$ in the second-round, and $\hat{A}$ sends
$t_A=MAC_{K_m}(0)$ in an additional third-round. For oOAKE with mutual authentication,
   the player $\hat{A}$ can
additionally send $t_A=MAC_{K_m}(0)$  in the first-round, and  the
player $\hat{B}$ responds back $MAC_{K_m}(1)$ in the subsequent
round. In practice, the message authentication code  MAC  can be instantiated
with HMAC \cite{BCK96}.

\section{More Discussions on the Specification of (s)OAKE}\label{morespecifications}
\textbf{Subgroup test vs. ephemeral DH-exponent leakage.}
We note that the damage caused by  ignoring the subgroup test of
peer's DH-component (but still  with  the supergroup $G^{\prime}$ membership
check) can be much   relieved (and even waived),   if the ephemeral
private values generated 
 within  the protocol
run are well-protected.
For example, even if an  adversary learns some partial information
about $db+ey$ by issuing a small subgroup attack against the
honest $\hat{B}$ (by setting $X$ to be in a small subgroup), it
still cannot derive the value  $b$ without compromising the
ephemeral value $y$. Also  note that the adversary actually
cannot derive the full value of $db+ey$ by small subgroup attacks,
as the DH-exponent $y$ is independent at random in
each session. In this case, we suggest that embedded subgroup
test is sufficient.
For presentation simplicity and unity, in the rest of this paper,
 it is assumed that $t=\frac{N}{q}$ for implementations with
embedded subgroup test, 
 and $t=1$ with explicit subgroup test. 



 \textbf{Ephemeral private values exposable to adversary.} 
The ephemeral private values exposable to adversary,  
 generated by the honest $\hat{B}$ (resp., $\hat{A}$) during
the protocol run,  are specified to be:  $y$ (resp., $x$) if $\hat{B}$ (resp.,
$\hat{A}$) does not  pre-compute $A^{cy}$ (resp., $B^{dx}$),  or
$(y, A^{cy})$ (resp., $(x, B^{dx})$) if $\hat{B}$ (resp.,
$\hat{A}$)  pre-computes  $A^{cy}$ (resp., $B^{dx}$). Other
ephemeral private values are erased promptly after use.  We remark
all ephemeral private values, except for the session-key in case
the session is successfully finished,  generated by an honest
party 
 within the protocol run 
 are
erased after the session is completed (whether finished or
aborted). For expired sessions,  the  session-keys are also
erased.

\section{More Discussions on the Security of (s)OAKE vs. HMQV}\label{AppExposedDH}

Assuming \emph{all}  the DH-components generated by \emph{all} uncorrupted players are \emph{not}
exposed  to the attacker prior to the sessions involving them (e.g., all honest players only generate fresh  ephemeral DH-components \emph{on the fly}, i.e., without pre-computation,
 in each session),
  and assuming \emph{all} the ephemeral DH-exponents generated during session runs are  unexposed to the attacker,
 the SK-security of HMQV can be  based on the CDH
 assumption, while we do not know how to prove this property with (s)OAKE. This is the only advantage of HMQV over (s)OAKE that we can see. 

 However, as already  stressed in \cite{K05}, security against exposed DH-exponents is deemed to be the main and prime concern  for any robust DHKE, and security against exposed offline pre-computed values (particularly, the DH-components) is important to both lower-power devices and to high volume servers \cite{K05}. The reason is, as pointed out in \cite{K05}, many applications in practice
will boost protocol performance by pre-computing and storing values  for later use in the
protocol.  In this case,
however, these stored values  are more vulnerable to leakage, 
particularly when DHKE is deployed in hostile environments with plagued
spyware or virus and in view of that the offline pre-computed DH-components are much less protected in practice as they are actually public values to be exchanged in plain.

 Also, for DHKE protocols running concurrently  in  settings like the 
Internet, we suggest it  is unreasonable or unrealistic to
 assume non-precomputation and non-exposure  of the \emph{public}  DH-components 
 for \emph{all} uncorrupted  parties in the system.
Note that, whenever there is an uncorrupted player whose  DH-component  is exposed  prior to the session in which the DH-component is to be used (the attacker can just set this session as the test-session), the security of HMQV relies on both the GDH assumption and the KEA assumption in most cases as clarified in Section \ref{SecWithinCK}.

 For the above reasons, we suggest that the security advantage of HMQV over (s)OAKE in this  special case is
insignificant in reality. Note that, even in this special case, (s)OAKE enjoys other security advantages: (1) stronger embedded subgroup test supported by offline pre-computability of $A^{cy}$ and $B^{dx}$; (2) resistance to more powerful  secrecy exposure of the additional pre-computed private values $A^{cy}$ and $B^{dx}$; (3) stronger resistance against collision attacks on the underlying hash function $h$; (4) tighter security reduction of sOAKE. 
  Further note that, in the case of pre-computed and exposed DH-components, 
 (s)OAKE is based upon weaker assumptions (i.e., only the GDH assumption)  than (H)MQV  (that is based on both the GDH assumption and the KEA assumption) for the most often case of $\hat{A}\neq \hat{B}$.


\textbf{(s)OAKE vs. robust (s)OAKE.}
Note that, in comparison with (s)OAKE that enjoys reasonable deniability,   the variant of robust (s)OAKE  proposed in Appendix \ref{NOvariants} loses the reasonable deniability property. But,  it seems that robust (s)OAKE may render  seemingly  stronger security, in the sense that even both the ephemeral DH-exponents $x$ and $y$ are exposed by an adversary the adversary still cannot compute the DH-secret $K_{\hat{A}}$ or $K_{\hat{B}}$.  We suggest that such a security advantage  of robust (s)OAKE over the plain (s)OAKE is 
not significant, based on  the following observation:

\begin{itemize}

\item If we assume a powerful  adversary that  can expose both ephemeral DH-exponents $x$ and $y$ for the test session, then it may  also be reasonable to assume that the adversary can expose one of the values $(K_{\hat{A}},K_{\hat{B}})$ for that  exposed session. Note that, from $(x,y)$ and one of the values $(K_{\hat{A}},K_{\hat{B}})$, the adversary can compute the value $g^{ab}$. As the value $g^{ab}$ is fixed and used in all sessions,  once the value $g^{ab}$ is gotten the adversary can compute the session-key for all other sessions with exposed both ephemeral DH-exponents.

\end{itemize}

 In the CK-framework, the test-session and its matching session (in case the matching session exists) are  assumed to be unexposed. That is, in the CK-framework, the adversary is only allowed to exposed ephemeral DH-exponents (and maybe other private values) for sessions other than the test-session and its matching session. Actually, as we show in  Appendix \ref{CKanalysis}, (s)OAKE is secure in the CK-framework assuming exposed DH-exponents $(x,y)$ and off-line computed values $(A^{cy},B^{dx})$.

Based on the above observations,   we suggest (s)OAKE achieves much better balance between security and privacy than robust (s)OAKE.

\section{Formulation and Analysis of (Session-Key) Computational\\ Fairness}\label{AppFairness}

In Section \ref{SecBeyondCK}, we introduced the new perspective of ``computational fairness" for DHKE by concrete  EDA attacks against (H)MQV, and showed that computational unfairness can cause some essential security damages to DHKE protocols.
We now consider how to formulate ``computational fairness" for DHKE protocols.

A first thought is to require that, to successfully finish a session (with session-key output) with an honest player (e.g., player $\hat{B}$), the computation of the malicious player (e.g., $\hat{A}$)  and that of its honest peer  should have the same computational complexity. But, such a formulation is imprecise and does not work. With (s)OAKE as an example, the honest player $\hat{B}$ has two ways  to compute $K_{\hat{B}}=A^{cy}X^{db+ye}$: one way is to use the simultaneous exponentiation techniques, which amounts to about 1.3 exponentiations; and another way is to compute two separate exponentiations $A^{cy}$ (that can be offline computed) and $X^{db+ey}$ and then multiply them to get $K_{\hat{B}}$. Moreover, there exist a number of different methods for simultaneous exponentiations with (slightly) varying computational complexity  \cite{MOV95,G98,DJM00}. Thus, simply requiring the computational complexity  of a malicious player and that of its honest peer to be the same is meaningless in general.

In this work, we focus on \emph{session-key} computational fairness, i.e., the computational fairness in computing the session-key, for implicitly authenticated DHKE protocols like (H)MQV and (s)OAKE as are the focus of this work (extension 
 to  general interactive  protocols  is discussed later). 
For any complete session-tag (e.g., $Tag=(\hat{A},A,\hat{B},B,X,Y)$ here for (H)MQV and (s)OAKE)) and $I\in\{\hat{A},\hat{B}\}$,  we first identify \emph{dominant-operation values} w.r.t. $Tag$ and $I$,  $(V^I_1, \cdots, V^I_{m_I})\in G_1\times\cdots \times G_{m_I}, m_I\geq 2$,   which are  specified  to compute the session-key $K$  by honest player $I\in\{\hat{A},\hat{B}\}$ for a  complete  session of DHKE specified  by  the complete session-tag $Tag$, where $G_{i}$, $1\leq i\leq m_I$ is the range of $V^I_i$.
Specifically,  $K=F_K(V^I_1,\cdots,V^I_{m_I},Tag)$,  where $K$ is the session-key output,   $F_K$ is some polynomial-time computable function (that is defined by the session-key computation specified for honest players).
  The dominant-operation values  of a complete session  are random variables defined over  the complete session-tag (as well as the choice of the random function  in the RO model).  We remark that  dominant operations are specific to  protocols, where for different key-exchange protocols the dominant operations can also be different.   For (s)OAKE and (H)MQV,  the dominant operation is defined  just to be  modular exponentiation. Then, roughly speaking, we say that a DHKE protocol enjoys  \emph{session-key computational fairness}, if for any complete session-tag $Tag$, the session-key computation involves the same  number of \emph{non-malleably independent}  dominant-operation values for both $I\in \{\hat{A},
  \hat{B}\}$.
Here, ``non-malleable independence" is defined in reminiscent of Definition \ref{DefComInd}. Specifically, we consider two notions of ``non-malleably  independence".

\begin{definition} [strong non-malleable independence]\label{DefSNMInd}

For the  dominant-operation values, \\ $(V^I_1, \cdots, V^I_{m_I})\in G_1\times\cdots \times G_{m}$, $m\geq 2$ and $I\in \{\hat{A},\hat{B}\}$, w.r.t. a complete session-tag $Tag$ on any sufficiently large security parameter $n$,
we say $V^I_1, \cdots, V^I_{m_I}$ are \emph{strongly} computationally (resp., perfectly)
 non-malleably  independent, if for   any polynomial-time computable (resp., any power unlimited) relation/algorithm $R$ (with components drawn from $G_1\times \cdots \times G_{m_I}\times \{0,1\}^*$)
   it holds that   the following quantity  is negligible in $n$ (resp., just $0$):

$$|\Pr[R(V^I_1,\cdots, V^I_{m_I},Tag)=1]-\Pr[R(U_1,\cdots, U_{m_I},Tag)=1],$$
 where $U_i,1\leq i\leq m_I$ is taken uniformly at random from $G_i$,
 and the probability is taken over the random coins of $R$ 
 (as well as the choice of the random function in the random oracle model).

\end{definition}

\textsf{Remark:} Note that  the above Definition \ref{DefSNMInd} is defined w.r.t. any complete session-tag, which does not explicitly  take the malicious player's ability into account. But, this definition ensures that, by the birthday paradox,  for any successfully finished session between a malicious player (e.g., player $I=\hat{B}$) and an honest player (e.g., player $\hat{A}$),
 no matter how the malicious player does (on the identity and DH-challenge of the honest player, i.e., $(A, X)$), it holds that: for any $(\alpha_1, \cdots, \alpha_{m_I})\in (G\setminus 1_G)^{m_I}$, the probability that $\Pr[V^I_i=\alpha_i]$ is negligible for any $i,1\leq i\leq m_I$. The reason is: for each concrete choice of $(B, Y)$ by $\hat{B}$ (which then determines a complete session-tag), the distribution of the values $(V^I_1,\cdots, V^I_{m_I})$ is indistinguishable from the uniform distribution. As the malicious player is polynomial-time (i.e., can make at most polynomial number of choices), by the birthday paradox it holds that the malicious player can set $V^I_i$ to be a predetermined value only with negligible probability.  This means that the malicious player cannot make the values $(V^I_1,\cdots, V^I_{m_I})$ maliciously correlated (under any pre-determined polynomial-time computable relation) with non-negligible probability. In this sense, the notion of  ``self-sealed computational independence" in accordance with Definition \ref{DefComInd} (which is defined specific to NMJPOK for proving the joint knowledge of  $b$ and $y$ w.r.t. a single DH-challenge $X$) can be viewed as a special and weaker case of strong non-malleable independence defined here.

\begin{definition} [general non-malleable  independence]

For the  dominant-operation values, \\ $(V^I_1, \cdots, V^I_{m_I})\in G_1\times\cdots \times G_{m_I}$, $m_I\geq 2$ and $I\in \{\hat{A},\hat{B}\}$, w.r.t.   a complete session-tag $Tag$ on any sufficiently large security parameter $n$,
 we say $V^I_1, \cdots, V^I_{m_I}$ are \emph{generally} computationally (resp., perfectly) non-malleably  independent, if there exists at most one $j,1\leq j\leq m_I$ such that for   any polynomial-time computable (resp., any power unlimited)  relation/algorithm  $R$ (with components drawn from $G_1\times \cdots \times G_{m_I}\times \{0,1\}^*$)
   it holds that    the following quantity  is negligible in $n$ (resp., just $0$):

$$|\Pr[R(V^I_1,\cdots,V^I_j,\cdots,  V^I_{m_I},Tag)=1]-\Pr[R(U_1,\cdots,U_{j-1},V^I_j,U_{j+1},\cdots, U_{m_I},Tag)=1],$$
 where $U_i,1\leq i\neq j \leq m_I$ is taken uniformly at random from $G_i$,  and the probability is taken over the random coins of $R$ 
  (as well as the choice of the random function in the random oracle model).

\end{definition}

\textsf{Remark:} 
The definition of general non-malleable independence says that the distribution of $(V^I_1,\cdots,\\ V^I_{j-1},V^I_j,V^I_{j+1},\cdots, V^I_{m_I})$ is computationally indistinguishable from $(U_1,\cdots,U_{j-1},V^I_{j},U_{j+1},\cdots, U_{m_I})$.  As the values $(U_1,\cdots,U_{j-1},V^I_{j},U_{j+1},\cdots, U_{m_I})$ are mutually independent, it then implies that the values of  $(V^I_1,\cdots,V^I_{j-1},V^I_j,V^I_{j+1},\cdots, V^I_{m_I})$ are also computationally  independent. This definition also ensures that, no matter how a malicious polynomial-time  player $I\in \{\hat{A},\hat{B}\}$ does, (by the birthday paradox) it holds that: (1) The malicious player cannot make the values $(V^I_1, \cdots, V^I_{j-1},V^I_{j+1},\cdots, V^I_{m_I})$ correlated to $V^I_j$ under any predetermined polynomial-time computable relation. In particular,  for any $(\alpha_1, \cdots, \alpha_{j-1},\alpha_{j+1},\cdots, \alpha_{m_I})\in (G\setminus 1_G)^{m_I-1}$,
 $\Pr[V^I_i=\alpha_i]$ is negligible  for  any $i, 1\leq i\neq j\leq m_I$.  (2) Any efforts of the malicious player in order to change the value $V^I_j$ (which then changes the session-tag) will cause all other values $(V^I_1, \cdots, V^I_{j-1},V^I_{j+1},\cdots, V^I_{m_I})$ changed (to some values indistinguishable from random ones).  Thus, the malicious player is also infeasible to set the value $V^I_j$ correlated to any of the values $(V^I_1, \cdots, V^I_{j-1},V^I_{j+1},\cdots, V^I_{m_I})$.
This also further implies that the value $V^I_j$ is committed to $V^I_i$ for any $i, 1\leq i\neq j\leq m_I$, 
in the sense that:  the malicious player cannot (with non-negligible probability by the birthday paradox)  output two different session tags on which the values $V^I_j$ are different but the value $V^I_i$ remains the same.  In this sense, the notion of ``committed computational independence" in accordance with Definition \ref{DefComInd} (which is defined specific to sNMJPOK)  can be viewed as a special and weaker case of general  non-malleable independence defined here. Finally, it is direct that strong non-malleable independence is stronger than general non-malleable independence.




\begin{definition}[(session-key) computational fairness]\label{DefGNMInd}

We say a DHKE protocol has session-key computational   fairness,  if for any complete session-tag $Tag$ on any sufficiently large security parameter $n$, the session-key computation involves the same  number of \emph{non-malleably independent}  dominant-operation values for any $I\in \{\hat{A},  \hat{B}\}$.
 %
That is, for any complete session-tag $Tag$ on sufficiently large security parameter and for each player $I\in \{\hat{A},\hat{B}\}$, it holds that: (1) the dominant-operation values $V^I_1,\cdots,V^I_{m_I}$ w.r.t. $Tag$, involved in computing the session-key via $F_K(V^I_1,\cdots,V^I_{m_I}, Tag)$,
  are (strong or general)  non-malleably independent, and (2)  $m_{\hat{A}}=m_{\hat{B}}$,
   where 
  $F_K$ is some predetermined polynomial-time computable function specified   to compute session-key (according to protocol specification).

\end{definition}

\textsf{Remark:}
Though session-key computational fairness is defined w.r.t. any complete session tag, according to the discussions following Definition \ref{DefSNMInd} and Definition \ref{DefGNMInd}, it particularly ensures that:  for any polynomial-time malicious player $I$, no matter how it does, (by the birthday paradox) it is infeasible to make the values $V^I_1,\cdots,V^I_{m_I}$ (involved in session-key computation) correlated under any predetermined polynomial-time computable relation.
   Note that we used the number of \emph{non-malleably independent} dominant-operation values \emph{involved} in session-key computation as the measurement for session-key computational fairness. The reason we require the dominant-operation values to be non-malleably independent is that, without such a requirement, as shown by  our EDA attacks on (H)MQV,  an adversary can potentially set these  values maliciously correlated such that the session-key can be  computed much more easily (than the ways specified for honest players) \emph{even without knowing any of the dominant-operation values}. The reason we only require the dominant-operation values \emph{involved} (rather than computed) in session-key computation is that,  there can be multiple different ways to compute the session-key from dominant-operation values. With the function $F_K(V_1,V_2)=H_K(V_1\cdot V_2)$ as an example, where $V_1$ and $V_2$ are non-malleably independent,  one can compute two separate exponentiations $V_1$ and $V_2$ and then compute the session-key, but one can also use the simultaneous exponentiations technique to compute $V_1\cdot V_2$ with only about 1.3 exponentiations. Furthermore, there are a number of different methods for simultaneous exponentiations with (slightly) varying computational complexities. But, with any computation way, the value of $F_K(V_1,V_2)=H_K(V_1\cdot V_2)$ has to be computed, with which  two non-malleably independent exponentiations are involved.

\textsf{Remark:} 
We  note  that the issue of  computational fairness can  apply  to interactive protocols in general, as long as the honest players have the same computational operations under protocol specifications.\footnote{In particular, most key-exchange protocols are protocols of such type, while key distribution protocols (e.g., via public-key encryption) are not. }
For implicitly authenticated DHKE protocols like (H)MQV and (s)OAKE, we only considered here the session-key computational fairness. In general, for  key-exchange protocols with explicit authentication (e.g., via signatures and/or MACs), besides session-key computational fairness, we need also consider authentication computational fairness. The formulation of session-key computational fairness is also instrumental in formulating authentication computational fairness, which is beyond the scope of this work. 

\begin{proposition} (s)OAKE is session-key computationally fair assuming $h: \{0,1\}^*\rightarrow G\setminus 1_G$ is a random oracle, while (H)MQV is not session-key computationally fair.

\end{proposition}

\noindent \textbf{Proof.}
For both (s)OAKE and (H)MQV, the dominant operation (involved in session-key computation) is defined to be modular exponentiation. A complete session-tag consists of $(\hat{A},A=g^a,\hat{B},B=g^b,X=g^x,Y=g^y)$.

For (s)OAKE and any complete session-tag $Tag$,  the dominant operation values specified for the  player $\hat{A}$ (resp., $\hat{B}$) are $V^{\hat{A}}_1=B^{dx}\in G\setminus 1_G$ and $V^{\hat{A}}_2=Y^{ca+ex}\in G\setminus 1_G$ (resp., $V^{\hat{B}}_1=A^{cy}$ and $V^{\hat{B}}_2=X^{db+ey}$), where $c=h(\hat{A},A,Y),d=h(\hat{B},B,X),e=h(X,Y)$ (resp., $c=d=1$ and $e=h(\hat{A},A,\hat{B},B,X,Y)$) for OAKE (resp., sOAKE). The function $F_K$ is specified to be  $F_K(V_1,V_2,str)=H_K(V_1\cdot V_2)$. It is clear that, similar  to the analysis of Proposition \ref{SSJPOKproof} and Proposition \ref{sSSJPOKproof},  the distribution of $(V^I_1,V^I_2)$, for both $I\in\{\hat{A},\hat{B}\}$,  is identical to that of $(U_1,U_2)$  for OAKE (resp., $(V^I_1,U_2)$ for sOAKE) in the random oracle model,  where $U_i, i\in \{1,2\}$ is taken uniformly at random from $G\setminus 1_G$. That is,  $(V^I_1,V^I_2)$ are strongly \emph{perfect} non-malleably independent for OAKE (resp., generally perfect non-malleably independent for sOAKE).  Thus, both OAKE and sOAKE enjoy session-key computational fairness.

For (H)MQV and any complete session-tag $Tag$,  the dominant operation values specified for the  player $\hat{A}$ (resp., $\hat{B}$) are $V^{\hat{A}}_1=Y^{x+da}\in G$ and $V^{\hat{A}}_2=B^{e(x+da)}\in G$ (resp., $V^{\hat{B}}_1=X^{y+eb}$ and $V^{\hat{B}}_2=A^{d(y+eb)}$), where $d=h(X,\hat{B}),e=h(Y,\hat{A})$ for HMQV  (resp., $d=2^l+(X \mod 2^l)$ and $e=2^l+(Y\mod 2^l)$ for MQV). The function $F_K$ is specified to be  $F_K(V_1,V_2,str)=H_K(V_1\cdot V_2)$. Our concrete EDA attacks presented in Section \ref{SecBeyondCK} demonstrate that both MQV and HMQV do not satisfy computational fairness. Specifically, consider the following specific relations (corresponding to the two  specific cases of our attack): (1) $R(V_1,V_2,Tag)=1$ iff $V_1\cdot V_2=1_G$; (2) $R(V_1,V_2,Tag)=1$ iff $V_1\cdot V_2=YB^e$, where $YB^e$ can be publicly computed from the session-tag  $Tag$. 
  For all these specific relations, there exist complete session-tags $Tag$ (corresponding to the sessions caused by the EDA attacks presented in Section \ref{SecBeyondCK}) such that 
  $\Pr[R(V^{\hat{A}}_1,V^{\hat{A}}_2,Tag)=1]=1$, while $\Pr[R(U_1,U_2,Tag)=1]=1$ or $\Pr[R(V^{\hat{A}}_1,U_2,Tag)=1]=1$ or $\Pr[R(U_1,V^{\hat{A}}_2,Tag)=1]=1$ is always negligible w.r.t. these specific relations (as each of  the values of $U_1\cdot U_2$, $V^{\hat{A}}_1\cdot U_2$ and $U_1\cdot V^{\hat{A}}_1$ is distributed uniformly  over  $G\setminus 1_G$).
\hfill $\square$

\textsf{Remark:} By the   session-key computational fairness property of (s)OAKE, the session-key computation involves two non-malleably independent values $A^{cy}$ and $X^{db+ey}$ no matter how a malicious $\hat{B}$ does (i.e., $\hat{B}$ is infeasible to make the values $A^{cy}$ and $X^{db+ey}$ correlated under any predetermined polynomial-time computable relation). If  we view each non-malleably independent exponentiation value as a proof-of-knowledge of the corresponding exponent, then to compute the session-key any PPT player has to "know" both $cy$ and $db+ey$, from which both the static secret-key $b$ and the ephemeral DH-exponent $y$ can be efficiently derived. In this sense, the session-key computation of (s)OAKE  itself can be viewed as a non-malleable join proof-of-knowledge of  both $b$ and $y$.  This further implies that a malicious player is infeasible to set the session-key to some values that can be  publicly computed from the session transcript.

\textbf{Comparisons with the fairness notions in secure multi-party computation (SMC). } The notion of ``fairness" was intensively studied in the literature of secure multi-party computation  (see \cite{GL05}  for an overview of the various fairness notions considered in SMC). Informally speaking,
a protocol is fair if either all the parties learn the output of the function, or no party
learns anything (about the output). This property is also referred to  as ``complete fairness" (along with many variants), which mainly deals with prematurely  adversarial aborting.
To bypass some  impossibility results on  achieving fair SMC protocols with a  majority of corrupted  players, the work \cite{GMPY11} introduced the notion of ``resource fair SMC". The resource fairness considered in \cite{GMPY11} is still a variant of ``complete fairness". Specifically, the ``resource fairness"  \cite{GMPY11} captures ``fairness through gradual release". Here,  protocols using gradual release consist of a ``computation" phase, where some
computation is carried out, followed by a ``revealing" phase, where the parties gradually
release their private information towards learning the protocol output. Then, roughly speaking, resource fairness requires that the honest players and the adversary run essentially the same number of steps in order to obtain protocol output.

 Casting ``fairness through gradual release" into DHKE, it means that: players $\hat{A}$ and $\hat{B}$ gradually release their DH-exponents $X$ and $Y$ in sequential steps, so that both parties can output the session-key or both cannot.
 Clearly, the notions of  ``complete fairness"  and ``resource fairness" considered in the literature of SMC are  significantly different from the session-key  computational fairness formulated and considered in this work.  Specifically, we assume both parties honestly send their DH-exponents, and  computational fairness is about the session-key computation complexity. That is, our  computational fairness is to capture the fairness between non-aborting players in computing session-key outputs (i.e., if both players do not abort,  they should invest essentially the same computational resources  in computing  the session-key output), while  ``complete fairness" and its variant  in the literature of  SMC mainly deal with prematurely  adversarial aborting. Also, the resource fairness considered in \cite{GMPY11} is relative to experiment in which the protocol is run  or the protocol needs to be aware of the computational power of the
adversary (up to a constant) \cite{GMPY11}.


\subsection{On Fixing HMQV to Achieve  Computational Fairness}\label{AppYZ-MQV}

In  \cite{OAKE08,OAKE07}, we proposed some variants of (H)MQV,  \emph{just in the spirit of (s)OAKE and NMJPOK} to prevent our EDA attacks and to render the property of session-key computational fairness. The key point is to put $A$ (resp.,  $B$) into the input of $d$ (resp., $e$).
Specifically,  we have the following fixing approaches, by setting     (1) $d=h(X, \hat{B}, A)$
and $e=h(Y, \hat{A}, B)$; or  (2) $d=h(\hat{A}, A, \hat{B},  B, X,
Y)$ and $e=h(d)$; or (3) $d=h(\hat{A},
A, X)$ and $e=h(\hat{B}, B, Y)$, etc.
Other components remain unchanged. 
  For the above  third fixing  solution, in order to get only one exponentiation  online efficiency, we can  make some further
modifications by setting 
$K_{\hat{A}}=(Y^eB)^{xd+a}$, $K_{\hat{B}}=(X^dA)^{ye+b}$, where
$d=h(\hat{A}, A, X)$ and $e=h(\hat{B}, B, Y)$; The session-key is
still $K=H_K(K_{\hat{A}})=H_K(K_{\hat{B}})$. For presentation simplicity, we refer to this solution as the fourth fixing solution  (this protocol variant
is named as OAKE-MQV in \cite{OAKE08,OAKE07}). 

 Unfortunately, we failed in providing the provable security for any of  the above HMQV variants  in the CK-framework. In particular, we do not know how to extend the  security proof of  HMQV in \cite{K05} to any of  the above four fixing solutions. Indeed, HMQV was very carefully designed to enjoy provable security in the CK-framework.  Below, we present some concrete obstacles in extending the proof of HMQV \cite{K05} to these HMQV variants. But, there can  be more obstacles.
 \begin{itemize}
 \item For the first  and the second solutions, we note that the proof of HMQV for the case of $A=B$ (specifically, the proof of Lemma 24 in Section 6.3) fails.  The underlying reason is: the inputs of $d$ and the inputs of $e$ share some common values,   such that in the repeated experiment of redefining $e$ the value $d$ will also be changed.

  \item For the third and   the fourth solutions, we do not know how to extend the proofs of Lemma 11 (to be more precise, Case-3 of Claim 13), Lemma 17 and  Lemma 29 to these two solutions. The underlying reason is:  the messages to be signed  by the signer $\hat{B}$ by the underlying XCR or DCR signatures (defined in accordance with the third and the fourth solutions) are  the fixed value $(\hat{B},B)$, while in HMQV the message to be signed is its peer's identity $\hat{A}$ that may be set by the adversary.  In addition, for the fourth solution, the proof of Lemma 27 
      also fails.  The underlying reason is about the order of $d$ and $e$ in order to compute the value $X^b$. Also, the third and the fourth solutions have the following disadvantage that, in case the intermediate private value $y+eb$ (computed by $\hat{B}$ in a session) is leaked, this leaked value allows an adversary to impersonate $\hat{B}$ in any other sessions (no matter what the values $(X, A)$ are).
 \end{itemize}
 Besides lacking provable security in the CK-framework, many other  advantageous features enjoyed by (s)OAKE (as clarified in Section \ref{SecYZvsHMQV}) are also lost with  the above fixing solutions.  
 To the best of our knowledge, we do not know how to achieve, besides the OAKE family,  implicitly authenticated DHKE protocols that enjoy all the following properties: (1) provable security in the CK-framework; (2) online optimal efficiency \emph{and/or} reasonable deniability; (3) session-key computational fairness. The surrounding issues are quite subtle and tricky, and indeed (s)OAKE was very carefully designed to achieve all these features (and much more as clarified in Section \ref{SecYZvsHMQV}).

%

\section{Security Analysis  of (s)OAKE in  the CK-Framework} \label{CKanalysis}



  One of main conceptual
contributions of the analysis of HMQV in the CK-framework
\cite{K05} is to cast the design of HMQV in terms of Hashed Dual
challenge-Response (HDR)
 signatures and Hashed Challenge-Response (HCR) signatures,
 which are in turn based  on Dual  Challenge-Response
(DCR) signatures  and eXponential Challenge-Response (XCR)
signatures  and can be traced back to Schnorr's identification
scheme \cite{S91}. 
 We show that OAKE and sOAKE 
 all can be casted in terms of
HDR signatures. Moreover, the HDR signatures implied by
the (s)OAKE protocols, referred to as OAKE-HDR and
sOAKE-HDR,  are both \emph{online
efficient} and \emph{strongly secure}. This provides extra security strength of the underlying building tools, say SSJOPK and sSSJPOK, used in (s)OAKE.  
To this end, we first demonstrate a divided forking lemma with a  new family of signature schemes,   which may itself be of independent interest. 








\subsection{A New Family of Signature Schemes, and Divided Forking Lemma} \label{online/offline} 
\textbf{Notation note:} \emph{ For presentation simplicity, in this subsection, we a bit abuse the notations  of $a, c, d, e,  f, k, s, z,\rho, \mathcal{C}$, which are  different from the notations used
outside this subsection.}

A common paradigm, known as the Fiat-Shamir paradigm \cite{FS86},
of obtaining signatures is to collapse a 3-round public-coin
honest-verifier zero-knowledge, known as  $\Sigma$-protocol, into a
non-interactive scheme with hash functions that are  modeled to
be random oracles \cite{BR93}.


\begin{definition}[$\Sigma$-protocol \cite{C96}] A three-round public-coin protocol $\langle P, V\rangle$ is said to
be a $\Sigma$-protocol for an $\mathcal{NP}$-relation $\mathcal{R}$ if the
following hold:
\begin{itemize}
\item Completeness. If $P$, $V$ follow the protocol, the verifier
always accepts.

\item Special soundness. From any common input $U$ of length $n$
and any pair of accepting conversations on input $U$, $(a, e, z)$
and $(a, e^{\prime}, z^{\prime})$ where $e\neq e^{\prime}$, one
can efficiently compute $w$ such that $(U, w)\in \mathcal{R}$ with
overwhelming probability. Here $a$, $e$, $z$ stand for the first,
the second and the third message respectively and $e$ is assumed
to be a string of length $l$ (that is polynomially related to $n$)
selected uniformly at random in $\{0, 1\}^l$. 

\item Perfect/statistical SHVZK (special honest verifier
zero-knowledge). There exists a probabilistic polynomial-time
(PPT) simulator $S$, which on input $U$ (where there exists an
$\mathcal{NP}$-witness $w$ such that $(U, w)\in \mathcal{R}$)  and a random
challenge string $\hat{e}$, outputs an accepting conversation of
the form $(\hat{a}, \hat{e}, \hat{z})$, with the same probability
distribution as that of the real conversation $(a, e, z)$ between the
honest $P(w)$, $V$ on input $U$.
\end{itemize}

\end{definition}

 The first $\Sigma$-protocol (for an
$\mathcal{NP}$-language) in the literature  can be traced back to
the honest verifier zero-knowledge (HVZK) protocol for Graph Isomorphism \cite{GMW91} (but  the
name of $\Sigma$-protocol is adopted much later in \cite{C96}),
 and a large number of $\Sigma$-protocols for various languages  have been developed
 now. $\Sigma$-protocols have been proved to be a very powerful
cryptographic tool, and are widely used in numerous important
cryptographic applications.  Below,   we briefly  recall the
 $\Sigma$-protocol examples for DLP and RSA. 

\textbf{$\Sigma$-Protocol for DLP \cite{S91}.}
 The following is a
$\Sigma$-protocol $\langle P, V\rangle$ proposed by Schnorr
\cite{S91} for proving the knowledge of discrete logarithm, $w$,
for a common input of the  form $(p, q, g, U)$ such that $U=g^w \
mod\ p$, where $p, q$ are primes  $g$  is an element in $Z_p^*$ of
order $q$. Normally, the length of $q$, $|q|$, is denoted as the
security parameter.

\begin{itemize}
\item $P$ chooses $r$ at random in $Z_q$  and sends $a=g^r \ mod \
p$ to $V$.

\item $V$ chooses a challenge $e$ at random in $Z_{2^l}$ and sends
it to $P$. Here, $l$ is fixed such that $2^l< q$.

\item $P$ sends $z=r+ew \ mod\  q$ to $V$, who checks that
$g^z=aU^e \ mod\ p$, that $p$, $q$ are prime and that $g$, $h$ are
of  order $q$, and accepts iff this is the case.
\end{itemize}

\textbf{$\Sigma$-Protocol for RSA \cite{GQ88}.} Let $n$ be an RSA
modulus and $q$ be a prime. Assume we are given some element $y\in
\textbf{\textsf{Z}}_n^*$, and $P$ knows an element $w$ such that
$w^q=y \mod n$. The following protocol is a $\Sigma$-protocol for
proving the knowledge of $q$-th roots modulo $n$.

\begin{itemize}
\item $P$ chooses $r$ at random in $Z_n^*$  and sends $a=r^q \mod
 n$ to $V$.

 \item $V$ chooses a challenge $e$ at random in
$Z_{2^l}$ and sends it to $P$. Here, $l$ is fixed such that $2^l<
q$.

\item $P$ sends $z=rw^e \mod n$ to $V$, who checks that $z^q=ay^e
\mod n$, that $q$ is a prime, that $gcd(a, n)=gcd(y,n)=1$,   and
accepts iff this is the case.
\end{itemize}

\textbf{The Fiat-Shamir paradigm and its provable security.} Given
\emph{any}  $\Sigma$-protocol $(a, e, z)$  on common input $U$
(which will be viewed as signing public-key), the Fiat-Shamir
paradigm collapse the $\Sigma$-protocol into a signature scheme as
follows: $(a, e=h(a, m), z)$, where $m$ is the message to be
signed and $h$ is a hash function. Note in actual signature scheme
with the Fiat-Shamir paradigm, the generated signature only
consists of $(e, z)$  as the value $a$  can be computed from $(e,
z)$. The provable security of the general Fiat-Shamir paradigm is
shown by Pointcheval and Stern \cite{PS00} in the random oracle
model (assuming $h$ to be an idealized random function). The core
of the security arguments of Pointcheval and Stern \cite{PS00} is
a forking lemma.

\textbf{On-line/off-line signature.} The notion of
\emph{on-line/off-line signature} is introduced in \cite{EGM89}.
The idea is to perform signature generation into two phases: the
off-line phase and the on-line phase. On-line/off-line signature
schemes are useful, since in many applications the signer (e.g., a
smart-card) has a very limited response time once the message is
presented (but it can carry out costly computations between
consecutive signing requests). The on-line phase is typically very
fast, and hence can be executed even on a weak processor.
On-line/off-line signature schemes are particularly remarkable in
smart-card based applications \cite{ST01}: the off-line phase can
be implemented  either during the card manufacturing process or as
a background computation whenever the  card is connected to power.

Note that for signature schemes obtained via the Fiat-Shamir
scheme, the signer can pre-compute and store a list of  values
$(a=g^r, r)$. Then, to sign a message $m$, it simply computes
$e=h(a, m)$ and $z$. With Schnorr's signature as an illustrative
example, in this case, the signer only needs to perform $z=r+h(m,
a)w$ online, where $a=g^r$ and $r$ are offline  pre-computed and stored.
Some general transformation from any signature scheme to secure
off-line/off-line signature scheme are know (e.g.,
\cite{EGM89,ST01}), but typically are not as efficient (for both
 computational complexity and  space complexity of the signer) as
 the signature resultant directly via the Fiat-Shamir paradigm.

\textbf{The Digital Signature Standard (DSS).} The DSS scheme
\cite{DSS} is a variant of Schnorr's signature \cite{S91} via the
Fiat-Shamir paradigm. The general structure of DSS is as follows:

\begin{itemize}

\item Public-key:  $U=g^w \in G^{\prime}$, where $w\in Z^*_q$.
Typically, $w$ is a 160-bit prime.

 \item Secret-key: $w$.

\item Signature generation: Let $m\in \{0, 1\}^*$ be the message
to be signed.

\begin{enumerate}

\item Compute $a=g^r \mod p$, where $r$ is taken randomly from
$Z_q$. Compute $d=f(a)$, where $f: G^{\prime} \rightarrow Z^*_q$
is a conversion function.

Typically, for DSS with $G^{\prime}=Z^*_p$, $f$ is just the ``mod
$q$" operation; for DSS with $G^{\prime}$ being some elliptic
curve group over a finite field (i.e., $a$ stands for an elliptic
curve point $(x, y)$), $f(a)$ is to take the $x$-coordinate of
$a$.

\item Compute $s$ from the equation $h(m)=sr-dw \mod q$, as
follows:

\begin{itemize}
\item Compute $\hat{r}=r^{-1}$.

\item Compute $s=(h(m)+dw)\hat{r}$, or $s=h(m)\hat{r}+dw\hat{r}$
with offline pre-computed $dw\hat{r}$, where $h$ is a hash
function.


 \end{itemize}

\item Output  $(d, s)$ as the signature.
\end{enumerate}

\item Signature verification: Given $(e=h(m), d, s)$ where $d, s
\in Z^*_q$, the verifier verifies the signature as follows:
\begin{itemize}

\item Compute $\hat{s}=s^{-1}$.

\item  Verify  $f(g^{e\hat{s}}U^{d\hat{s}})=d$, where $e=h(m)$.

\end{itemize}
\end{itemize}

 Recall that in the DSS scheme, the signature is generated as: $(d,
 s=er^{-1}+dwr^{-1})$, where $e=h(m)$, $d=f(a)$ and $a=g^r$,  $w$ is the
 secret-key. In general, the conversion $f: G^{\prime}\rightarrow
 Z^*_q$ also can be viewed as RO. Observe that the value $m$
 (i.e., the message to be signed) and the value $a=g^r$ are not
 put into the input of a single  RO in the DSS scheme, contrary to signature schemes
 via the Fiat-Shamir scheme where $(m, a)$ is put into the single
 RO $h$. The separation of $m$ and $a$ in the inputs of  ROs and the
 way of signature generation of DSS bring the following advantage
to DSS.

Specifically, the signer can pre-compute a list of values $a$'s
(just as in signature schemes via the Fiat-Shamir paradigm), but
contrary to signature schemes via the Fiat-Shamir paradigm, the
signer of DSS does not need to store these pre-computed values.
Specifically, for each pre-computed value $a=g^r$, the DSS signer
can off-line compute $d=f(a)$, $r^{-1}$  and $dwr^{-1}$, and only
stores $(d, r^{-1}, dwr^{-1})$ (note that it is unreasonable to
assume the message to be signed is always known beforehand).
\emph{Actually, for smart-card based applications, the values $(d,
r^{-1}, dwr^{-1})$'s
 can be stored during the card
manufacturing process}. 
Note that $d, r^{-1}, dwr^{-1} \in Z_q$ while $a\in G^{\prime}$.
Suppose $G^{\prime}=Z^*_p$ (where $p$ is typically of 1024 bits
while $q$ is of 160 bits) and the signer pre-computes $k$ values
of $a$, then in comparison with Schnorr's
signature scheme 
 the space complexity (of storing pre-computed values) is reduced
 from $(|p|+|q|)k$ to 
 $3|q|k$. 
 But, we remark that for implementations of DSS based on elliptic curves, such an advantage is insignificant.





\textbf{Challenge-divided $\Sigma$-protocols and   challenge-divided Fiat-Shamir
paradigm.} Next, we show a modified Fiat-Shamir paradigm, named
\emph{challenge-divided Fiat-Shamir paradigm}, that is applicable to a variant of  
$\Sigma$-protocol with \emph{divided} random challenges (that is
referred to as \emph{challenge-divided} $\Sigma$-protocol). 
Below, we first describe the challenge-divided $\Sigma$-protocols for DLP
and RSA.

\textbf{Challenge-divided $\Sigma$-Protocol for DLP.}
 The common input is the same as that of Schnorr's protocol for DLP: 
  $(p, q, g, U)$ such that $U=g^w \
mod\ p$.

\begin{itemize}
\item $P$ chooses $r$ at random in $Z_q$  and sends $a=g^r \ mod \
p$ to $V$.

\item $V$ chooses a \emph{pair} of challenges $d, e$ at random in
$Z_{2^l}\times Z_{2^l}$
 and sends $(d, e)$ to $P$. Here, $l$ is fixed such that
$2^l< q$.

\item $P$ sends $z=er+dw \ mod\  q$ (resp., $z=dr+ew$)  to $V$,
who checks that $g^z=a^eU^d \ mod\ p$ (resp., $g^z=a^dU^e \ mod\
p$), that $p$, $q$ are prime and that $g$, $h$ are of  order $q$,
and accepts iff this is the case.
\end{itemize}

\textbf{Challenge-divided $\Sigma$-Protocol for RSA.} Let $n$ be an RSA
modulus and $q$ be a prime. The common input is $(n, q, y)$, and
the private input is $w$ such that $y=w^q \mod n$.

\begin{itemize}
\item $P$ chooses $r$ at random in $Z_n^*$  and sends $a=r^q \mod
 n$ to $V$.

 \item $V$ chooses a \emph{pair} of  challenges $d,e$ at random in
$Z_{2^l}\times Z_{2^l}$ and sends $(d, e)$ to $P$. Here, $l$ is
fixed such that $2^l< q$.

\item $P$ sends $z=r^dw^e \mod n$ (resp., $z=r^ew^d \mod n$) to
$V$, who checks that $z^q=a^dy^e \mod n$ (resp., $z^q=a^dy^e \mod
n$), that $q$ is a prime, that $gcd(a, n)=gcd(y,n)=1$,   and
accepts iff this is the case.
\end{itemize}

\textbf{The challenge-divided Fiat-Shamir paradigm for challenge-divided $\Sigma$-protocols.} Let $F$ be a one-way function (OWF)
admitting challenge-divided $\Sigma$-protocols, i.e., the range of the OWF
has a challenge-divided $\Sigma$-protocol for proving the knowledge of the
corresponding preimage w.r.t.  the $\mathcal{NP}$-relation $\{(U,
w)| U=F(w) \}$. Let the random challenge be of length $Len$. Denote by $d, e$ the (divided) random challenges,
and let $U=F(w)$ be signer's public-key and $w$ the secret-key. To
sign a message $m$, the signer computes $a$, $d=\tilde{f}(a)$, $e=\tilde{h}(m)$,
and  $z$,  and  then outputs $(d, z)$ as the signature on $m$,
where $\tilde{h}$ and $\tilde{f}$ are conversion functions from $\{0, 1\}^*$
to $\{0,1\}^{Len}$. 
In security analysis in the RO model, we assume both $\tilde{h}$ and $\tilde{f}$ are hash functions that are modeled to be random oracles.

\textbf{\emph{Challenge-divided} Schnorr  signature scheme.} With Schnorr's  $\Sigma$-protocol for
DLP as an illustrative instance, the transformed signature via the
above challenge-divided Fiat-Shamir paradigm is called \emph{challenge-divided}
Schnorr signature.  Note that for signatures from the above  challenge-divided  Schnorr's $\Sigma$-protocol for DLP, we have that  $\tilde{f}=f$ and $\tilde{h}=h$ are conversion functions from $\{0, 1\}^*$
to $Z^*_q$. In practice, $f$  can  simply be the ``mod $q$" operation for $G^{\prime}=Z^*_p$ 
 or the operation of taking input's  $x$-coordinate when $G^{\prime}$ is some elliptic
curve group over a finite field.
In the
following, we directly describe the online/offline version of
challenge-divided Schnorr's signature.

\begin{itemize}
\item Public-key:  $U=g^{-w} \in G^{\prime}$, where $w\in Z^*_q$.
\item Secret-key: $w$.

\item Message to be signed: $m$. 

\item Offline pre-computation: the signer pre-computes and stores
$(r,d, dw)$ (resp., $(d, rd)$), where  $r$ is taken randomly  by
the signer from $Z^*_q$,  $a=g^r$, $d=f(a)$. The signature
verifier can pre-compute $e=h(m)$ and $\hat{e}=e^{-1}$, in case it
knows $m$ before receiving the
signature. 

\item Online signature generation: After receiving the message $m$
to be signed, the signer computes $e=h(m)$, retrieves the
pre-stored value $(r, d, dw)$ (resp., $(d, dr)$), and computes
$z=er+dw$ (resp., $z=dr+ew$). The signer outputs
 $(d, z)$ as the signature on $m$. 

\item Signature verification: given a signature $(e=h(m), d, z)$
where $d, z \in Z^*_q$, check that  $d, z \in Z^*_q$ and $f(g^{z\hat{e}}U^{d\hat{e}})=d$
(resp., $f(g^{z\hat{d}}U^{e\hat{d}})=d$), where $\hat{e}=e^{-1}$
(resp., $\hat{d}=d^{-1}$). Note that $\hat{e}=e^{-1}$  can be
offline pre-computed by the verifier, in case it knows the message
$m$ before receiving the
signature. 

\end{itemize}

\begin{theorem}
Assuming $h, f: \{0, 1\}^*\rightarrow \{0, 1\}^l/\{0\} \subseteq Z^*_q$
are random oracles where $l$ is the security parameter (for
presentation simplicity, we assume the range of ROs is $\{0,
1\}^l$ rather than $\{0, 1\}^l/\{0\}$), the challenge-divided Schnorr scheme
is existentially  unforgeable against adaptive chosen message attacks
under the DLP assumption.
\end{theorem}

\noindent \textbf{Proof.} We mainly provide the proof for
challenge-divided Schnorr with $z=er+dw$, the proof for the case of
$z=dr+ew$ is similar.

Given a polynomial-time and successful forger $\mathcal{F}$, i.e.,
$\mathcal{F}$ successfully outputs (after polynomially many
adaptively chosen queries to the signing oracle and random
oracles), with non-negligible probability in polynomial-time, a
valid signature on a new message that is different from those
queried to the signing oracle,   we build an efficient solver
$\mathcal{C}$ for the DLP problem, namely, $\mathcal{C}$ gets as
input a random element
$U=g^{-w}$ in $G$ and outputs 
the corresponding discrete logarithm $w$  also with non-negligible
probability. \emph{For presentation simplicity, we assume the random
oracles $h, f$ are identical, namely   we use the unique RO $h$ to
handle all RO queries $e=h(m)$ and $d=h(a)$.}  The algorithm
$\mathcal{C}$ is presented in Figure \ref{dS}.

 \begin{figure}[!p]
\begin{center}

\begin{tabular} {|c|}
 \hline 
\\
 \textbf{Building the DLP solver $\mathcal{C}$ from the challenge-divided Schnorr forger $\mathcal{F}$}\\ 

\begin{minipage}[t] {6.4in} \small
\vspace{0.1cm}

\textbf{Setup:} The input to $\mathcal{C}$ is a random element
$U=g^{-w}$ in $G$, and its goal is to compute $w$. 
To this end, $\mathcal{C}$  provides $\mathcal{F}$ with a random
tape, and  runs the forger $\mathcal{F}$ 
  as the
challenge-divided Schnorr signer  of public-key $U$.

\textbf{RO queries:} $\mathcal{C}$  provides random answers to
queries to  the
random oracle $h$,  
under the limitation that if the same $h$ query is presented more
than once, $\mathcal{C}$ answers it with the same response as in
the first time.

\textbf{Signature query simulation:} Each time $\mathcal{F}$
queries the signing oracle  for a challenge-divided Schnorr  signature on
message $m_i$,
$1\leq i \leq R$, 
chosen by $\mathcal{F}$ adaptively, where $m_i$ denotes the message in the $i$-th query,  $\mathcal{C}$ answers the
query  as follows (note that $\mathcal{C}$ does not know the
secret-key $w$ corresponding to the public-key $U=g^w$):

\begin{description}

\item [S1.] Chooses $z_i \in_{\textup{R}} Z^*_q$  and $d_i
\in_{\textup{R}} \{0, 1\}^l \subseteq Z^*_q$ where $l$ is the
output length of the
RO $h$. 
  If $h(m)$ has
been defined by previous query to $h$, then sets $e_i=h(m)$,
otherwise chooses $ e_i \in_{\textup{R}} \{0, 1\}^l$ and defines
$h(m)=e_i$.

\item [S2.] Computes $a_i=g^{z_ie_i^{-1}}U^{d_ie_i^{-1}}$.

\item [S3.]  If  $h(a_i)$ has been previously defined, 
 $\mathcal{C}$ aborts its run and outputs
``fail". Otherwise, sets $h(a_i)=d_i$. Recall that, for
presentation simplicity,  we have assumed $f=h$.


\item [S4.] $\mathcal{C}$ responds to $\mathcal{F}$'s signing
query $m_i$ with the simulated  signature $(d_i, z_i)$.

\end{description}

When $\mathcal{F}$ halts, $\mathcal{C}$ checks whether the
following conditions hold:
\begin{description}
\item [F1.] $\mathcal{F}$ outputs  $(m,  d, z)$ such that $(d, z)$
is a valid signature on $m$. That is, $d, z$ are in $Z^*_q$,
$e=h(m)$ $a=g^{ze^{-1}}U^{de^{-1}}$, and $d=h(a)$.

\item [F2.] $m$ was not queried by $\mathcal{F}$ to the signing
oracle previously, i.e., $m\neq m_i$ for all $i, 1\leq i\leq R$.

\item  [F3.] The values 
 $h(m)$  and $h(a)$ were 
queried 
 from the RO $h$.

\end{description}

If these three conditions hold, $\mathcal{C}$ proceeds to the
``repeat experiments" below; in all other cases $\mathcal{C}$
halts and outputs ``fail".

\textbf{The repeat experiments.} $\mathcal{C}$ runs $\mathcal{F}$
again for a second time, under the same public-key $U$ and using
the same coins for 
$\mathcal{F}$. 
 There are two cases according to the
order of the RO queries of $h(m)$  and $h(a)$:

\begin{description}
\item [C1.] $h(m)$ posterior to $h(a)$: $\mathcal{C}$ rewinds
$\mathcal{F}$ to the point of making the RO query $h(m)$, responds
back a new independent value $e^{\prime} \in_{\textup{R}} \{0,
1\}^l$. All subsequent actions of $\mathcal{C}$ (including random
answers to subsequent RO
queries) are independent of the first run. 
 If in
this repeated run $\mathcal{F}$ outputs a valid signature $(d,
z^{\prime})$ for the message $m$, i.e., $e^{\prime}=h(m)$,
$d=h(a)$ and $a=g^{z^{\prime}e^{\prime -1}}U^{de^{\prime -1}}$,
$\mathcal{C}$ computes $w=(z^{\prime}e^{\prime
-1}-ze^{-1})/(de^{\prime -1}-de^{-1})  \mod q$.


\item [C2.] $h(a)$ posterior  to  $h(m)$: $\mathcal{C}$ rewinds
$\mathcal{F}$ to the point of making the RO query $h(a)$, responds
back a new independent value $d^{\prime} \in_{\textup{R}} \{0,
1\}^l$. All subsequent actions of $\mathcal{C}$ (including random
answers to subsequent RO
queries) are independent of the first run. 
 If in
this repeated run $\mathcal{F}$ outputs a valid signature
$(d^{\prime}, z^{\prime})$ for the message $m$, i.e., $e=h(m)$,
$d^{\prime}=h(a)$ and
$a=g^{z^{\prime}e^{-1}}U^{d^{\prime}e^{-1}}$, $\mathcal{C}$
computes $w=(z^{\prime}-z)/(d^{\prime}-d) \mod q$.


\end{description}

\end{minipage}
\\
\\
\hline
\end{tabular}
\caption{\label{dS} Reduction from DLP to challenge-divided Schnorr
forgeries}
\end{center}

\end{figure}

For the description of $\mathcal{C}$ in Figure \ref{dS}, suppose
$\mathcal{F}$ makes $Q$ RO queries and $R$ signing oracle queries
(where $Q$ and $R$ are some polynomials in the security parameter
$l$), we have the following proposition:

\begin{proposition}\label{ProbProp}
With probability at most $(RQ+R^2/2)/(q-1)$ (that is negligible),
$\mathcal{C}$ fails in one of Step S3 of signature simulations
(note that, assuming $\mathcal{F}$ never fails at Step S3  in
signature simulations, the signature simulations are perfect).
$\mathcal{C}$ fails at Step F3  with probability at most $(2Q+3)2^{-l}$.
\end{proposition}

\textbf{Proof} (of Proposition of \ref{ProbProp}).  It is easy to
check that suppose $\mathcal{C}$ never fails at Step S3, the
signature simulations by $\mathcal{C}$ are of identical
distribution with that of real signatures by using the secret-key
$w$.

Next, we limit the upper-bound of Step S3 failure. Note that for
each $a_i$ generated by $\mathcal{C}$ at Step S2, it is
distributed uniformly in $G\setminus 1_G$. In the RO model, there are two
cases for $\mathcal{C}$ fails at Step S3:

\textbf{Case 1.} For \emph{some} $i$, $1\leq i\leq R$,
$\mathcal{F}$ ever successfully guessed the value $a_i$  in one of
its $Q$ random oracle queries. Thus, the probability that
$\mathcal{C}$ fails in Case 1  is at most $RQ/(q-1)$.

\textbf{Case 2.} For some $i$, $1\leq i\leq R$, the value $a_i$
has ever been generated in dealing with the $j$-th  signing oracle
query, $j<i$. The probability that $\mathcal{C}$ fails in Case 2
is at most $C_R^2/(q-1)\leq (R^2/2)/(q-1)$, where $C_R^2$ is the combination number of selecting two elements from a set of $R$ elements.  

Finally, it is easy to check that $\mathcal{C}$  fails at Step F3
with probability at most $(2Q+3)2^{-l}$. To see this, first note that there are two possibilities for $\mathcal{F}$ to output   $d=h(a)$ 
  without making RO query with $a$: 
  (1) $\mathcal{F}$ directly guesses the value  $d=h(a)$, 
  which occurs with probability 
  $2^{-l}$. (2) The value $d=h(a)$ 
  collides with some other values from the RO answers (i.e., $h(a)=h(a^\prime)$
   for some $a^\prime$ queried by $\mathcal{F}$ to RO). As $\mathcal{F}$ makes at most $Q$ RO queries,
the latter case can  occur with probability at most $Q2^{-l}$. Thus, with probability at least $1-(Q+1)2^{-l}$,  $\mathcal{F}$
 knows 
 $a$ 
 (i.e., queries the RO with $a$). 
Note that from $(a, d, z)$  the value  $log_a^{g^zU^d}$ is  (which should be equal to  $h(m)=e$) is then
 determined. Conditioned on this,  the probability $e=h(m)=log_a^{g^zU^d}$
 is $2^{-l}$,  as $e$ is distributed uniformly over $\{0,
 1\}^l$. Thus, $\mathcal{F}$ does not query $h(m)$ with probability at
 most $(Q+1)2^{-l}+(1-(Q+1)2^{-l})\cdot 2^{-l}< (Q+2)\cdot2^{-l}$. \hfill $\square$



Thus, suppose the forger $\mathcal{F}$ succeeds (i.e., outputs a
valid signature $(d, z)$ for a new message $m$ different from
those queried) with non-negligible probability  in its real attack
against the signer of public-key $U$, $\mathcal{F}$ succeeds in
the first run of $\mathcal{C}$ in Figure \ref{dS} 
  also with
non-negligible probability (up to a gap at most
$(QR+R^2/2)/(q-1)$). Then,  with non-negligible probability (with
a gap at most $(QR+R^2/2)/(q-1)+(2Q+3)2^{-l}$ to the success
probability of $\mathcal{F}$ in its real attack), $\mathcal{C}$
does the repeated second run.

For presentation simplicity, we write the signature of
challenge-divided Schnorr on a message $m$ as $(m, e=h(m), a, d=h(a), z)$.
Note that given a pair of  different signatures on the same $m$
(and $a$): $\{(m, e, a, d, z), (m, e^{\prime}, a, d,
z^{\prime})\}$ that corresponds to Case C1 in Figure \ref{dS}, or,
$\{(m, e, a, d, z), (m, e, a, d^{\prime}, z^{\prime})\}$ that
corresponds to Case C2 in Figure \ref{dS}, the value $w$ computed
by $\mathcal{C}$ is correct. Thus, to finish the theorem, what left is to show that
conditioned $\mathcal{F}$ succeeds  in outputting the valid $(m,
e, a, d, z)$ in the first run of $\mathcal{C}$, with
non-negligible probability $\mathcal{F}$ will also  succeed in
Case C1 or Case C2 of the repeated second run. We  note that this
can be shown by a straightforwardly  \emph{extended} version of the
Pointcheval-Stern forking lemma \cite{PS00} (that was originally
developed to argue the security of digital signature schemes via
the Fiat-Shamir paradigm). 
  For
completeness, we reproduce the forking lemma tailored for
signature schemes via the challenge-divided Fiat-Shamir paradigm,
 referred to as \emph{divided} forking lemma. 

Suppose $\mathcal{F}$ produces, with probability
$\varepsilon^{\prime}$, a valid signature $(m, e, a, d, z)$,
within  the time bound $T$ 
 in its real attack against the signer of
public-key $U$, then with probability at least
$\varepsilon=(\varepsilon^{\prime}-(QR+R^2/2)/(q-1)-(2Q+3)2^{-l})/2$ $\mathcal{F}$ outputs a valid signature $(m, e, a, d,
z)$ in the first run of $\mathcal{C}$ described in Figure \ref{dS}
such that $\mathcal{F}$ made both $h(m)=e$ and $h(a)=d$ queries to
the RO with the order of \emph{$h(m)$ being posterior to $h(a)$} or the order of \emph{$h(a)$ being posterior to $h(m)$}.
 Without loss of
generality, we assume it is the former case, i.e., the RO query
$h(m)$ is posterior to $h(a)$ (the analysis of the case of $h(a)$
being posterior to $h(m)$ is similar). We have the following
lemma, from which the theorem is then established.

\begin{lemma}[divided forking lemma]\label{dividedforking} Suppose $\mathcal{F}$ produces, with probability
$\varepsilon$, a valid signature $(m,e,a,d,z)$ within  the
time bound $T$ in the first run of $\mathcal{C}$  such
that $\mathcal{F}$ made both $h(m)=e$ and $h(a)=d$ RO queries with
the order \emph{$h(m)$ being posterior to $h(a)$}, then within
time $T^{\prime}\leq (2/\varepsilon +
(\varepsilon/4Q-2^{-l})^{-1})\cdot T$ and with probability at
least $\frac{1}{9}$, 
a replay of
$\mathcal{F}$ outputs a valid signature $(m, e^{\prime}, a, d,
z^{\prime})$ for $e^{\prime}\neq e$.

\end{lemma}

\textbf{Proof} (of Lemma \ref{dividedforking}). The proof of Lemma
\ref{dividedforking} is essentially identical to that of Lemma 2
in  \cite{PS00}, which we re-produce here for completeness. We mention that, as in \cite{PS00}, although the divided forking lemma is presented here  w.r.t. the challenge-divided Schnorr's signature (based on the challenge-divided Schnorr's $\Sigma$-protocol for DLP), it  can be directly extended and applied to signatures derived from other challenge-divided $\Sigma$-protocols.


  Denote by $\omega$ the random tape of
$\mathcal{F}$, and assume $\mathcal{F}$ makes at most $Q$ RO
queries $\mathcal{Q}_1, \cdots, \mathcal{Q}_Q$ (for presentation
simplicity, we assume all RO queries are distinct), and denote by
$\rho=(\rho_1, \cdots, \rho_Q)$ the $Q$ RO answers. It is clear a
random choice of the random function $h$ (i.e., the RO)
corresponds to a random choice of $\rho$.

Define $\mathcal{S}$ to be the set of $(\omega, h)$ such that
$\mathcal{F}^h(\omega)$ outputs a valid signature $(m, e, a, d,
z)$ in the first run of $\mathcal{C}$, such that $\mathcal{F}$
made both $h(m)$ and $h(a)$ RO queries with the order of $h(m)$
being posterior to $h(a)$. That is,
$\Pr[\mathcal{S}]=\varepsilon$. Define $Ind(\omega, h)$ to be the
index of the RO query $h(m)$, i.e., $m=\mathcal{Q}_{Ind(\omega,
h)}$. Define $\mathcal{S}_i$ be the subset of $\mathcal{S}$ such
that $Ind(\omega, h)=i$ for $1\leq i\leq Q$. That is, the set
$\{\mathcal{S}_1, \cdots, \mathcal{S}_Q\}$ is a partition of
$\mathcal{S}$. Define $I=\{i| Pr[\mathcal{S}_i|\mathcal{S}]\geq
1/2Q\}$, i.e., $\Pr[\mathcal{S}_i|i\in I]\geq \varepsilon/2Q$. 
For each $i\in I$, define by $h_i$ the restriction of $h$ to
queries of index strictly less than $i$, they by applying the
Splitting Lemma (Lemma 1, page 12  in \cite{PS00}),  there exists
a subset $\Omega_i$ (of $\mathcal{S}$) such that: (1) for any
$(\omega, h)\in \Omega_i$, $\Pr_{h^{\prime}}[(\omega,
h^{\prime})\in \mathcal{S}_i|h^{\prime}_i=h_i]\geq
\varepsilon/4Q$; (2) $\Pr[\Omega_i|\mathcal{S}_i]\geq
\frac{1}{2}$. As all the subsets $\mathcal{S}_i$ are disjoint, it
is calculated that $\Pr_{\omega, h}[\exists i (\omega, h)\in
\Omega_i\cap \mathcal{S}_i|\mathcal{S}]\geq \frac{1}{4}$ (for more
details, the reader is referred to \cite{PS00}).

By the Lemma 3 (page 14) in \cite{PS00}, we get $\Pr[Ind(\omega,
h)\in I|\mathcal{S}]\geq \frac{1}{2}$. Now, run $\mathcal{F}$
$2/\varepsilon$ times with random $\omega$ and $h$, with
probability $1-(1-\varepsilon)^{2/\varepsilon}\geq \frac{4}{5}$ we
get one successful pair $(\omega, h)\in \mathcal{S}$. Denote by
$\beta$ the index $Ind(\omega, h)$ corresponding to the successful
pair. We know with probability at least $\frac{1}{4}$, $\beta \in
I$ and $(\omega, h)\in \mathcal{S}_{\beta} \cap \Omega_{\beta}$.
Consequently, with probability at least $\frac{1}{5}$, the
$2/\varepsilon$ runs have provided a successful pair $(\omega, h)
\in \mathcal{S}_{\beta} \cap \Omega_{\beta}$ where
$\beta=Ind(\omega, h)$. As $\Pr_{h^{\prime}}[(\omega,
h^{\prime})\in
\mathcal{S}_{\beta}|h^{\prime}_{\beta}=h_{\beta}]\geq
\varepsilon/4Q$ in this case, we get $\Pr_{h^{\prime}}[(\omega,
h^{\prime})\in \mathcal{S}_{\beta} \wedge \rho_{\beta}\neq
\rho^{\prime}_{\beta}|h^{\prime}_{\beta}=h_{\beta}]\geq
\varepsilon/4Q-2^{-l}$, where
$\rho_{\beta}=h(\mathcal{Q}_{\beta})$ and
$\rho^{\prime}_{\beta}=h^{\prime}(\mathcal{Q}_{\beta})$. Now, we
replay $\mathcal{F}$ with fixed $\omega$ but randomly chose
$h^{\prime}$ such that $h^{\prime}_{\beta}=h_{\beta}$, for
$(\varepsilon/4Q-2^{-l})^{-1}$ times, with probability at least
$\frac{3}{5}$, we will get another success. That is, after less
than $2/\varepsilon + (\varepsilon/4Q-2^{-l})^{-1}$ repetitions of
$\mathcal{F}$'s attack, with probability at lease $\frac{1}{5}
\times \frac{3}{5}\geq \frac{1}{9}$, we have obtained two valid
signatures $(m, e, a, d, z)$ and $(m, e^{\prime}, a, d,
z^{\prime})$ for $e\neq e^{\prime}$. \hfill $\square$ $\square$







\textbf{Challenge-divided Schnorr vs. DSS.} We note all performance
advantages of DSS (recalled in Appendix \ref{CKanalysis}) are essentially preserved with the
challenge-divided Schnorr scheme.  We also note the techniques proposed in
\cite{NMVR94} for improving the performance of DSS in certain
scenarios, e.g., signature batch verification and compression,
etc, are also applicable to challenge-divided Schnorr.
 In addition,  challenge-divided Schnorr
 has the following advantages over DSS:

 \vspace{-0.3cm}
\begin{itemize}

\item Same or better offline space complexity than DSS (much better than Schnorr scheme for implementation based $Z^*_p$. Suppose $k$ values of
$a$'s are pre-computed, the offline space complexity of
challenge-divided Schnorr with $z=er+dw$ is $3k|q|$ (which is the same as
that of DSS); But, for challenge-divided Schnorr with $z=dr+ew$, the offline
space complexity is only $2k|q|$.

Note that, for Schnorr signature scheme, suppose $G^{\prime}=Z^*_p$ (where $p$ is typically of 1024 bits
while $q$ is of 160 bits) and the signer pre-computes $k$ values
of $a$, then in  Schnorr's
signature scheme 
 the space complexity (of storing pre-computed values) is  $(|p|+|q|)k$.  

 \vspace{-0.3cm}
\item More efficient  signature generation in total. To compute the value $s$ in the DSS-signature (recalled in Appendix \ref{CKanalysis}), the signer of DSS performs 
1 modular inverse (i.e., $\hat{r}=r^{-1}$) and 2 modular
multiplications 
 in total. In comparison, to compute the value $z$ in the
challenge-divided Schnorr signature, the signer only  performs 2
modular multiplications in total (without performing the modular
inverse operation). We remark that modular inverse is a relatively
expensive operation (which is typically performed by the Euclid
algorithm), and  is thus much preferable to dispense with
(particularly for smart-card-based deployment).
 \vspace{-0.2cm}
\item More efficient offline pre-computation.  Besides the same
other pre-computations, the signer of DSS needs to perform 1
modular inverse $r^{-1}$ and 2 modular multiplications for computing $dwr^{-1}$,
but the signer of challenge-divided Schnorr  needs to offline perform only 1
modular
multiplication $dw$ or $dr$. 

 \vspace{-0.3cm}
\item More efficient online signature verification (for the case
of $z=er+dw$).  For verifying
 a DSS-signature $(d, s)$, the verifier has to compute
 $\hat{s}=s^{-1}$ online (which is a relatively expensive operation), as the value $s$ is  known to the
 verifier \emph{only when  the signature  comes to it}.
  In comparison,
 for verification of  challenge-divided Schnorr  with $z=er+dw$, the verifier only
 needs to compute the inverse  $\hat{e}=e^{-1}$ where $e=h(m)$.
 In case the verifier learns the message to be signed
 prior to receiving the signature from the signer (which is quite common in certain scenarios),
  the  values     $e$ and $e^{-1}$ can both  be offline pre-computed by the
 verifier of challenge-divided Schnorr. For challenge-divided Schnorr with $z=dr+ew$,
 signature verification is of the same computational complexity as
 that of DSS.


 \vspace{-0.3cm}
\item Provable security in the random oracle model.  We show that,
assuming both $h$ and $f$ are random oracles, the  challenge-divided Schnorr
scheme is existentially  unforgeable against adaptive chosen message
attacks \cite{GMR88} under the DLP assumption in the RO model.


\end{itemize}

\textbf{Challenge-divided Schnorr vs. Schnorr.} For implementations of challenge-divided Schnorr and Schnorr based over  order $q$ subgroups of   $Z^*_p$, where $p$ is typically of 1024 bits and $q$ is of about 160 bits,  similar to DSS in this case, challenge-divided Schnorr enjoys much better offline space efficiency than Schnorr. However, for elliptic curve based  implementations of both challenge-divided Schnorr and Schnorr, such offline space efficiency advantage disappears. \emph{As mentioned, the introduction of challenge-divided Schnorr is mainly to introduce the divided forking lemma to be used in the analysis of (s)OAKE in the CK-framework. }
 \vspace{-0.3cm}

\subsection{Casting (s)OAKE in Terms of \emph{Online Efficient} and \emph{Strongly Secure} HDR Signatures} \label{AppHDRsecurity}


 Informally speaking, a HDR signature scheme is an
\emph{interactive} signature scheme between two parties in the
public-key model. The two parties generate the \emph{same}
signature, which is actually  a \emph{hashed} value of the
DH-secret shared between the two parties,  with the \emph{dual}
roles of signer and challenger: each party generates the signature
with private values of its static secret-key and the secret
DH-exponent with respect to  its peer's DH-component and
public-key as the challenges.
With a HDR signature, we are only interested  to ensure
verifiability of the signature by the two intended parties,  and
thus we make no assumptions or requirements regarding the
transferability or verifiability of the signature by a third
party. Roughly speaking, a HDR signature scheme is secure if the
signature cannot be generated by any other parties  other than the
two intended (honest) parties.  

\begin{definition}  [(s)OAKE-HDR signature schemes] Let $\hat{A}$,
$\hat{B}$ be two parties with public-keys $A=g^a$, $B=g^b$,
respectively.  Let $m_{\hat{A}}$, $m_{\hat{B}}$ be two messages.
The \textsf{\textup{OAKE-HDR, sOAKE-HDR}} 
signatures of $\hat{B}$ on messages $(m_{\hat{A}}, m_{\hat{B}},
\hat{A}, A, \hat{B}, B,  X, Y)$ 
 are defined as a vector  of values
(the signatures of $\hat{A}$ 
 are defined
straightforwardly):

\begin{description}

\item [OAKE-HDR.] $\{\hat{A}, A, m_{\hat{A}}, m_{\hat{B}}, X, Y,
HSIG^{OAKE}_{\hat{A},\hat{B}}(m_{\hat{A}}, m_{\hat{B}}, X,
Y)=H_K(A^{yc}X^{bd+ye})\}$, where $X=g^x$, $Y=g^y$ are chosen by
$\hat{A}$, $\hat{B}$ respectively as the random \emph{challenge}
and \emph{response}, $x, y \in_{\textup{R}} Z^*_q$,
$c=h(m_{\hat{A}}, \hat{A}, A, Y)$, $d=h(m_{\hat{B}}, \hat{B}, B,
X)$ and $e=h(X, Y)$.

Another form of OAKE-HDR is to set $c=h(\hat{A}, A, Y)$,
$d=h(\hat{B}, B, X)$ and $e=h(m_{\hat{A}}, m_{\hat{B}} , X, Y)$.
Both of these two versions are secure.


\item [sOAKE-HDR.] $\{\hat{A}, A, m_{\hat{A}},  m_{\hat{B}}, X, Y,
HSIG^{sOAKE}_{\hat{A},\hat{B}}(m_{\hat{A}},  m_{\hat{B}}, X,
Y)=H_K(A^{yc}X^{bd+ye})\}$, where $c=d=1$,  $e=h(m_{\hat{A}},
m_{\hat{B}}, \hat{A}, A,
\hat{B}, B,  X, Y)$. 



\end{description}

\end{definition}

Note that the online efficiency of
(s)OAKE-HDR  can be only one
exponentiation for each player. In comparison,  each player of  HMQV-HDR performs
about  1.3  online exponentiations.
For presentation simplicity, in the above HDR
signature description  we assume the CA in the underlying PKI will
check the membership $G\setminus 1_G$ of registered public-keys, and each player
checks the membership $G\setminus 1_G$ of  its peer's DH-component. These subgroup
tests may not be necessary for the security of HDR in general,
assuming no ephemeral private state is exposed,   and thus can  be
relaxed in some scenarios (see \cite{K05,M05} for more
details). 


\textbf{(s)OAKE in a nutshell.} Actually, the above
OAKE-HDR/sOAKE-HDR can be viewed as a general structure of the
(s)OAKE protocols. Specifically, OAKE and sOAKE are instantiated
with OAKE-HDR and sOAKE-HDR respectively, with the special $m_{\hat{A}}$
and $m_{\hat{B}}$ that are set to be the empty string. In general,
$m_{\hat{A}}$ (resp., $m_{\hat{B}}$) can include some values
 sent to $\hat{A}$ (resp., $\hat{B}$)  from  $\hat{B}$ (resp., $\hat{A}$), which  does
not affect the pre-computability of (s)OAKE. In particular, in
practice with pre-computed and reused DH-components,
$m_{\hat{A}}$ (resp., $m_{\hat{B}}$) can include a random nonce
generated and sent by $\hat{B}$ (resp., $\hat{A}$).

 In the following, we show the
security of OAKE-HDR, sOAKE-HDR with off-line pre-computed
DH-exponents, DH-components, and  the values $A^{yc}$ or $B^{xd}$
(that may be potentially exposed to the forger \emph{even  prior
to the session
involving these pre-computed values}),   
  on which the security of OAKE and
sOAKE 
 in the CK-framework will be  based. In particular, we show that
 our OAKE-HDR and sOAKE-HDR 
 satisfy a  stronger
 security definition (than the definition given in \cite{K05}) in accordance with Definition \ref{DCRdef-SEC}. 

\textbf{On the \emph{strong} security of HDR.}  The \emph{strong}
security of our definition for HDR lies in that:
\begin {itemize}

\item  We assume $(y, Y, A^{cy})$ are off-line pre-computed, and
the forger can get them prior to the session run  involving them.

This particularly renders stronger capability to the attacker  to
perform colliding (birthday) attacks against the hash function $h$
(that is of length $|q|/2$ for HMQV). To deal with this subtlety,
the actual HMQV implementation  needs some changes in practice (to
be clarified later).

\item  In the forging game defined in Figure \ref{HDR}, the
successful forgery requires that the \emph{whole} vector
$(\hat{A}, A, m_1, m_0,  X_0, Y_0)$ did not appear in any of the
responses of $\hat{B}$ to $\mathcal{F}$'s queries.  The definition
for the security of HCR in \cite{K05} only requires that the pair
$(Y_0, m_0)$ did not appear in responses from the signer. 
 As we shall see, the HMQV-HDR scheme may not be  strongly
secure in general.
\end{itemize}

\textbf{OAKE-HDR vs. HMQV-HDR.}  In \cite{K05}, the HMQV-HDR (of
$\hat{B}$) is defined to be $\{X, Y, \\
DSIG^{HMQV}_{\hat{A},\hat{B}}(m_{\hat{A}}, m_{\hat{B}}, X,
Y)=H_K((XA^d)^{y+be})\}$,
  where $d=h(m_{\hat{A}}, X)$,
$e=h(m_{\hat{B}}, Y)$. For building HMQV with HMQV-HDR,
$m_{\hat{B}}$ (resp., $m_{\hat{A}}$) is set to be its
\emph{peer's} identity  $\hat{A}$ (resp., $\hat{B}$).
 The underlying HMQV-XCR-signature is
defined to be $X^{y+be}$,
  where 
$e=h(m_{\hat{B}}, Y)$. The following are some brief comparisons
between OAKE-HDR and HMQV-HDR:

\begin{itemize}

\item One notable  advantageous feature of OAKE-HDR and sOAKE-HDR is
the online efficiency. Specifically, the online efficiency of
OAKE-HDR and sOAKE-HDR, for each player,  can be only one
exponentiation. In comparison,  each player of  HMQV-HDR performs
about  1.3  online exponentiations.


\item As we shall see, the OAKE-HDR and  sOAKE-HDR are
\emph{strongly} secure in accordance with Definition \ref{DCRdef-SEC}.
We note that the   HMQV-XCR underlying HMQV-HDR is not \emph{strongly}
secure. For example, to forge a HMQV-XCR signature $(X, Y,
\sigma=X^{b+ey})$ on message $m$, where $e=h(m, Y)$, the forger
can first query the signer with $(m, X^{\prime}=X^2)$, gets back
$(X^{\prime}, Y, \sigma^{\prime}=X^{\prime b+ey})$, and then
outputs $(X, Y, \sigma=\sigma^{\prime \frac{1}{2}})$ as the XCR
signature on $m$. Note that the triple $(X, Y, \sigma)$ did not
appear in  any one  of the responses from the HMQV-XCR signer $\hat{B}$. We
note that one way to remedy this security vulnerability of
HMQV-XCR is to commit $X$ also to $e$ by defining $e=h(m, X, Y)$.

\item The security of OAKE-HDR/sOAKE-HDR against uncorrupted players
\emph{other than the signer itself}, with offline pre-computed $(y, Y,
A^{cy})$ that can be exposable to the adversary even prior to the
session involving  $(y, Y, A^{cy})$, is based on the gap
Diffie-Hellman (GDH) assumption.

The security of HMQV-HDR against
uncorrupted players \emph{other than the signer itself}, with offline
pre-computed DH-component $Y$,  is based on both
the GDH assumption and the non-standard KEA assumption \cite{D91},  even if the pre-computed DH-exponent $y$ is not exposable and only the pre-computed DH-component $Y$ is exposable.
Furthermore, for robust security of HMQV-HDR with pre-computed
DH-components, when the number of messages in the system is large,
HMQV-HDR needs to make the following modifications \cite{K05}: (1)
Increase the output length, i.e., $l$, of the hash function $h$,
e.g., from $|q|/2$ to $|q|$, which may bring  negative impact on
the performance  of HMQV. (2) Add random nonces into the input of
$d$ and $e$, or, put the message to be signed also into $H_K$,
which may  increase the system
complexity. 

\item The generation of the sOAKE-HDR signature uses minimal (i.e.,
only one) random oracle  (in computing the value of $e$).


\item  The HMQV-HDR signature is actually an XCR signature w.r.t.
the challenge $XA^d$. 
In comparison, 
OAKE-HDR and sOAKE-HDR in general cannot be viewed as a structure of
XCR w.r.t. some
challenge $f(X, A)$ for some function $f$. 

 \item  As we shall see,  the special protocol structure of
OAKE-HDR and sOAKE-HDR  also  much simplifies, in certain scenarios,
the security analysis of OAKE and sOAKE in the CK-framework. 

\end{itemize}

Next, we show the strong security of OAKE-HDR, sOAKE-HDR under the
Gap Diffie-Hellman (GDH) assumption in the random oracle model.

\begin{theorem}\label{HDR-analysis}
 Under the GDH
assumption, OAKE-HDR and sOAKE-HDR signatures of $\hat{B}$, with
offline pre-computed and exposable  $(y, Y, A^{cy})$,  are \emph{strongly}
secure in the random oracle model, 
 with respect to any
uncorrupted player other than the signer $\hat{B}$ itself
\emph{even if the forger is given the private keys of all
uncorrupted players in the system other than $b$ of
$\hat{B}$} 
\end{theorem}


\noindent \textbf{Proof} (of Theorem \ref{HDR-analysis}).  Given
an efficient and successful forger $\mathcal{F}$ against OAKE-HDR
or sOAKE-HDR, i.e., $\mathcal{F}$ wins the forgery game in Figure
 \ref{HDR} with respect to some uncorrupted player $\hat{A}\neq \hat{B}$ with non-negligible probability, we build
an efficient solver $\mathcal{C}$ for  GDH problem 
also with non-negligible probability. 
 The algorithm $\mathcal{C}$ for OAKE-HDR is
presented in Figure \ref{OAKE-DCR} (page \pageref{OAKE-DCR}), and the algorithm $\mathcal{C}$
for sOAKE-HDR is presented in Figure \ref{sOAKE-DCR} (page \pageref{sOAKE-DCR}).

For the description of $\mathcal{C}$ in Figure \ref{OAKE-DCR},
suppose $\mathcal{F}$ makes $Q_h$ RO queries to $h$,  $Q_H$
queries to $H_K$,  $Q_s$ signing oracle queries, 
 where $Q_h$, $Q_H$,  $Q_s$
 are  polynomial in the security parameter $l$ (i.e., the output length of $h$).  
 We have the following observations:

\begin{itemize}

\item The signature simulation at steps S1-S3 is perfect.


\item Now, suppose $\mathcal{F}$ outputs a successful forgery
$(\hat{A}, A, m_1, m_0, X_0, Y_0, r_0)$,
which particularly implies that $r_0$ should be  $H_K(\sigma_0)$, where 
$\sigma_0=A^{y_0c_0}X_0^{bd_0+y_0e_0}$, $X_0=U$, $Y_0=g^{y_0}$,
$c_0=h(m_1, \hat{A}, A, Y_0)$, $d_0=h(m_0, \hat{B}, B,  X_0)$ and
$e_0=h(X_0, Y_0)$. We investigate the probability that
$\mathcal{C}$ aborts at step F3. We have the following
observations:

\begin{itemize}
\item With probability at most $\frac{1}{2^l-1}+2^{-k}+Q_H/2^k$,
$\mathcal{F}$ can succeed with undefined 
any one of $\{c_0, d_0, e_0\}$. 
Here,
 $\frac{1}{2^l-1}$ is  the probability 
 that
 $\mathcal{F}$ guesses $\sigma_0$ with undefined $c_0$ or $d_0$ or $e_0$,  $2^{-k}$ is the probability that
 $\mathcal{F}$ simply guesses the value $r_0$, and $Q_H/2^k$ is the  probability upper-bound that   $r_0=H_K(\sigma_0)$ collides with some   $H_K$-answers. 

 \item With defined $c_0$ and $d_0$ and $e_0$, there are two cases for $\mathcal{F}$ to succeed
 without  querying $H_K(\sigma_0)$:

\begin{description}
\item [Case-1.]
 $\mathcal{F}$ simply
 guesses the value $r_0$. This probability is $2^{-k}$.

  \item [Case-2.] $r_0$ is  the value $r$ set  by
  $\mathcal{C}$ at one of S3.1 steps, where $r$ is supposed to be
  $H_K(\sigma)$ w.r.t. a stored vector  $(\hat{Z}, Z, m_{\hat{Z}}, m_{\hat{B}}, X,  y,
  Y, Z^{cy},
  r)$. 
  Recall that for the value $r$ set at   step S3.1,
  $\mathcal{C}$ does not know $\sigma$ (as it does not know $b$),
  and thus in this case both $\mathcal{C}$ and $\mathcal{F}$ may
  not   make  the RO-query $H_K(\sigma_0)=H_K(\sigma)$. In this
  case,  by the birthday paradox with probability at least
   $1-Q^2_H/2^{-k}$,
  $\sigma_0=\sigma$, i.e., $A^{c_0y_0}X_0^{d_0b+e_0y_0}=Z^{cy}X^{db+ey}$,
  where $c=h(m_{\hat{Z}}, \hat{Z}, Z, Y)$,  $d=h(m_{\hat{B}}, \hat{B}, B,  X)$,
  $e=h(X, Y)$, $c_0=h(m_1, \hat{A}, A, Y_0)$,  $d_0=h(m_0, \hat{B}, B,
  X_0)$,      $e_0=h(X_0, Y_0)$, and $(m_0, m_1, \hat{A}, A, \hat{B}, B, X_0, Y_0)\neq
(m_{\hat{A}}, m_{\hat{B}}, \hat{Z}, Z, \hat{B}, B,  X, Y)$.

By the
NMJPOK and TBSS  properties  of OAKE, for any value $\sigma\in
G\setminus 1_G$ and any  $(m_1, m_0, \hat{A}, A,  \hat{B},  B, X_0,
Y_0)$, the probability
$\Pr[A^{c_0y_0}X_0^{d_0b+e_0y_0}=\sigma]\leq \frac{1}{2^l-1}$,
where $X_0$ is the given random element in $G\setminus 1_G$, $\hat{A}$ and
$\hat{B}$ are uncorrupted players. This is true,  even if  the
public-key $A$ (resp., $B$) is removed from $c_0$ (resp.,
$d_0$), as the public-keys $A$ and $B$ are generated by the
uncorrupted players  $\hat{A}$ and $\hat{B}$ independently at
random. Then, by straightforward calculation, we can get that $\mathcal{F}$ succeeds in Case-2
with probability at most $O(\frac{Q^2_h}{2^l-1}+\frac{Q_s+Q^2_H}{2^k})$. 



\textbf{Note:} To rule out the possibility of Case-2, the analysis
of HMQV-HCR requires the   KEA assumption \cite{D91}. Furthermore,
  to resist
to birthday attacks in Case-2, some modifications of HMQV are
recommended in \cite{K05}: (1) increase the output length, i.e.,
$l$, of $h$, e.g.,  from $|q|/2$ to $|q|$. 
 (2) Add random
and \emph{fresh} nonces (which cannot be offline pre-computed) to
the input of $h$, or put the messages to be signed $m_{\hat{A}},
m_{\hat{B}}$ into the input of $H_K$. 
   \end{description}

  \item With probability at most $\frac{1}{2^l-1}$, the query
  $H_K(\sigma_0)$ is prior to  any one of  the queries  $\{c_0, d_0, e_0\}$. 

\end{itemize}

\item It is easy to check that, in case the forger $\mathcal{F}$
successfully outputs another different forge satisfying the
conditions F1-F3 in the repeat experiment C1 or C2, the output of
$\mathcal{C}$ is the correct value of $CDH(X_0, B)$.

\end{itemize}
The similar observations  can be easily checked for the algorithm
$\mathcal{C}$ for sOAKE-HDR  described in Figure \ref{sOAKE-DCR}.
Putting all together, we have that: suppose for some uncorrupted
player $\hat{A}\neq \hat{B}$,   the forger $\mathcal{F}$
 provides, with non-negligible probability,  a successful forgery w.r.t. $\hat{A}$
  in its real interactions with the signer of OAKE-HDR/sOAKE-HDR, then with also
 non-negligible probability (up to a negligible gap specified by
 the above observations) $\mathcal{F}$ succeeds under the run of
 $\mathcal{C}$. Then, by applying  the forking lemma, specifically, the  divided forking
lemma (Lemma \ref{dividedforking}) for OAKE-HDR and the normal
forking lemma of \cite{PS00} for sOAKE-HDR, the theorem is 
established. \hfill $\square$

 \begin{figure}[!p]
\begin{center}

\begin{tabular} {|c|}
 \hline 
\\
 \textbf{Building the CDH solver $\mathcal{C}$ from the sOAKE-HDR forger $\mathcal{F}$}\\ 

\begin{minipage}[t] {6.6in} \small
\vspace{0.1cm}

\textbf{Setup:} $\mathcal{C}$ does the same as it does  for the
forger $\mathcal{F}$ against  OAKE-HDR.


\textbf{Signature query simulation:} Each time $\mathcal{F}$
queries $\hat{B}$ for a signature on values $(\hat{Z}, Z,
m_{\hat{B}}, m_{\hat{A}})$, $\mathcal{C}$ answers the query for
$\hat{B}$ as follows (note that $\mathcal{C}$ does not know $b$):

\begin{description}
\item [S1.]  $\mathcal{C}$ generates $y\in_{\textup{R}}Z^*_q$,
$Y=g^y$  
and  $Z^{y}$. 
Again, $(y, Y, Z^{y})$ can be pre-computed by $\mathcal{C}$ and
leaked to $\mathcal{F}$ prior to the session. Then, $\mathcal{C}$
responds  $(y, Y=g^y, Z^{y})$ to $\mathcal{F}$, and stores the
vector $(\hat{Z}, Z, m_{\hat{Z}}, m_{\hat{B}}, y, Y, A^{y})$ as an
``incomplete session".

\item [S2.]  $\mathcal{F}$  presents $\mathcal{C}$ with $(\hat{Z},
Z, m_{\hat{Z}}, m_{\hat{B}}, Y)$, and a \emph{challenge} $X$.

\item [S3.]  $\hat{B}$ checks that $X\in G\setminus 1_G$ (if not, it
aborts) and that $(\hat{Z}, Z, m_{\hat{Z}}, m_{\hat{B}}, Y)$ is in
one of its incomplete sessions (if not, it ignores the query). 
$\mathcal{C}$ checks  for every value $\sigma\in G\setminus 1_G$
\emph{previously used by $\mathcal{F}$} as input to $H_K$ whether
$\sigma=Z^{y}X^{b+ye}$, where $e=h(m_{\hat{Z}}, m_{\hat{B}},
\hat{Z}, Z,  \hat{B}, B,  X, Y)$ (in case of undefined $e$,
$\mathcal{C}$ defines it with the RO $h$). It does so using the
DDH-oracle $\mathcal{O}$, specifically, by checking whether
$CDH(X, B)=(\sigma/Z^{y}X^{ye})$. If the answer is positive, then
$\mathcal{C}$ sets $r$ to the already determined value of
$H_K(\sigma)$.

\begin{description} \item [S3.1.] In any other cases, 
 $r$ is set to be a random value in $\{0, 1\}^k$, where
$k$ is the output length of $H_K$. Note that, in this case,
$\mathcal{C}$ does not know $\sigma=Z^{y}X^{b+ey}$, as it does not
know $b$, which also implies that $\mathcal{C}$ does not make
(actually realize) the RO-query $H_K(\sigma)$ \emph{even if the
value $\sigma$ has been well-defined (with predetermined $d$ and
$e$) and known to $\mathcal{F}$}.
\end{description}
Finally, $\mathcal{C}$ marks the vector  $(\hat{Z}, Z,
m_{\hat{Z}}, m_{\hat{B}}, X,  y, Y, Z^{y})$ as a ``\emph{complete
session}", stores $(\hat{Z}, Z, m_{\hat{Z}}, m_{\hat{B}}, X,  y,
Y, Z^{y}, r)$ and  responds  $(\hat{Z}, Z, m_{\hat{Z}},
m_{\hat{B}}, X,
 Y,  r)$ to $\mathcal{F}$.

\end{description}

\textbf{RO queries:} $\mathcal{C}$  provides random answers to
queries to  the
random oracles $h$ and $H_K$ (made by $\mathcal{F}$),  
under the limitation that if the same RO-query is presented more
than once, $\mathcal{C}$ answers it with the same response as in
the first time. But, for each \emph{new} query $\sigma$ to $H_K$,
$\mathcal{C}$ checks whether  $\sigma=Z^{y}X^{b+ey}$ for any one
of the stored  vectors $(\hat{Z}, Z, m_{\hat{Z}}, m_{\hat{B}}, X,
y, Y, Z^{y},  r)$ 
 (as before, this check is
done using the DDH-oracle). If equality holds then the
corresponding $r$ is returned as the predefined $H_K(\sigma)$,
otherwise a random $r$ is returned.

\textbf{Upon $\mathcal{F}$'s termination.} When $\mathcal{F}$
halts, $\mathcal{C}$ checks whether the following conditions hold:
\begin{description}

\item [F1.] $\mathcal{F}$ outputs a valid HDR-signature $(\hat{A},
A,  m_1, m_0, X_0, Y_0, r_0)$, where $\hat{A}\neq \hat{B}$ is an
uncorrupted player. In particular, it implies
that $r_0$ should be  $H_K(\sigma_0)$, where 
$\sigma_0=A^{y_0}X_0^{b+y_0e_0}$, 
 $Y_0=g^{y_0}$ (chosen by $\mathcal{F}$),  and $e_0=h(m_1, m_0,  \hat{A}, A, \hat{B}, B, X_0
 Y_0)$.

\item [F2.]  
  $(\hat{A}, A, m_1, m_0, X_0, Y_0)$ did
not appear in any of the above responses of the simulated sOAKE-HDR
signatures. 



\item [F3.] The value $e_0=h(m_1, m_0,  \hat{A}, A, \hat{B}, B,
X_0    Y_0)$ was
queried 
 from the RO $h$, and the value $H_K(\sigma_0)$ was queried  
 from  $H_K$ \emph{being posterior
 to
 the query $e_0$}. 
  Otherwise,  $\mathcal{C}$ aborts.

\end{description}

If these three  conditions hold, $\mathcal{C}$ proceeds to the
``repeat experiment" below, else it aborts. 

\textbf{The repeat experiment.} $\mathcal{C}$ runs $\mathcal{F}$
again for a second time, under the same input $(B, X_0)$ and using
the same coins for 
$\mathcal{F}$. $\mathcal{C}$ rewinds   $\mathcal{F}$ to the point
of making the RO query $h(m_1, m_0,  \hat{A}, A, \hat{B}, B, X_0
Y_0)$, responds back a new independent value $e^{\prime}_0
\in_{\textup{R}} \{0, 1\}^l$. All subsequent actions of
$\mathcal{C}$ (including random answers to subsequent RO
queries) are independent of the first run. 
If in this repeated run $\mathcal{F}$ outputs a successful forgery
 $(\hat{A}^{\prime}, A^{\prime}, m^{\prime}_1, m_0, X_0, Y_0,  r^{\prime}_0)$ satisfying  the conditions F1-F3
 (otherwise, $\mathcal{C}$
 aborts), which particularly
 implies that $r^{\prime}_0=H_K(\sigma^{\prime}_0)$,
 $\sigma^{\prime}_0=A^{\prime y_0}X_0^{b+y_0e^{\prime}_0}$,
 $\mathcal{C}$ computes
 $CDH(X_0, Y_0)=g^{x_0y_0}=
 [(\sigma_0/Y_0^{a})/(\sigma^{\prime}_0/Y_0^{a^{\prime}})]^{(e_0-e^{\prime}_0)^{-1}}$,
 where $a$ and $a^{\prime}$ are the private keys of the uncorrupted   $\hat{A}$ and $\hat{A}^{\prime}$
 (different from $\hat{B}$, which are assumed to be known to
 $\mathcal{C}$).  Note that $(\hat{A}^{\prime}, A^{\prime},
 m^{\prime}_1)$ need not necessarily to
  equal  $(\hat{A}, A, m_1)$. Finally, $\mathcal{C}$ computes   $CDH(U, V)=CDH(X_0, B)=
\sigma_0/((g^{x_0y_0})^{e_0}\cdot
Y_0^{a})$.



\end{minipage}
\\

\hline
\end{tabular}
\caption{\label{sOAKE-DCR} Reduction from GDH to sOAKE-HDR
forgeries}
\end{center}

\end{figure}

\textbf{On the role of putting  players' public-keys into the inputs of $c,
d$ for OAKE-HDR and $e$ for sOAKE-HDR.} We remark that the players'
public-keys in the inputs of $c, d, e$ for OAKE-HDR/sOAKE-HDR 
actually play \emph{no} role in the above security analysis. That
is, the above security analysis is actually with respect to a
(\emph{public-key free}) variant of OAKE-HDR/sOAKE-HDR,  with
public-keys are removed from the inputs of $c, d, e$. Recall that,
players' public-keys are only used for arguing the TBSS property
of OAKE-HDR/sOAKE-HDR. Specifically, for
any value $\sigma\in G\setminus 1_G$ and any  $(m_1, m_0, \hat{A}, A, 
\hat{B},  B, X_0, Y_0)$, the probability
$\Pr[\sigma_0=A^{c_0y_0}X_0^{d_0b+e_0y_0}=\sigma]\leq
\frac{1}{2^l-1}$, where $c_0=h(m_1, \hat{A}, A, Y_0)$, $d_0=h(m_0,
\hat{B}, B,
  X_0)$,      $e_0=h(X_0, Y_0)$ and the probability is taken over  only the choice of the random function $h$. But, as we assume $\hat{A}$ and
  $\hat{B}$ are both uncorrupted players, their public-keys are
  generated independently at random. Also, the value $X_0$ is the given random DH-component (not generated
  by the attacker). To affect the distribution of
  $\sigma_0$, the only freedom of the attacker is to maliciously  choose
  $(Y_0, m_0, m_1)$, which however does not change the distribution of
  $\sigma_0$. In particular, for
any value $\sigma\in G\setminus 1_G$  and  for any  $(Y_0, m_0, m_1)$ chosen maliciously by the attacker  w.r.t. the  fixed
  $(\hat{A}, A, \hat{B}, B, X_0)$, it still holds that $\Pr[\sigma_0=\sigma]\leq
  \frac{1}{2^l-1}$.

\textbf{Security of OAKE-HDR/sOAKE-HDR against the signer itself.}
The above security analysis considers the security of
OAKE-HDR/sOAKE-HDR against any other uncorrupted players other than
the signer itself, i.e., the (in)feasibility of outputting a
successful forgery  $(m_1, m_0, \hat{A}, A, 
\hat{B},  B, X_0,  Y_0, r_0)$ where $\hat{A}$ is an uncorrupted
player and $\hat{A} \neq \hat{B}$. But, the forger $\mathcal{F}$
may also be against the signer $\hat{B}$ itself. That is,
$\mathcal{F}$ may
output a successful forgery of the form: $(m_1, m_0, \hat{B}, B, 
\hat{B},  B, X_0, Y_0, r_0)$ (i.e., $\hat{A}=\hat{B}$). Here, we
further investigate the feasibility of successful forgeries of
this form. We distinguish two cases: (1) $Y_0=X_0$, i.e., the
successful forgery is of the form $(m_1, m_0, \hat{B}, B, 
\hat{B},  B, X_0, X_0, r_0)$. For this case, we show OAKE-HDR and
sOAKE-HDR are still secure under the traditional  CDH assumption  (not the
stronger GDH assumption) in the RO model; (2) $Y_0\neq X_0$. For
this case, we show OAKE-HDR and sOAKE-HDR are secure under the GDH
assumption, and additionally \emph{the KEA assumption}, in the RO
model. We remark that the KEA assumption is only used to rule out
the feasibility of successful forgeries in  this case of $Y_0\neq X_0$ and $\hat{A}=\hat{B}$.

\begin{corollary}\label{HDR-signerXX}
 \emph{Under the computational Diffie-Hellman (CDH)
assumption}, (public-key free) OAKE-HDR and sOAKE-HDR signatures of $\hat{B}$, with
offline pre-computed and exposable  $(y, Y, A^{cy})$, are \emph{strongly}
secure in the random oracle model, 
 with respect to the signer $\hat{B}$ itself with $Y_0=X_0$.

\end{corollary}

\noindent \textbf{Proof.} This case implies that the forger
$\mathcal{F}$ can output, with non-negligible probability, a
successful forgery of the form:  $(m_1, m_0, \hat{B}, B, 
\hat{B},  B, X_0, X_0, r_0)$, where $r_0=H_K(\sigma_0)$,
$\sigma_0=B^{c_0x_0}X_0^{d_0b+e_0x_0}=(X_0^{c_0}X_0^{d_0})^bX_0^{e_0x_0}$,
$c_0=h(m_1, \hat{B}, B, X_0)$, $d_0=h(m_0, \hat{B}, B,
  X_0)$,      $e_0=h(X_0, X_0)$ for OAKE-HDR (for sOAKE-HDR, $c_0=d_0=1$ and
  $e_0=h(m_1, m_0, \hat{B}, B, \hat{B}, B, X_0, X_0)$).
Note that from $\sigma_0$ and $\hat{B}$'s secret-key $b$, we can
compute $X_0^{x_0}$. But, as mentioned, 
 the hardness of computing  $X^x$ from random $X$
is equivalent to
that of the CDH problem \cite{MW96,NR04}. 

 With the above observations, we modify the algorithm $\mathcal{C}$
 depicted in Figure \ref{OAKE-DCR} and Figure \ref{sOAKE-DCR} as
 follows:

 \begin{itemize}
\item $\mathcal{C}$ knows (sets) also the private key $b$ for
$\hat{B}$. By knowing the private key $b$, $\mathcal{C}$ dispenses
with the DDH-oracle in order to make the answers to RO-queries to
be consistent.

\item After $\mathcal{F}$ outputs a successful forgery of the form $(m_1, m_0, \hat{B}, B, 
\hat{B},  B, X_0, X_0, r_0)$, satisfying the conditions F1-F3,
$\mathcal{C}$ simply computes out $X_0^{x_0}$ from $\sigma_0$ and
the private-key $b$.  Note that $\mathcal{C}$ does not need to
perform the rewinding experiments at all in this case.

 \end{itemize}

The analysis show that, in case of successful forgery against the
signer itself with $Y_0=X_0$, the security  not only is  based on
the weaker hardness assumption (say, the CDH assumption rather than the GDH assumption), but also of tighter security reduction
(to the underlying hardness assumption, say the CDH assumption
here).  \hfill $\square$

Now we consider the case of $Y_0\neq X_0$. As mentioned, it is the
only place we need to additionally use the KEA assumption.

 \begin{definition}\label{DefKEA} [Knowledge-of-Exponent Assumption (KEA)] Let $G$ be a cyclic group of prime order $q$
generated by an element $g$, and consider algorithms that on input
a triple $(g, C=g^c, z)$ output a pair   $(Y, Z)\in G^2$,  where $c$ is taken uniformly at random from $Z^*_q$ and
\emph{$z\in \{0, 1\}^*$ is an arbitrary string that is
generated independently of $C$}. Such an algorithm $\mathcal{A}$
is said to be a KEA algorithm if with non-negligible probability
(over the choice of $g, c$ and $\mathcal{A}$'s random coins)
$\mathcal{A}(g, g^c, z)$ outputs $(Y, Z)\in G^2$ such that $Z=Y^c$. Here,
$C=g^c$ is the random challenge to the KEA algorithm $\mathcal{A}$, and
$z$ captures the auxiliary input of $\mathcal{A}$ that is
independent of the challenge $C$.

We say that the KEA assumption holds over $G$, if for every
probabilistic polynomial-time (PPT) KEA algorithm
$\mathcal{A}$ for $G$ there exists another efficient algorithm
$\mathcal{K}$, referred to as the KEA-extractor,  for which the following property holds except for a negligible probability: let $(g, g^c, z)$ be an input to
$\mathcal{A}$ and $\rho$ a vector of random coins for
$\mathcal{A}$ on which $\mathcal{A}$ outputs $(Y, Z=Y^c)$, then, on
the same inputs and random coins, $\mathcal{K}(g, C, z, \rho)$
outputs the triple $(Y, Z=Y^c, y)$ where $Y=g^y$.

\end{definition}
\vspace{-0.3cm}



\begin{corollary}\label{HDR-signerXY}
 Under the GDH assumption, and additionally the KEA
 assumption, (public-key free) OAKE-HDR and sOAKE-HDR signatures of $\hat{B}$, with
offline pre-computed and exposable  $(y, Y, A^{cy})$, are \emph{strongly}
secure in the random oracle model, 
 with respect to the signer $\hat{B}$ itself with $Y_0\neq X_0$.

\end{corollary}

\noindent \textbf{Proof.} The proof of Corollary  \ref{HDR-signerXY}
follows the same outline of that of Theorem \ref{HDR-analysis}. We
 highlight the main differences, and how the KEA assumption comes
 into force in the security analysis. The analysis is mainly
 w.r.t. OAKE-HDR (the similar,  and actually simpler,  hold also for sOAKE-HDR).

 The main difference between the proof of Corollary  \ref{HDR-signerXY}
 and that of Theorem \ref{HDR-analysis} is that, here, the forger
 outputs with non-negligible probability a successful forgery of
 the form: $(m_1, m_0, \hat{B}, B, 
\hat{B},  B, X_0,\\ Y_0, r_0)$, where $r_0=H_K(\sigma_0)$,
$\sigma_0=B^{c_0y_0}X_0^{d_0b+e_0y_0}$,  $c_0=h(m_1, \hat{B}, B,
Y_0)$, $d_0=h(m_0, \hat{B}, B,
  X_0)$,      $e_0=h(X_0, Y_0)$. The key point is that, by
  performing the rewinding experiments, we cannot directly output
  the $CDH(B, X_0)$, as we do not know the private key $b$ of
  $\hat{B}$ (recall that we are going to compute $CDH(B, X_0)$ by running the forger $\mathcal{F}$). Note that in the security analysis of Theorem
  \ref{HDR-analysis}, we heavily relied on the fact that we know
  the private key of any uncorrupted player other than the signer
  itself.

  We modify the algorithm $\mathcal{C}$ depicted in Figure
  \ref{OAKE-DCR} and Figure \ref{sOAKE-DCR} as follows: the actions
  of $\mathcal{C}$ remain unchanged until the rewinding
  experiments; $\mathcal{C}$ performs the rewinding experiments
  according to the order of the RO-queries $c_0, d_0, e_0$.

  \begin{description}

\item [$d_0$ posterior to $c_0, e_0$.] In this case, by rewinding
$\mathcal{F}$ to the point of making the query $d_0=h(m_0,
\hat{B}, B,   X_0)$, and redefines $h(m_0, \hat{B}, B,   X_0)$ to
be a new independent $d^{\prime}_0$,  $\mathcal{C}$ will get
$\sigma^{\prime}_0=B^{c_0y_0}X_0^{d^{\prime}_0b+e_0y_0}$. Then,
from $\sigma_0$ and $\sigma^{\prime}_0$, $\mathcal{C}$ gets that
$CDH(B,
X_0)=(\sigma/\sigma^{\prime}_0)^{(d_0-d^{\prime}_0)^{-1}}$.
\emph{Note that, in this case, $\mathcal{C}$ does not rely on the
KEA assumption for breaking the CDH assumption} (but still with
the DDH-oracle).

\item [$c_0$ posterior to $d_0, e_0$.] In this case, by rewinding
$\mathcal{F}$ to the point of making the query $c_0=h(m_1,
\hat{B}, B,   Y_0)$, and redefines $h(m_1, \hat{B}, B,   Y_0)$ to
be a new independent $c^{\prime}_0$,  $\mathcal{C}$ will get
$\sigma^{\prime}_0=B^{c^{\prime}_0y_0}X_0^{d_0b+e_0y_0}$. Then,
from $\sigma_0$ and $\sigma^{\prime}_0$, $\mathcal{C}$ gets
$CDH(B,
Y_0)=B^{y_0}=(\sigma/\sigma^{\prime}_0)^{(c_0-c^{\prime}_0)^{-1}}$.
That is, given $B$, $\mathcal{C}$ can output $(Y_0, B^{y_0})$. By
the KEA assumption, it implies that $\mathcal{F}$ knows $y_0$
(which can be derived from the internal state of $\mathcal{F}$).
More formally, there exists an algorithm that, given $B$ and $X_0$ and the random coins of $\mathcal{C}$ and
$\mathcal{F}$ can successfully output $y_0$. Now, with the
knowledge of $y_0$, $CDH(B, X_0)$ can be derived from $\sigma_0$
(or $\sigma^{\prime}_0$).

\item [$e_0$ posterior to $c_0, d_0$.] In this case, by rewinding
$\mathcal{F}$ to the point of making the query $e_0=h(X_0, Y_0)$,
and redefines $h(X_0,   Y_0)$ to be a new independent
$e^{\prime}_0$,  $\mathcal{C}$ will get
$\sigma^{\prime}_0=B^{c_0y_0}X_0^{d_0b+e^{\prime}_0y_0}$. Then,
from $\sigma_0$ and $\sigma^{\prime}_0$, $\mathcal{C}$ gets
$CDH(X_0,
Y_0)=X_0^{y_0}=(\sigma/\sigma^{\prime}_0)^{(e_0-e^{\prime}_0)^{-1}}$.
Then, by the KEA assumption, the knowledge of $y_0$ can be
derived, with which $CDH(X_0, B)$ can then be computed  from
either $\sigma_0$ or $\sigma^{\prime}_0$.  \hfill $\square$


  \end{description}

\subsubsection{Extension to Robust (s)OAKE-HDR Signatures}\label{robustHDR}

In this section, we show that the security analysis of (s)OAKE-HDR signatures can be extended to  robust (s)OAKE-HDR signatures. We first re-describe the robust (s)OAKE-HDR signatures:

\begin{definition}  [robust (s)OAKE-HDR signatures] Let $\hat{A}$,$\hat{B}$ be two parties with public-keys $A=g^a$, $B=g^b$, respectively. 
 Let  $m_{\hat{A}}$, $m_{\hat{B}}$ be two messages.
The \textsf{\textup{robust (s)OAKE-HDR}} 
signatures of $\hat{B}$ on messages
$(m_{\hat{A}},m_{\hat{B}},\hat{A},A, \hat{B},B,X,Y)$ 
 are defined as a vector  of values
(the signatures of $\hat{A}$ 
 are defined similarly):

\begin{description}

\item [Robust OAKE-HDR.] $\{\hat{A}, A, m_{\hat{A}}, m_{\hat{B}}, X, Y,
HSIG^{OAKE}_{\hat{A},\hat{B}}(m_{\hat{A}}, m_{\hat{B}}, X,
Y)=H_K(A^{b+yc}X^{bd+ye})\}$, where $X=g^x$, $Y=g^y$ are chosen by
$\hat{A}$, $\hat{B}$ respectively as the random \emph{challenge}
and \emph{response}, $x, y \in_{\textup{R}} Z^*_q$,
$c=h(m_{\hat{A}}, \hat{A}, A, Y)$, $d=h(m_{\hat{B}}, \hat{B}, B,
X)$ and $e=h(X, Y)$.


\item [Robust sOAKE-HDR.] $\{\hat{A}, A, m_{\hat{A}},  m_{\hat{B}}, X, Y,
HSIG^{sOAKE}_{\hat{A},\hat{B}}(m_{\hat{A}},  m_{\hat{B}}, X,
Y)=H_K(A^{b+yc}X^{bd+ye})\}$, where $c=d=1$,  $e=h(m_{\hat{A}},
m_{\hat{B}}, \hat{A}, A,
\hat{B}, B,  X, Y)$. 



\end{description}

\end{definition}
For the  security analysis of the robust (s)OAKE variant, the exposed values  $A^{cy}$ and $B^{dx}$ for (s)OAKE are changed to be $A^{b+cy}$ and $B^{a+dx}$.

\textbf{Security analysis extension for the case of $\hat{A}\neq \hat{B}$.} We note that the proof of Theorem \ref{HDR-analysis} can be straightforwardly extended to robust (s)OAKE-HDR signatures, by the following observations:

\begin{itemize}
\item In Step S3 and for answering RO queries, to ensure the consistency of RO queries with each $\sigma$ previously queried by $\mathcal{F}$ to the RO $H_K$,  the challenger $\mathcal{C}$ checks whether
$\sigma=Z^{b+cy}X^{bd+ye}$ by checking whether
     $CDH(B,X^dZ)=\sigma/Z^{cy}X^{ye}=Z^bX^{db}=(X^dZ)^b$ via its DDH oracle.

 \item The repeat experiments can still go through because that:  $\hat{B}\neq \hat{A}$,  $\hat{A}$ is an uncorrupted player and the challenger knows the secret-key $a$. Thus the value $A^{b+cy}=(BY^c)^a$ can be removed from $\sigma$.
\end{itemize}

\textbf{Security analysis extension for the case of $\hat{A}=\hat{B}$ and $X=Y$.} The security analysis of robust (s)OAKE-HDR signatures for this case  is essentially the same as in the analysis of Corollary \ref{HDR-signerXX}.

\textbf{Security analysis extension for the case of $\hat{A}=\hat{B}$ and $X\neq Y$.}  The key differences, in comparison with the proof of Corollary \ref{HDR-signerXY}, are that:

 \begin{itemize}
\item For robust OAKE-HDR signature,  the output of the challenger $\mathcal{C}$ during the rewinding  experiments is $CDH(B,X_0)$ for the case of $d_0$ posterior to $c_0,e_0$, and is $CDH(X_0^{d_0}B,B)=X_0^{bd_0}B^b$ in the rest two cases.

\item For robust sOAKE-HDR signature, as $c=d=1$, the output of the challenger $\mathcal{C}$ during the rewinding  experiments is always  $CDH(X_0^{d_0}B,B)=X_0^{bd_0}B^b$.

\end{itemize}
But, either case contradicts the CDH assumption, by the following proposition:

\begin{proposition}\label{CDHv}
Given random elements $B=g^b,X=g^x\in G\setminus 1_G$, where $b,x$ are taken independently at random from $Z^*_q$,  the hardness of computing $CDH(B,X)$ is equivalent to  that of computing $CDH(X^dB,B)=(X^dB)^b$, where $d=h(\hat{B},B,X)$.

\end{proposition}

\noindent \textbf{Proof} (of Proposition \ref{CDHv}). First recall that 
 the hardness of computing  $B^b$ from random $B=g^b$
is equivalent to
that of the CDH problem \cite{MW96,NR04}. Thus,  the ability of computing $CDH(B,X)$ (given $(B,X)$)  is equivalent to the ability of computing $B^b$ (given $B$ only), which then  implies the ability of computing $CDH(X^dB,B)=X^{bd}B^b$.

Suppose there exists an efficient algorithm $\tilde{A}$  that can compute $CDH(X^d,B)=X^{db}B^b$ (from $B$ and $X$) with non-negligible probability, then there exists another efficient algorithm $\tilde{B}$ that can breaks the CDH assumption with also non-negligible probability. The input of $\tilde{B}$ is a random element $B\in G\setminus 1_G$, and its goal is to break the CDH assumption by computing  $CDH(B,B)=B^b$.  Towards this goal, $\tilde{B}$ generates $X=g^x$ where $x$ is taken uniformly at random from $Z^*_q$, and then runs $\tilde{A}$ on input $(B,X)$. After getting $CDH(X^dB)=X^{db}B^b=B^{xd}B^b$ from the output of $\tilde{A}$, $\tilde{B}$ computes $B^b=CDH(X^dB,B)/B^{xd}$.

According to the above discussions, given random elements $(B,X)$, under the CDH  assumption no efficient algorithm can compute either $CHD(B,X)$ or $CDH(X^dB,B)$ with non-negligible probability. \hfill $\square$

In addition, in view of the fact that  $c=d=1$ for  robust sOAKE-HDR signature, there is another analysis method for robust sOAKE-HDR signature. 
Specifically, given random elements $U,V$, the challenger $\mathcal{C}$ sets (in the Setup procedure) that: $B=V$ and $X_0=(U/B)$ (rather than $X_0=U$).  Note that, in this case, the output of the challenger $\mathcal{C}$ during the rewinding experiments is $CDH(X_0^{d_0}B,B)=X_0^{b}B^b=\frac{U^b}{B^b}B^b=U^b=CDH(U,B)$, which directly violates the GDH assumption.

%
%

\subsection{Analysis of  (s)OAKE with Offline  Pre-Computation  in the
CK-Framework}\label{basicanalysis}
\textbf{Brief description of the CK-framework.} In the CK-framework for a DHKE protocol, a CMIM adversary $\mathcal{A}$ controls all the communication channels among concurrent session runs of the  KE protocol. In addition, $\mathcal{A}$ is allowed access to secret information via the following three types of queries: (1) state-reveal queries for ongoing incomplete sessions; (2) session-key queries for completed sessions; (3) corruption queries upon which all information in the memory of the corrupted parties will be leaked to $\mathcal{A}$. A session $(\hat{A},\hat{B},X,Y)$ is called \emph{exposed},  if it or its matching session $(\hat{B},\hat{A},Y,X)$ suffers from any of these three queries.

The session-key security (SK-security) within the CK-framework is captured as follows: for any complete session $(\hat{A},\hat{B},X,Y)$ adaptively selected by $\mathcal{A}$, referred to as the \emph{test session}, as long as it is unexposed it holds with overwhelming probability  that (1) the session-key outputs of the test session and its matching session are identical; (2) $\mathcal{A}$ cannot distinguish the session-key output of the test session from a random value.
At a high level,
the SK-security essentially says that a party that completes a
session has the following guarantees \cite{CK01}: (1) if the peer to the
session is uncorrupted then the session-key is unknown to anyone
except this peer; (2) if the \emph{unexposed} peer
completes a matching session then the two
parties have the same shared key.

Next, we present the
 analysis of OAKE and sOAKE protocols in the CK-framework \emph{with pre-specified peers}, with
 offline pre-computed and exposable  DH-exponents, DH-components, and DH-secrets derived from 
  one's DH-component and its peer's public-key (say, $A^{cy}$ and $B^{dx}$) which may be exposed to the
 adversary prior to the session involving these pre-computed
 values. The analysis can also be straightforwardly extended to that of the robust (s)OAKE variant, where the exposed value $A^{cy}$ and $B^{dx}$ are changed to be $A^{b+cy}$ and $B^{a+dx}$.

Using the terminology of HDR signatures, 
a session of OAKE (resp., sOAKE), for the basic protocol version
without explicit mutual identifications and key confirmations,
between two parties $\hat{A}$ and $\hat{B}$ consists of a basic
Diffie-Hellman exchange of DH-components $X=g^x$ and $Y=g^y$; And
 the session-key $K$ is then  computed as the corresponding HDR-signatures,
 specifically,   
$K=HSIG^{OAKE}_{\hat{A}, \hat{B}}(m_{\hat{A}}, m_{\hat{B}}, X, Y)$
for OAKE and (resp., $K=HSIG^{sOAKE}_{\hat{A}, \hat{B}}(m_{\hat{A}},
m_{\hat{B}}, X, Y)$ for sOAKE), where \emph{ $m_{\hat{A}}$ and
$m_{\hat{B}}$ are the  empty string for both OAKE and sOAKE.}

During a session of (s)OAKE
within the CK-framework,  with offline pre-computation, a party can be activated with three types
of activations (for presentation simplicity, we assume $\hat{A}$
denotes the identity of the party being activated and $\hat{B}$
the identity of the intended peer to the session):

\begin{description}
\item \emph{Initiate}($\hat{A}$, $\hat{B}$) (i.e., $\hat{A}$ is
activated as the initiator): $\hat{A}$ generates a value $X=g^x$,
$x\in_{\textup{R}} Z^*_q$, creates a local session of the protocol
which it identifies as (the incomplete) session $(\hat{A},
\hat{B}, X)$, and outputs the DH-component $X$ as its outgoing
message.

Here $(X, x, B^{dx})$, where $d=h(\hat{B}, B, X)$ for OAKE or $d=1$
for sOAKE can be offline pre-computed by $\hat{A}$, which may be
exposed to the adversary prior to the session involving them.

\item \emph{Respond}($\hat{A}, \hat{B}, Y$) (i.e., $\hat{A}$ is
activated as the responder): $\hat{A}$ checks $Y\in G\setminus 1_G$, if so
it generates a value $X=g^x$, $x\in_{\textup{R}} Z^*_q$, outputs
$X$, computes the session-key and then  completes the session
$(\hat{A}, \hat{B}, X, Y)$.

Again,  $(X, x, B^{dx})$  can be offline pre-computed by
$\hat{A}$, which may be exposed to the adversary prior to the
session involving them.

\item \emph{Complete}($\hat{A}, \hat{B}, X, Y$) (i.e., the
initiator $\hat{A}$ receives $Y$ from the responder peer
$\hat{B}$): $\hat{A}$ checks that $Y\in G\setminus 1_G$ and that it has an
open session with identifier $(\hat{A}, \hat{B}, X)$. If any of
these conditions fails $\hat{A}$ ignores the activation, otherwise
it computes the session-key and completes the session $(\hat{A},
\hat{B}, X, Y)$.

\end{description}

With the above notation, it is ensured that if $(\hat{A}, \hat{B},
X, Y)$ is a complete session at $\hat{A}$,  then its matching
session (if it exists) is unique, which is $(\hat{B}, \hat{A}, Y, X)$ owned by the player $\hat{B}$.  
In the following analysis,  we specify that the values,
exposable to the adversary via session-state query (against an incomplete session), include the DH-component and DH-exponent and the DH-secret of one's DH-component and its peer's
public-key, e.g., $(Y, y, A^{cy})$. 

\begin{theorem} \label{YZanalysis}
Under the GDH assumption in the RO model, the OAKE and sOAKE
protocols (actually, the variants with public-keys  removed
from the inputs of $c, d, e$), with offline pre-computed
DH-components, DH-exponents, and the DH-secrets of one's
DH-component and its peer's public-key (say $A^{cy}$ and
$B^{dx}$), are SK-secure  in the CK-framework w.r.t. any
test-session between a pair of  different players.
\end{theorem}

\noindent \textbf{Proof.} According to the SK-security definition
in the CK-framework, we need to prove OAKE and sOAKE satisfy the
following two requirements:

\begin{description}
\item [Requirement-1.]  If two parties $\hat{A}, \hat{B}$ complete
matching sessions, then their session-keys are the same.

\item [Requirement-2.]  Under the GDH assumption, there is no
feasible adversary that succeeds in distinguishing the session-key
of an unexposed session with non-negligible probability.

\end{description}

 The Requirement-1 can be  trivially checked for both OAKE and
 sOAKE. In the following, we focus on establishing the
 Requirement-2.

 Denote by $(\hat{A}, \hat{B}, X_0, Y_0)$  the unexposed
 test-session between a pair of \emph{uncorrupted} players $\hat{A}$ and $\hat{B}$  where $\hat{A}\neq \hat{B}$, and by $H_K(v)$  the session-key of the
 test-session that is referred to as the \emph{test} HDR-signature, where $v=A^{cy}X^{db+ey}=B^{dx}Y^{ca+ex}$. As $H_K$
 is a random oracle, there are only two
 strategies for the adversary $\mathscr{A}$ to distinguish
 $H_K(v)$ from a random value:


 \begin{description}
\item [Key-replication attack.] $\mathscr{A}$ succeeds in forcing
the establishment of a session (other than the  test-session or
its matching session) that has the same session-key output as the
test-session. In this case, $\mathscr{A}$ can learn the
test-session key by simply querying the session to get the same
key (without having to learn the value of the test HDR-signature).

\item [Forging attack.] At some point in its run, $\mathscr{A}$
queries  the RO $H_K$ with the value $v$. This implies that
$\mathscr{A}$ succeeds in computing or learning the test
HDR-signature (i.e., the session-key of the test-session) via its
attacks. For presentation simplicity, we assume $\mathscr{A}$
directly outputs the session-key of the test-session, referred to as  the
test-signature, via a successful forging attack.

 \end{description}

\label{tighter} The possibility of key-replication attack is trivially ruled out \emph{unconditionally} in the RO model,
 by the NMJPOK and TBSS property of OAKE and sOAKE.
 Specifically, for any session-tag $(\hat{A}, A, \hat{B}, B, X, Y)$ and  for any value $\sigma\in
G\setminus 1_G$, the probability
$\Pr[K_{\hat{A}}=K_{\hat{B}}=\sigma]\leq \frac{1}{2^l-1}$ holds for both OAKE and sOAKE, where the probability is taken over only the choice of the random function $h$.
Then, by the birthday paradox (as done in the previous NMJPOK and computational fairness analysis), any efficient attacker can succeed in the key-replication attack only with negligible probability.
  Actually, as the test-session and its matching session are defined without taking public-keys into account in the CK-framework, the possibility of  key-replication attack is trivially ruled out \emph{unconditionally} in the RO model also for the public-key free variant of (s)OAKE. Specifically,  for any  test-session $(\hat{A},\hat{B},X,Y)$ and any session $(\hat{A}^\prime, \hat{B}^\prime, X^\prime, Y^\prime)$ that is unmatched to the test-session (which implies that at least of the following inequalities holds: $\hat{A}\neq \hat{A}^\prime$, $\hat{B}\neq \hat{B}^\prime$, $X\neq X^\prime$ and $Y\neq Y^\prime$), it holds that $\Pr[K_{\hat{A}}=K_{\hat{A}^\prime}]=\frac{1}{2^l-1}$. As the attacker is polynomial-time, it cannot make two unmatched  sessions to output the same session-key with non-negligible probability.

 \textbf{Note on security reduction tightness.}
   We
 note that, however, the analysis of HMQV to rule out
 key-replication attack in \cite{K05} is quite complicated, and is still reduced to the
 underlying hardness assumptions
 (to be precise, to the unforgeability of HMQV-HDR). That
 is, the analysis of (s)OAKE in order to rule out the key-replication
 attacks is not only much simpler, but also does not go through  costly security reductions.
  Also, as we shall see,  sOAKE is at least as tight as HMQV in other parts of the security analysis. We did not try to make a direct comparison on  the security reduction tightness between OAKE and HMQV, as they use different forking lemma.

 Then,
 in the following analysis, we only focus on ruling out the forging attack.
Recall that $\hat{A}\neq \hat{B}$ for the test-session  $(\hat{A},
\hat{B}, X_0, Y_0)$ held by $\hat{A}$.
 In the rest,  we make analysis mainly
 with respect to the OAKE protocol, the similar and actually simpler hold also for sOAKE.


 Now,  suppose there is an efficient  KE-attacker $\mathscr{A}$
 who succeeds,  by  forging attacks,  against
 the test-session $(\hat{A}, \hat{B}, X_0, Y_0)$ with $\hat{A}\neq
 \hat{B}$ (particularly, $A\neq B$), we present  an efficient
 forger $\mathcal{F}$  against the underlying OAKE-HDR signature, which contradicts
 the security of the underlying OAKE-HDR signature scheme (that is
 based on the GDH assumption), and thus establishing the theorem.
 $\mathcal{F}$ works as follows, by running $\mathscr{A}$ as a
 subroutine.

\begin{enumerate}

\item We assume $\mathcal{F}$ successfully guessed the
\emph{unexposed} test-session $(\hat{A}, \hat{B}, X_0, Y_0)$ held at
$\hat{A}$, where $\hat{A}\neq \hat{B}$.

\item The inputs of $\mathcal{F}$ are $(B, X_0)$, and
$\mathcal{F}$ has  oracle access to the OAKE-HDR signer $\hat{B}$
of public-key $B$.

\item $\mathcal{F}$ sets the inputs to all parties other than
$\hat{B}$, and thus can perfectly emulate these parties. In
particular, $\mathcal{F}$ can deal with state-reveal queries,
session-key queries by $\mathscr{A}$ on any session other than the
test-session and its matching session, and  party corruption
queries on any party other than $\hat{A}$ and $\hat{B}$.

\item 
 When $\mathscr{A}$
activates a session at $\hat{B}$, either as a responder or
initiator, with peer identity $\hat{P}$ of public-key $P$  and
incoming message $X$, then $\mathcal{F}$  feeds $\hat{B}$ the
value $(\hat{P}, P, X)$. In response, $\mathcal{F}$ gets  values
$(y, Y, P^{cy})$ from $\hat{B}$, and then $\mathcal{F}$ hands
$\mathscr{A}$ the value  $Y$   as the outgoing message from
$\hat{B}$. 
 Actually, the
values $(y, Y, P^{cy})$ 
 can be offline pre-computed by $\hat{B}$,
 and leaked to $\mathcal{F}$ (and $\mathscr{A}$) prior to the
session involving them.

\item When $\mathscr{A}$ issues a state-reveal query against an
incomplete  session $(\hat{B}, \hat{P}, Y)$ (\emph{not matching
to the test-session}) held at $\hat{B}$, then $\mathcal{F}$
returns the values $(Y, y, P^{cy})$ to $\mathscr{A}$.


\item When $\mathscr{A}$ issues a session-key query to a session
$(\hat{B}, \hat{P}, Y, X)$ (\emph{not matching to the
test-session}) held at $\hat{B}$, then $\mathcal{F}$ queries the
session-signature from its signing oracle $\hat{B}$ by presenting
the signing oracle with $(\hat{P}, P, X, Y)$, and returns the
HDR-signature from $\hat{B}$ to $\mathscr{A}$.

\item When $\mathscr{A}$ halts with a valid test-signature,
denoted $\sigma_0$, $\mathcal{F}$ stops and outputs $\sigma_0$.

\end{enumerate}

Suppose there are $n$ parties in total in the system, and each
party is activated at most $m$ times (where $n$ and $m$ are
polynomials in the security parameter), in actual analysis
$\mathcal{F}$ guesses the test-session by choosing uniformly at
random a triple $(\hat{P}_i, \hat{P}_j, t)$ (hoping that
$\hat{P}_i=\hat{A}$ and $\hat{P}_j=\hat{B}$ and the test-session
is the $t$-th session activated at $\hat{A}$ with peer $\hat{B}$),
where $1\leq i\neq j \leq n$ and $1\leq t \leq m$. Thus, with
probability $(n^2m)^{-1}$, $\mathcal{F}$ successfully guesses the
test-session. It is easy to check that, conditioned on
$\mathcal{F}$ successfully checks the test-session, the view of
$\mathscr{A}$ under the run of $\mathcal{F}$ is identical to that
in the  real run of $\mathscr{A}$. Suppose $\mathscr{A}$ successfully
outputs, with non-negligible probability $\varepsilon$, the valid
test-signature via forging attack in its real run, with still non-negligible
probability $(n^2m)^{-1}\varepsilon$ 
 $\mathscr{A}$ (and thus $\mathcal{F}$) outputs the
valid test-signature under the run of $\mathcal{F}$.

We need then to check whether the valid test HDR-signature
outputted by $\mathcal{F}$  is a \emph{successful} OAKE-HDR
forgery. As the test-signature  output by $\mathscr{A}$ is valid,
according to Definition \ref{DCRdef-SEC}, we only need to show the
vector $\{\hat{A}, A,  X_0, Y_0\}$ did not appear in any one of the  responses
from the signing oracle $\hat{B}$.   We distinguish three cases,
according to the appearance of $Y_0$:


\begin{description}

\item [Case-1.] $Y_0$ was never output in any one of the signatures
issued by $\hat{B}$. In this case, the test HDR-signature output
by $\mathscr{A}$ (and thus $\mathcal{F}$)  is clearly a successful
forgery against OAKE-HDR.

\item [Case-2.] $Y_0$ was output in one of the signatures issued
by $\hat{B}$ in a session \emph{non-matching} to the test-session.
Denote by $(\hat{B}, \hat{P}, Y_0, X)$ this non-matching session,
we have that $\hat{P}\neq \hat{A}$ or $X\neq X_0$. That is,
$(\hat{P}, P, X)\neq (\hat{A}, A, X_0)$. As  $\hat{B}$ uses  random
and independent DH-components in each session, the   value $Y_0$ is
only used in this  non-matching session $(\hat{B}, \hat{P}, Y_0,
X)$, and thus does not appear (except for a negligible probability
of accidental repetition) in any other signatures issued by
$\hat{B}$ in other sessions different from $(\hat{B}, \hat{P},
Y_0, X)$. Putting all together, we get that $\{\hat{A}, A,  X_0,
Y_0\}$ did not appear in any of the HDR-signatures issued by
$\hat{B}$, and thus the test HDR-signature output by $\mathcal{F}$
is a successful forgery against OAKE-HDR.

\item [Case-3.] $Y_0$ was generated by $\hat{B}$ in the
\emph{matching} session $(\hat{B}, \hat{A}, Y_0, X_0)$.  However,
this matching session was never queried by $\mathscr{A}$ via
session-key query or session-state query  (recall we assume the
test-session and its matching session are unexposed in the
CK-framework), which in turn implies that $\mathcal{F}$ never
queries $\hat{B}$ for the HDR-signature of this matching session.
Also,  the random value $Y_0$ was used by $\hat{B}$ only for
this matching session (except for a negligible probability of
accidental repetition). This implies that, in Case-3,  the values
$\{\hat{A}, A,  X_0, Y_0\}$  also did not appear in any one of the responses from
the signing oracle $\hat{B}$, and thus the test HDR-signature
output by $\mathcal{F}$ is a successful forgery against OAKE-HDR.
\hfill $\square$

\end{description}

\textbf{Notes on the security analysis of (s)OAKE in the CK-framework.} For the above security analysis of (s)OAKE in the
CK-framework, we have the following observations and
notes:

\begin{itemize}

\item For the same security level (actually,  whenever the
DH-component is offline pre-computed and exposable, no matter whether the secret
DH-exponent is exposable or not), the security of HMQV in the
CK-framework relies on both the GDH assumption and the KEA
assumption. In contrast, for the
security of (s)OAKE \emph{even with the additional powerful exposure of DH-exponents and
$A^{cy}$ or $B^{dx}$},  the KEA assumption is dispensed with.

\item The security reduction (from the security of sOAKE to the
security of the underlying HDR signatures) is 
tighter than that of HMQV.

We remind  problems with security reduction in the random oracle model \cite{CGH98,N02,P03,CGH04}. Here, we only aimed to highlight the relative advantage of reduction tightness of sOAKE over HMQV, as  both HMQV and (s)OAKE are proved in the random oracle  model.


\item Note that the above security analysis is actually w.r.t. the
\emph{public-key free} variants of (s)OAKE, with players'
public-keys  removed from the inputs of the functions of $c, d,
e$. The reason is that the security of the underlying OAKE-HDR/sOAKE-HDR
signatures does not rely on them.

\item The analysis shows that OAKE and sOAKE remain their security in the
CK-framework, even if the attacker $\mathscr{A}$ exposes the private  values $(y, A^{cy})$  of the  matching session (but not the
session-key itself). This provides extra security guarantee of (s)OAKE that is beyond the CK-framework.  The reason is that, 
 even if these
pre-computed private values are used by $\hat{B}$ in the matching session
$(\hat{B}, \hat{A}, Y_0, X_0)$ and exposed to $\mathscr{A}$, the forger $\mathcal{F}$  never queries the \emph{full}
HDR-signature corresponding to this matching session as the underlying attacker $\mathscr{A}$ is not allowed to make the session-key query against the matching session (note that $\mathcal{F}$ queries the HDR signer for  a full session-signature only when $\mathscr{A}$ makes the session-key query against this session),  
and
thus $(\hat{A}, A, X_0, Y_0)$ still did not appear in any one of the
signatures issued by $\hat{B}$. 

\end{itemize}

Using Corollary \ref{HDR-signerXX} and Corollary \ref{HDR-signerXY}, we
have the following corollaries about the security of (s)OAKE in the
CK-framework w.r.t. any  test-session between the identical players
$\hat{A}=\hat{B}$. The proofs are straightforward adaptations  of the proof of Theorem \ref{YZanalysis}, and details are omitted here.

\begin{corollary}
Under the CDH assumption in the RO model, the OAKE and sOAKE
protocols (actually, the variants with public-keys  removed
from the inputs of $c, d, e$), with offline pre-computed and exposable DH-components, DH-exponents,
and the DH-secrets of one's
DH-component and its peer's public-key (say $A^{cy}$ and
$B^{dx}$),  are SK-secure  in the CK-framework w.r.t. any
test-session of identical peer and identical DH-component (i.e.,
$\hat{A}=\hat{B}$ and $X=Y$).
\end{corollary}

\begin{corollary}
Under the GDH assumption and additionally the KEA assumption in
the RO model, the OAKE and sOAKE protocols (actually, the variants
with public-keys  removed from the inputs of $c, d, e$), with
offline pre-computed and exposable  DH-components, DH-exponents, and the DH-secrets of one's
DH-component and its peer's public-key (say $A^{cy}$ and
$B^{dx}$),  are SK-secure  in the CK-framework w.r.t.
any test-session of identical peer but different  DH-components
(i.e., $\hat{A}=\hat{B}$ but $X\neq Y$).
\end{corollary}

\textbf{Notes on some inherent security limitations.} The reader should
beware of some inherent security limitations for any one-round and
two-round implicitly-authenticated DHKE protocols, e.g., the PFS
vulnerability for any  two-round implicitly-authenticated DHKE and the KCI
vulnerability for any one-round 
 DHKE
 (more
details are referred to \cite{K05}). Even for the three-round
version of OAKE (as well as HMQV) with explicit mutual authentications, there are also some inherent
limitations. For  example,  the protocol responder may not be able
to get  deniability in a fair way, in case  the malicious protocol
initiator just aborts after receiving the second-round message;
Also, both the three-round OAKE and (H)MQV suffer from the
cutting-last-message attack \cite{M04}, etc.  We remark that
losing 
 deniability fairness to protocol responder and lacking correct delivery guarantee of the last message 
are inherent to the protocol structure of OAKE and (H)MQV and do
not violate the definition of the SK-security in the CK-framework, which though can be easily
remedied but at the price of ruining the performance advantages and/or adding additional system complexity.

\section{Security of (s)OAKE Beyond the CK-framework}\label{beyondCK}

Following Section \ref{SecBeyondCK}, in this section we make some further investigations on the  security properties of (s)OAKE
not captured by the CK-framework, which further strengthens the security guarantee of the (s)OAKE protocols. 
 The first observation is:   the security analysis of (s)OAKE in the CK-framework also implies that (s)OAKE is resistant to reflection attacks. 

\subsection{Security with Public
Computations}\label{publiccomputation}

The work of \cite{KP06} considers a new attack scenario for
key-exchange protocols with public computations, where it is
convenient to split an entity (performing a run of KE-protocol)
into two parts:  a trusted  authentication device, and an
untrusted  computing device. The authentication device enforces
the confidentiality of the authentication data, while some
computing operations required by the protocol are \emph{publicly}
carried out by the (possibly untrusted)  computing device. This
allows to use an authentication device with little computing
power, and to make computing devices independent from users
\cite{KP06}.

The work \cite{KP06} gives some concrete applications that might
be benefited from public computations: (1)  Mobile phones include
smart cards which store the user authentication data; the handsets
themselves are the computing devices. (2) PCs (corresponding to
the computing device) equipped with a crypto token (corresponding
to the authentication device)  have a lot more computing power
than the token itself, but may be plagued by spyware or virus. For
more details, the reader is referred to \cite{KP06}.

\textbf{(H)MQV with public computations.} With the computation of
$\hat{B}$ as an example (the same holds for $\hat{A}$), a natural
split of authentication computation and public computation  is as
follows \cite{KP06}: The authentication device generates $(y, Y)$, forwards
$Y$ to the computation device; After getting $(\hat{A}, X)$ from
the computation device, the authentication device computes
$s=y+eb$, where $e=h(Y, \hat{A})$, and then forwards $s$ to the
computation device; After getting $s$ from the authentication
device, the computation device computes $K_{\hat{B}}=(XA^d)^s$,
and then the session-key, and then communicate with $\hat{A}$ with
the session-key.

One key point is: as we assume the computation device may not be
trustful, once the value $s$ is leaked to an attacker (who may
compromise the computation device), then the attacker can definitely
impersonate $\hat{B}$ to $\hat{A}$ in any sessions. Note that, by
only compromising the computation device, the attacker does not
learn the DH-exponent $y$ and the private-key $b$. This shows that
(H)MQV does not well support deployment in the  public computation model.

\textbf{(s)OAKE with public computations.} For applications in
such scenarios,   the natural split of authentication computation
and public computation for (s)OAKE is as follows, with the
computation of $\hat{B}$ as an example (the similar hold for
$\hat{A}$): (1)  The authentication device generates  $(y, Y)$ and
possibly $A^{cy}$ (in case the authentication device has learnt the peer identity  $\hat{A}$)
 where $c=1$ for sOAKE or $c=h(\hat{A}, A,  Y)$
for OAKE, and then forwards $Y$ and possibly $A^{cy}$ to the
computation device; (2) After getting $X$ from the computation
device, the authentication device computes $s=db+ey$, where
$d=h(\hat{B}, B,  Y)$ and $e=h(X, Y)$  for OAKE (resp., $d=1$ and
$e=h(\hat{A}, A, \hat{B}, B, X, Y)$ for sOAKE), and then forwards
$s$ to the computation device; (3) After getting $s$ from the
authentication device, the computation device computes
$K_{\hat{B}}=A^{cy}X^s$, and then the session-key, and then
communicate with $\hat{A}$ with the session-key. \emph{Note that
$y, Y, c, d, A^{cy}, db$ can be offline pre-computed by the
authentication device, and the authentication device can only
online compute $ey$ and $s$. Also, the computation device
essentially needs to compute only one exponentiation $X^s$.}

Below, we make some discussions about the security of sOAKE and OAKE
in the public computation model.\footnote{We note that some  modifications to (s)OAKE may be needed to give  a formal proof in the public computation model, in accordance with the work of \cite{KP06}. Here, we stress that (s)OAKE, particularly sOAKE, very well  supports the public-computation model even without such modifications.}

\textbf{Discussion on security of sOAKE with public computations.} We note that,
under the DLP assumption,  the knowledge of $(A^{y}, s)$ of a
session of sOAKE, learnt by the adversary by compromising the
computation device, is essentially useless for the attacker to violate other
sessions other than the matching session $(\hat{B}, \hat{A}, Y,
X)$. The reason is that $s=b+ey$ for sOAKE, where $e=h(\hat{A}, A,
\hat{B}, B,  X, Y)$ commits to the whole session-tag. Thus, the
value $s$ cannot be used by the attacker to violate a non-matching
session, unless it can compute  $y$ from $A^y$ (and thus $b$
from $s$) which however is infeasible by the DLP assumption.

\textbf{Discussion on security of OAKE with public computations.} The knowledge
$(A^{cy}, s)$  of a session of OAKE, where $s=db+ey$, $d=h(\hat{B},
B,  Y)$ and $e=h(X, Y)$, is essentially  useless under the DLP assumption  for
the attacker to violate other sessions other than sessions of the
tag $(\hat{A}^*, A^*, \hat{B}, B, X, Y)$ where $(\hat{A}^*, A^*)$
may be different from $(\hat{A}, A)$. As the DH-component $X$ is
generated by uncorrupted players randomly and independently, it
implies that the knowledge of $(A^{cy}, s)$ can only help the
attacker to violate the security of \emph{at most one unexposed
non-matching session}.

For example, consider that  the attacker interacts concurrently with
$\hat{A}$ (in the name of $\hat{B}$) and $\hat{B}$ (in the name of
$\hat{A}^*\neq \hat{A}$ but of the same public-key $A$); the attacker
faithfully relays the DH-components $X$ and $Y$ in the two
sessions; in case the    attacker learns both $s$ and the
private-key $a$ of $\hat{A}$, then it  can impersonate $\hat{B}$
to $\hat{A}$ in \emph{the unique  session in which $\hat{A}$ sends
$X$}.

We remark this weakness is at the price of supporting the
advantageous post-ID computability offered by OAKE.
 Though this weakness can be
trivially remedied (by putting $\hat{A}$ into $d$ and $\hat{B}$
into $c$), but at the price of  sacrificing the advantage of
post-ID computability.
Even with this (seemingly  unreasonable)
weakness in the public computation model  for OAKE in mind, the
potential damage caused is still much mitigated in
comparison with that  of (H)MQV in such scenarios.

\subsection{Resistance to KCI,  and Weak PFS}

  Recall that the security of DHKE protocols in the
  CK-framework is w.r.t. an unexposed  test-session $(\hat{A}, \hat{B}, X_0,
  Y_0)$, where $\hat{A}$ and $\hat{B}$ are uncorrupted parties
  (which implies both the private-keys $a, b$ are not exposed to the
  attacker) but the value $Y_0$ may be generated by the attacker impersonating $\hat{B}$ (in this case,
  the matching session does not exist). 
In this       section, we consider the security damage caused by
compromising static secret-keys of players, i.e., one or both of
the secret-keys $a, b$ of the test-session are exposed to the
attacker.

Firstly, we note that if both the peer $\hat{B}$ (in the
test-session) is corrupted and the value $Y_0$ is generated by the
attacker itself, then no security can be guaranteed for the
test-session within the CK-framework (as the attacker can now
compute  the session-key by itself). In this section, we mainly
investigate the resistance against key-compromise impersonation
(KCI) attacks, and perfect
forward security (PFS).  
 Roughly speaking, a key-compromise impersonation attack
is deemed successful if the attacker, knowing the private key $a$
of a party $\hat{A}$ (which of course allows the attacker to
impersonate $\hat{A}$), is able to impersonate another
\emph{different} uncorrupted party $\hat{B}\neq \hat{A}$ (for
which the attacker does not know the secret-key $b$)  to
$\hat{A}$. Note that for KCI attacks, the attacker still can
generate the DH-component $Y_0$ for the test-session (without the
matching session then). The PFS property says that  the leakage of
the static secret-key of a party should not  compromise the
security of session-keys ever established by that party, and
erased from memory before the leakage occurred. 

\begin{definition} [clean session
\cite{K05}] We say that a complete session of a key-exchange
protocol is \textsf{clean}, if the attacker did not have access to
the session's state at the time of session establishment (i.e.,
before the session is complete), nor it issued a session-key query
against the session after completion.

\end{definition}

Note that, for a \emph{clean} session at an uncorrupted  party,
the attacker did not issue a state-reveal query while the session
was incomplete or a session-key query after completion; Moreover,
the attacker was not actively controlling or impersonating the
party during the session establishment (neither by making any
choices on behalf of that party in that session or eavesdropping
into   the session's state). 

\begin{definition} \cite{K05}
We say that a KE-attacker $\mathscr{A}$ that has learned the
static secret-key of $\hat{A}$ succeeds in a KCI attack against
$\hat{A}$, if $\mathscr{A}$ is able to distinguish from random the
session-key of  a complete session at $\hat{A}$ for which the
session peer $\hat{B}\neq \hat{A}$ is uncorrupted (which implies
the private-key of $\hat{B}$ is not exposed to $\mathscr{A}$)  and
the session and its matching session (if it exists) are clean.
\end{definition}

In other words, the definition says that, as long as the attacker
is not actively controlling or observing the secret choices
(particularly the ephemeral DH-exponent $x$) of the test-session,
then even the knowledge of $\hat{A}$'s private-key still does not
allow $\mathscr{A}$ to compromise the session-key. In particular,
in such a protocol $\mathscr{A}$ cannot impersonate an uncorrupted
party $\hat{B}$ to         $\hat{A}$ in a way  that allows
$\mathscr{A}$ to learn any information about the resultant
session-key \cite{K05} (even if the attacker impersonates
$\hat{B}$ and generates the DH-component, say $Y_0$, by itself).

\begin{proposition}
 Under the GDH assumption in the random oracle model, the OAKE and sOAKE protocols (actually, their public-key free variants),
  with offline pre-computation,  resist KCI
 attacks in the CK-framework.

\end{proposition}

The resistance of (s)OAKE to KCI attacks is essentially implied
by the proof of Theorem \ref{HDR-analysis} and the proof of
Theorem \ref{YZanalysis}, from the observations that: for KCI
attacks the test-session is of different uncorrupted  peers
$\hat{A}\neq \hat{B}$, and the security of the underlying
OAKE-HDR/sOAKE-HDR hold even if the forger learns the private-key of
the  uncorrupted peer (the party $\hat{A}$ here).


\textbf{Weak PFS (wPFS).} It is clarified  in \cite{K05} that, no
2-round DHKE protocols \emph{with implicit key confirmation} can fully
render PFS security (the 3-round versions of HMQV and (s)OAKE,
with explicit key-confirmation and mutual authentications,  do
fully  provide PFS property). The work \cite{K05} formulates a
weak notion of PFS, named weak PFS (wPFS), and shows that HMQV
satisfies this wPFS property. Roughly speaking, wPFS property says
that \emph{if the attacker is not actively involved with the
choices of $X, Y$ at a session} (particularly if it does not get
to choose or learn the DH-exponent $x$ or $y$), then the resultant
session-key does enjoy forward security. Formally,

\begin{definition}\cite{K05}  A key-exchange protocol provides
wPFS,  if an attacker $\mathscr{A}$ cannot distinguish from random
the key of any clean  session $(\hat{A}, \hat{B}, X, Y)$, where
$Y$ is also generated by an uncorrupted party in a clean session,
even if $\mathscr{A}$ has learned the private keys of both
$\hat{A}$ and $\hat{B}$. 

\end{definition}


\begin{proposition}
 Under the CDH assumption (rather than the stronger GDH assumption), the OAKE and sOAKE protocols  provide wPFS property in the random oracle model.

\end{proposition}

For establishing the  wPFS property for (s)OAKE,  we do not need
here  to construct a OAKE-HDR/sOAKE-HDR forger from the attacker
violating the wPFS property. Actually,  we can directly reduce the
loss of wPFS to the CDH assumption, from the following
observations: given the knowledge of both $a$ and $b$, the
computation of $K_{\hat{A}}$ or $K_{\hat{B}}$ is reduced to the
computation of $g^{xye}$ from the random DH-components $X, Y$.
Recall that, for wPFS property, we assume the attacker is not
actively involved with the  choices of $X, Y$. Then, we can simply
guess the test-session, and set the DH-components as some random
elements $X, Y$, and then reduce the ability of the attacker to
violate wPFS \emph{directly} to the CDH assumption. More details
are omitted here.

\end{document}